\documentclass[iop,apj,numberedappendix]{emulateapj}
\usepackage{natbib}
\usepackage{times}
\usepackage{comment}
\usepackage{graphicx}
\usepackage{color}
\usepackage{gensymb} 

\newcommand{\eos}{EoS}
\newcommand{\msun}{\ensuremath{M_\sun}}
\newcommand{\msuns}{\ensuremath{M_\sun\,{\rm s}^{-1}}}
\newcommand{\tpb}{\ensuremath{t_{\rm pb}}}
\newcommand{\tadv}{\ensuremath{\tau_{\rm adv}}}
\newcommand{\theat}{\ensuremath{\tau_{\rm heat}}}
\newcommand{\tconv}{\ensuremath{\tau_{\rm conv}}}

\newcommand{\nue}{\ensuremath{\nu_{e}}}
\newcommand{\nuebar}{\ensuremath{\bar \nu_e}}

\newcommand{\numt}{\ensuremath{\nu_{\mu\tau}}}
\newcommand{\numtbar}{\ensuremath{\bar \nu_{\mu\tau}}}
\newcommand{\numu}{\ensuremath{\nu_{\mu}}}
\newcommand{\nutau}{\ensuremath{\nu_{\tau}}}
\newcommand{\numubar}{\ensuremath{\bar \nu_{\mu}}}
\newcommand{\nutaubar}{\ensuremath{\bar \nu_{\tau}}}
\newcommand{\mev}{\mbox{MeV}}

\newcommand{\rgain}{\ensuremath{r_{\rm gain}(\theta)}}
\newcommand{\dshock}{\ensuremath{d_{\rm shock}}}
\newcommand{\Rshock}{\ensuremath{R_{\rm shock}}}
\newcommand{\rshock}{\ensuremath{r_{\rm shock}(\theta)}}
\newcommand{\Ediag}{\ensuremath{E^{+}}}
\newcommand{\Ediagapprox}{\ensuremath{E^{+}_{\rm est}}}
\newcommand{\Eth}{\ensuremath{E_{\rm th}}}

\newcommand{\Ediagov}{\ensuremath{E^{+}_{\rm ov}}}
\newcommand{\Erecom}{\ensuremath{E_{\rm rec}}}
\newcommand{\Ediagovrec}{\ensuremath{E^{+}_{\rm ov, rec}}}

\newcommand{\alp}{\ensuremath{\alpha}}
\newcommand{\kms}{\ensuremath{\rm km \; s^{-1}}}
\newcommand{\kboltz}{\ensuremath{\rm k_B}}
\newcommand{\kbbar}{\ensuremath{\rm k_B \; baryon^{-1}}}

\newcommand{\Lnue}{\ensuremath{L_{\nue}}}
\newcommand{\Lnuebar}{\ensuremath{L_{\nuebar}}}
\newcommand{\Enue}{\ensuremath{\epsilon_{\nue}}}
\newcommand{\Enuebar}{\ensuremath{\epsilon_{\nuebar}}}

\newcommand{\pardirt}[1]{\frac{\partial #1 }{\partial t}}
\newcommand{\be}{\begin{eqnarray}}
\newcommand{\ee}{\end{eqnarray}}

\newcommand{\isotope}[2]{\ensuremath{\mathrm {^{#2}#1}}}
\newcommand{\gcc}{\ensuremath{{\mbox{g~cm}}^{-3}}}

\newcommand{\kmps}{\ensuremath{\mbox{km~s}^{-1}}}

\newcommand{\Bethes}{\ensuremath{{\mbox{B~s}}^{-1}}}
\newcommand{\Rnu}{\ensuremath{R_{\nu}}}
\newcommand{\omegaBV}{\ensuremath{\omega_{\rm BV}}}

\newcommand{\Mni}{\ensuremath{M_{{\rm Ni56}}}}
\newcommand{\Mzams}{\ensuremath{M_{\rm ZAMS}}}
\newcommand{\pdV}{\ensuremath{p\,dV}}

\newcommand{\chimera}{{\sc Chimera}}
\newcommand{\vertex}{{\sc Vertex}}
\newcommand{\prometheusvertex}{{\sc Prometheus-Vertex}}
\newcommand{\coconutvertex}{{\sc CoCoNuT-Vertex}}
\newcommand{\castro}{{\sc Castro}}
\newcommand{\zidsa}{{Zeus+IDSA}}
\newcommand{\zelmani}{{\sc {Zelmani}}}

\newcommand{\specen}[1]{\ensuremath{\varepsilon_{\rm #1}}}
\newcommand{\etherm}{\specen{th}}

\newcommand{\eint}{\specen{int}}

\newcommand{\Edint}{\ensuremath{\Ediag_{\rm th}}}
\newcommand{\Edgrav}{\ensuremath{\Ediag_{\rm grav}}}
\newcommand{\Edkin}{\ensuremath{\Ediag_{\rm kin}}}

\newcommand{\Evol}[1]{\ensuremath{E^{\rm #1}}}
\newcommand{\Evolc}[2]{\ensuremath{E^{\rm #1}_{\rm #2}}}
\newcommand{\dotEvol}[1]{\ensuremath{\dot{E}^{\rm #1}}}
\newcommand{\dEvol}[1]{\ensuremath{\Delta E^{\rm #1}}}

\newcommand{\Etot}{\ensuremath{\Evol{V}}}
\newcommand{\Qnu}{\ensuremath{Q_\nu}}
\newcommand{\Qnuc}{\ensuremath{Q_{\rm nuc}}}

\newcommand{\eden}[1]{\ensuremath{e_{\rm #1}}}
\newcommand{\edint}{\ensuremath{\eden{int}}}
\newcommand{\edth}{\ensuremath{\eden{th}}}
\newcommand{\edkin}{\ensuremath{\eden{kin}}}
\newcommand{\edgrav}{\ensuremath{\eden{grav}}}
\newcommand{\edbind}{\ensuremath{\eden{bind}}}
\newcommand{\edtot}{\ensuremath{\eden{tot}}}
\newcommand{\qdot}{\ensuremath{\dot{q}}}
\newcommand{\qdotn}{\ensuremath{\dot{q}_{\rm n}}}
\newcommand{\qdotnu}{\ensuremath{\dot{q}_{\nu}}}

\newcommand{\fluxX}[1]{\ensuremath{\mathbf{f}_{\rm #1}}}

\newcommand{\fluxkin}{\ensuremath{\fluxX{kin}}}

\newcommand{\fluxenth}{\ensuremath{\fluxX{enth}}}
\newcommand{\fluxbind}{\ensuremath{\fluxX{bind}}}
\newcommand{\fluxgrav}{\ensuremath{\fluxX{grav}}}

\newcommand{\FluxCum}[2]{\ensuremath{\mathcal{F}^{\rm #2}_{\rm #1}}}

\newcommand{\divv}[1]{\ensuremath{\nabla \cdot #1}}

\newcommand{\dSurf}{\ensuremath{d\mathbf{S}}}

\newcommand{\intvol}[1]{\int_V #1 \, dV}
\newcommand{\intsurf}[1]{\oint_{\dSurf} #1 \cdot \dSurf}

\newcommand{\uvec}{\ensuremath{\mathbf{u}}}
\newcommand{\Rdiag}{\ensuremath{{\cal V}^{\rm diag}}}
\newcommand{\Rsph}[1]{\ensuremath{{\cal V}^{\rm #1}}}

\slugcomment{ApJ in press.}

\shorttitle{Axisymmetric Core-Collapse Supernova Simulations}
\shortauthors{Bruenn et al.}

\begin{document}

\title{The Development of Explosions in Axisymmetric \emph{Ab Initio} Core-Collapse Supernova Simulations of 12--25 \msun\ Stars}

\author{
Stephen W. Bruenn\altaffilmark{1},
Eric J. Lentz\altaffilmark{2,3},
W. Raphael Hix\altaffilmark{3,2},
Anthony Mezzacappa\altaffilmark{2,4},
J. Austin Harris\altaffilmark{2},\\
O. E. Bronson Messer\altaffilmark{5,3},
Eirik Endeve\altaffilmark{6,2,4},
John M. Blondin\altaffilmark{7},
Merek Austin Chertkow\altaffilmark{2},\\
Eric J. Lingerfelt\altaffilmark{6,3},
Pedro Marronetti\altaffilmark{8}, and
Konstantin N. Yakunin\altaffilmark{2,3,4}
}
\email{bruenn@fau.edu}

\altaffiltext{1}{Department of Physics, Florida Atlantic University, 777 Glades Road, Boca Raton, FL 33431-0991, USA}
\altaffiltext{2}{Department of Physics and Astronomy, University of Tennessee, Knoxville, TN 37996-1200, USA}
\altaffiltext{3}{Physics Division, Oak Ridge National Laboratory, P.O. Box 2008, Oak Ridge, TN 37831-6354, USA}
\altaffiltext{4}{Joint Institute for Computational Sciences, Oak Ridge National Laboratory, P.O. Box 2008, Oak Ridge, TN 37831-6173, USA}
\altaffiltext{5}{National Center for Computational Sciences, Oak Ridge National Laboratory, P.O. Box 2008, Oak Ridge, TN 37831-6164, USA}
\altaffiltext{6}{Computer Science and Mathematics Division, Oak Ridge National Laboratory, P.O. Box 2008, Oak Ridge, TN 37831-6164, USA}
\altaffiltext{7}{Department of Physics, North Carolina State University,  Raleigh, NC 27695-8202, USA}
\altaffiltext{8}{Physics Division, National Science Foundation, Arlington, VA 22207 USA}

\begin{abstract}

We present four \emph{ab initio} axisymmetric core-collapse supernova simulations initiated from 12, 15, 20, and 25 \msun\ zero-age main sequence progenitors. 
All of the simulations yield explosions and have been evolved for at least 1.2 seconds after core bounce and 1 second after material first becomes unbound. 
These simulations were computed with our \chimera\ code employing ray-by-ray spectral neutrino transport, special and general relativistic transport effects, and state-of-the-art neutrino interactions. 
Continuing the evolution beyond 1 second after core bounce allows the explosions to develop more fully and the processes involved in powering the explosions to become more clearly evident. 
We compute explosion energy estimates, including the negative gravitational binding energy of the stellar envelope outside the expanding shock, of 0.34, 0.88, 0.38, and 0.70 Bethe (B $\equiv 10^{51}$ ergs) and increasing at 0.03, 0.15, 0.19, and 0.52 \Bethes, respectively, for the 12, 15, 20, and 25 \msun\ models at the endpoint of this report.  
We examine the growth of the explosion energy in our models through detailed analyses of the energy sources and flows. 
We discuss how the explosion energies may be subject to stochastic variations as exemplfied by the effect of the explosion geometry of the 20 \msun\ model in reducing its explosion energy.
We compute the proto-neutron star masses and kick velocities. 
We compare our results for the explosion energies and ejected \isotope{Ni}{56} masses against some observational standards despite the large error bars in both models and observations. 
\end{abstract}


\section{Introduction}
\label{sec:intro}

It has been nearly fifty years since \citet{CoWh66} first proposed that core-collapse supernovae may be driven by neutrino energy deposition in the mantle from neutrinos produced and transported from a gravitationally collapsing core, and most current investigations of the core collapse supernova (CCSN) explosion mechanism still center around this idea.
Colgate and White's seminal work was followed by increasingly sophisticated one-dimensional (1D) models that eventually included multi-frequency (spectral) neutrino transport, the state of the art in weak interaction physics of the day (the impact of the newly discovered weak neutral currents was part of the important progress made through these models), a nuclear equation of state, and general relativistic treatment of gravity \citep{Arne66,Wils71,Wils74,Brue75,WiCoCo75,Arne77}.
The culmination of 1D models was not to occur until almost forty years after Colgate and White's original proposal, with simulations that implemented general relativistic Boltzmann neutrino transport and hydrodynamics coupled to general relativistic gravity, using state of the art weak interactions and equations of state \citep{LiMeTh01,LiMeMe04}.
The fundamental conclusion drawn from these and other 1D studies is that neutrinos alone cannot power CCSNe in spherical symmetry, yet 1D simulations do remain a vital guide to the remaining CCSN physics at low computational cost \citep[e.g.,][]{ThBuPi03,  SuYaSu05, LeMeMe12a,LeMeMe12b, MuJaMa12}.

Two decades ago, the first two-dimensional (2D) simulations \citep{HeBeCo92,HeBeHi94} transformed the field.
The fundamental feedback between mass accretion, against which the shock must work, and the neutrino luminosities powering the explosion was finally unchained in these models and others \citep[e.g.,][]{BuHaFr95,JaMu96}, allowing continued accretion and explosion to coexist, something impossible in 1D models.
Though two decades have passed, 2D models are only now beginning to fill in the CCSN landscape in a fashion done previously in 1D.
Two-dimensional simulations including spectral neutrino transport; general relativistic corrections to 2D Newtonian self gravity or approximate 2D general relativistic self gravity; all relevant neutrino weak interactions, including neutrino-energy-coupled scattering and electron capture on nuclei modeled to include nucleon--nucleon interactions; and sophisticated nuclear equations of state, have been performed with two codes thus far, \vertex\ \citep{BuRaJa06,BuJaRa06,MaJa09,MuJaMa12,MuJaHe12} and \chimera\ \citep{BrDiMe06,MeBrBl07a,BrMeHi09b,BrMeHi13}, though these codes use the ``ray-by-ray'' approximation to neutrino transport rather than a fully 2D transport implementation.
(We will discuss the ray-by-ray approximation in Sections~\ref{sec:methods} and \ref{sec:rbr}.)
While these models have demonstrated from first principles that neutrino-driven explosions are possible when aided by multidimensional effects, especially neutrino-driven convection and the Standing Accretion Shock Instability \citep[SASI;][]{BlMeDe03}, only a few have been carried out sufficiently long after bounce to determine important characteristics of the explosion, such as the final explosion energy and ejecta nucleosynthesis.

CCSN simulations for multiple progenitors have been performed with three additional codes that include multi-frequency neutrino transport coupled to 2D hydrodynamics.
The \zidsa\ code \citep{SuKoTa10,SuTaKo13,SuYaTa14,TaKoSu14,NaTaKu14}, the VULCAN code \citep{DeBuLi06,BuLiDe07,OtBuDe08}, and the \castro\ code \citep{DoBuZh15} have been used to perform 2D simulations with Newtonian gravity, a reduced set of neutrino interactions, no relativistic neutrino transport corrections (e.g., gravitational redshift), and, in the case of \zidsa, the exclusion of muon and tau neutrinos and antineutrinos.
Among all of the above efforts, VULCAN and its successor \castro\ are the only codes that implement fully 2D, rather than ray-by-ray, neutrino transport.
We discuss the strengths and weakness of these approaches in Section~\ref{sec:discussion}.

Ultimately, the restriction to axisymmetry must be lifted and three-dimensional (3D) simulations with all of the physics mentioned above will have to be performed without some of the numerical approximations used, particularly the ray-by-ray neutrino transport approximation.
Three-dimensional simulations with spectral, ray-by-ray neutrino transport, approximate general relativity, and state-of-the-art weak interactions are underway, and the results from early stages of the post-bounce dynamics have been reported by \citet{HaMuWo13}, using \vertex, and \citet{LeBrHi15}, using \chimera. \citet{TaKoSu12,TaKoSu14} also report 3D Newtonian simulations using \zidsa. 
Fully general relativistic simulations with reduced neutrino physics (a leakage scheme for neutrino transport with a reduced set of neutrino interactions), have been also performed using the \zelmani\ code \citep{OtAbMo13,MoRiOt14}.

In \citet[][ hereafter Paper~1]{BrMeHi13} we presented the first 500~ms of the simulations presented here.
In this paper we present in detail the complete neutrino powered phase of these explosions.
Section~\ref{sec:methods} describes the methodology and configuration of our simulations.
Section~\ref{sec:results} details the development of explosions in our models and examines the driving physics and the general properties of the models.
In Section~\ref{sec:Obs_CCSNe} we compare the outcomes of our simulations with observations.
Section~\ref{sec:discussion} places our results in the context of the results obtained by other groups and discusses this in the context of the included physics and methodologies.
We summarize our findings in Section~\ref{sec:summary}.

\section{Methodology}
\label{sec:methods}
\chimera\footnote{\url{ChimeraSN.org}} is a multi-physics code built specifically for multidimensional simulation of CCSNe that has been under development for more than a decade \citep{HiMeLi01,Brue05,BrDiMe06,MeBrBl07a,MeBrBl08,BrMeHi09b,BrMeHi09a,BrMeHi13}.
It is a combination of separate codes for hydrodynamics and gravity; neutrino transport and opacities; and nuclear EoS and reaction network, coupled by a layer that oversees data management, parallelism, I/O, and control.
Hydrodynamics are evolved via a dimensionally-split, Lagrangian-plus-remap Newtonian scheme with piecewise parabolic reconstruction \citep[PPMLR;][]{CoWo84} as implemented in VH1 \citep{HaBlLi12}.
Self-gravity is computed by multipole expansion \citep{MuSt95} replacing the Newtonian monopole  with a GR monopole \citep[][Case~A]{MaDiJa06}.
Neutrino transport is computed in the ``ray-by-ray'' (RbR) approximation \citep{BuRaJa03}, where an independent, spherically symmetric transport solve is computed for each radial ``ray'' (i.e., all radial zones at a fixed latitude, $\theta$).
Neutrinos are advected laterally (in the $\theta$-direction) with the fluid and contribute to the lateral pressure gradient where $\rho>10^{12}\,\gcc$.
The transport solver is an improved and updated version of the multi-group (frequency) flux-limited diffusion (MGFLD) transport solver of \citet{Brue85}, enhanced for GR \citep{BrDeMe01}, with an additional geometric flux limiter to prevent the over-rapid transition to free streaming of the standard flux-limiter.
 
All $\mathcal{O}$(v/c) observer corrections in the transport equation are included.
We solve for all three flavors of neutrinos and anti-neutrinos using four coupled species: \nue, \nuebar, $\numt=\{\numu,\nutau\}$, $\numtbar=\{\numubar,\nutaubar\}$, with 20 logarithmically spaced energy groups each covering $\alpha\epsilon = 4$--250~\mev, where $\alpha$ is the lapse function and $\epsilon$  the comoving-frame group-center energy.
The neutrino--matter interactions include emission, absorption, and non-iso-energetic scattering on free nucleons \citep{RePrLa98} with weak magnetism corrections \citep{Horo02}; emission/absorption (electron capture) on nuclei \citep{LaMaSa03,HiMeMe03}; iso-energetic scattering on nuclei, including ion-ion correlations; non-iso-energetic scattering on electrons and positrons; and pair emission from $e^+e^-$-annihilation \citep{Brue85} and nucleon-nucleon bremsstrahlung \citep{HaRa98}.
These neutrino interactions are similar to those used in the \vertex\ simulations, except that unlike \vertex, \chimera\ does not include the effective mass corrections at high density nor the $\nue\nuebar\rightarrow\numt\numtbar$ pair-conversion process.

We utilize the $K = 220$~\mev\ incompressibility version of the \citet{LaSw91} EoS for  $\rho>10^{11}$~\gcc\ and an enhanced version of the \cite{Coop85} EoS for  $\rho<10^{11}$~\gcc\ where nuclear statistical equilibrium (NSE) applies.
At lower temperatures, where NSE is not applicable, we implement a 14-species \alp-network (\alp, \isotope{C}{12}-\isotope{Zn}{60}) with the integrated XNet  nuclear network code \citep{HiTh99b}.
In these non-NSE regions, we additionally track, but do not react, the abundance of neutrons, protons, and an auxiliary heavy species.
The initial abundances of these non-reactive species are built from the composition given for each progenitor by \citet{WoHe07} with properties of the auxiliary heavy species chosen to conserve the electron fraction of material that is not in the \alp-network.
When material must advect from an NSE region into a non-NSE region or when the temperature of a zone falls so that NSE is no longer appropriate, the composition of the advected or transitioned material must be determined.
The 4-component nuclear composition used by traditional supernova EoSs does not match the needs of the \alp-network, so we modified the NSE to use 17-species (14 \alp-species, free neutrons and protons, and \isotope{Fe}{56}).
When advected into non-NSE regions \isotope{Fe}{56} is mapped to the auxiliary species in the network.
This 17-species NSE computation allows the network to be filled with a representative composition without unphysically large numbers of free nucleons or locking most of the composition in the inert auxiliary nucleus.
The Cooperstein electron--photon EoS with an extension to non-degenerate electron gases is used throughout the simulation.

During the evolution, the radial zones are gradually and automatically repositioned during the remap step to follow changes in the radial structure.
To minimize restrictions on the time step from the Courant limit,  we ``freeze'' the lateral hydrodynamics for the inner 8~zones during collapse, and after prompt convection fades we expand the laterally frozen region to the inner $\sim$8--10~km.
In the ``frozen'' region we set $v_\theta = 0$ and skip the lateral hydrodynamic sweep.
The full radial hydrodynamics and neutrino transport are always computed to the center of the simulation for all radial rays.
At about 800~ms after core bounce in each simulation we switch to shell averaging of the fluid properties in the inner spherical region to avoid potential problems near the center.
A more extensive description of \chimera\ is under preparation.

\defcitealias{BrMeHi13}{Paper~1}
In this paper, we present the continuation of the four non-rotating axisymmetric models (designated B12-WH07, B15-WH07, B20-WH07, B25-WH07, corresponding to progenitor zero-age main sequence masses of 12, 15, 20, and 25~\msun),  presented in \citetalias{BrMeHi13}.
The simulations were initialized without applied perturbation from the inner 30000, 20000, 21000 and 23000~km, respectively of the pre-supernova progenitors of \citet{WoHe07}.
A grid of 512 non-equally spaced radial zones covers from the stellar center into the oxygen-rich layers.
The 2D models employ 256 uniformly-sized angular zones from 0 to $\pi$ for an angular resolution of 0.70\degree.
The simulations were carried out in full 2D from the onset of collapse with the very small post-bounce roundoff errors supplying the perturbations for the growth of fluid instabilities.
For post-processing nucleosynthesis and other analyses we have included 4000--8000 passive Lagrangian tracer particles per simulation. Analyses utilizing these tracers will be reported in later papers (J.~A. Harris et al., in prep.).
These simulations began in February 2012 and ran intermittently for approximately two years on the `Jaguar' (OLCF) and `Kraken' (NICS) computers located in Oak Ridge and on the `Hopper' and `Edison' computers at NERSC in Oakland.
Each simulation was computed using 256 parallel MPI tasks on 256 processor cores with one task/core for each radial ray.
In \citetalias{BrMeHi13} we reported the first 800~ms of B12-WH07 and 500~ms for the other three models.
We have now evolved to 1400~ms for three of the models (B12-WH07, B20-WH07, and B25-WH07), while B15-WH07 was terminated about 1200~ms after bounce when its shock crossed the outer boundary of the grid.

We have designated this group of models as `Series-B' with the `Series-A' appellation retroactively applied to the first group of production simulations made with \chimera\ \citep{BrMeHi09b} using the same four progenitors, but with only 256 radial zones covering  a reduced portion of the original progenitor.
The `Series' designation signifies a group of simulations using the same code, numerical methods, and base input physics.
In addition to the progenitor and resolution changes from Series-A to Series-B, improvements were made to the handling of the transition at the NSE boundary, control of the odd-even `carbuncle' numerical instability for grid aligned shocks, as well as many smaller code improvements.
These simulations complete Series-B and the first models of the more extensive `Series-C' are now underway using an updated version of \chimera\ that will be reported separately when completed.

\section{Results}
\label{sec:results}

As discussed in \citetalias{BrMeHi13}, all models acquire a negative lepton gradient unstable to convection immediately after the shock breaks out of the \nue-sphere, but this is stabilized by the positive entropy gradient laid down by the shock as it gathers strength. 
By $\tpb\sim12$~ms, however, an extended unstable entropy gradient develops farther out behind the shock as it weakens, extending down to $\sim$60~km and driving a brief episode of convection (Figure~\ref{fig:entropy}; 12~ms panel in \citetalias{BrMeHi13}). This episode of ``prompt'' convection has little appreciable effect on the shock, and by $\tpb\sim40$~ms convective activity has ceased in all models.

We additionally compared the shock trajectories of our 2D models with their 1D counterparts in \citetalias{BrMeHi13} and showed that prior to $\sim$100~ms after bounce these trajectories are quite similar, but thereafter the shock trajectories for the 2D models begin to diverge from their 1D counterparts.
This result echoes many previous studies \citep[e.g.][]{MaJa09}.
Whereas the radii of the 1D models begin to decline after 100~ms, having reached a plateau of about 190~km, the mean shock radii of the 2D simulations begin to increase.
We remarked in \citetalias{BrMeHi13} on the similarity of the shock trajectories for the different progenitors during the quasi-stationary accretion phase, which lasts until about 150~ms after bounce.
We attributed this to the fact that while the separate factors responsible for the radius of the shock during this phase, as given by Equation~(1) of \citet{Jank12}, are different for the different progenitors, their combinations are similar (upper panel of Figure~3 in \citetalias{BrMeHi13}).
The expansion of the shock in the 2D models after 100~ms, in contrast to their 1D counterparts, is due to the onset of fluid instabilities below the shock.
Here we provide a detailed account of the development of asphericity from convection and the SASI; the build-up to explosion; the operation of the explosion mechanism; the development of explosion energy; the impact of accretion and large-scale morphology on the explosions; and the growth and deflection of the proto-NS.

\subsection{Development of Instabilities and Asphericity}
\label{sec:instability}

\begin{figure}
\includegraphics[width=\columnwidth,clip]{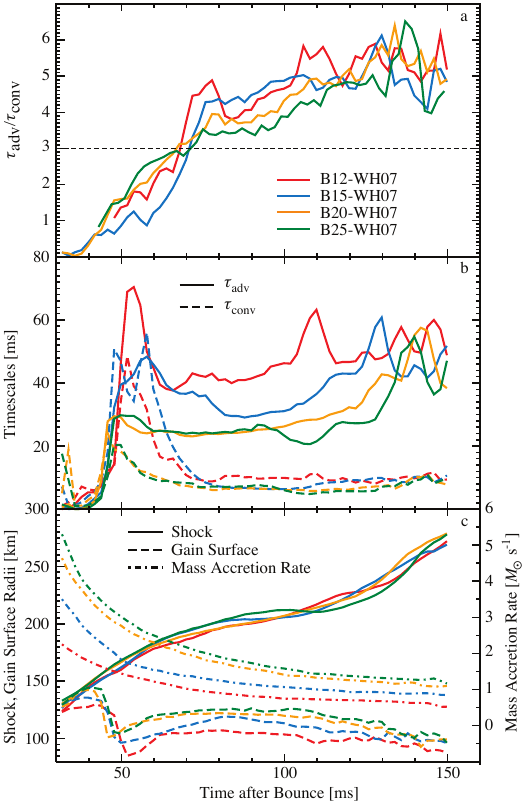}
\caption{\label{fig:early}
Properties of the pre-explosion phase for models B12-WH07 (red lines), B15-WH07 (blue lines), B20-WH07 (orange lines), and B25-WH07 (green lines) .
Plotted as a function of time after core bounce with a 5~ms boxcar smoothing are (a) the ratio of advective and convective growth timescales $\tadv/\tconv$ (Equation~\ref{tadv_tconv});
(b) the advective (\tadv; solid lines) and convective growth (\tconv; dashed lines) timescales; and
(c) (left ordinate) the shock (solid lines) and gain (dashed lines) radii and
(right ordinate) the mass accretion rate computed through a stationary spherical surface at a radius of 300~km (dash-dotted lines).
}
\end{figure}

\begin{figure}
\includegraphics[width=\columnwidth,clip]{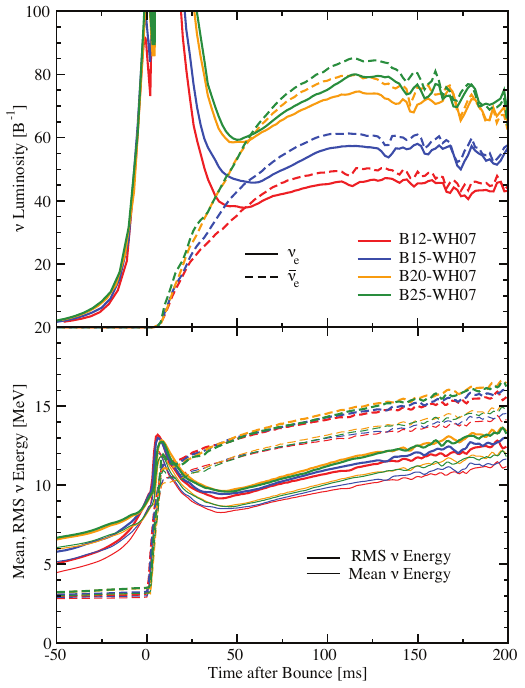}
\caption{\label{fig:Lum_Mean_RMS} The fluid-frame luminosities (top) and the mean and RMS energies (bottom) for \nue\ (solid lines) and \nuebar\ (dashed lines) at 1000~km during the early evolution of the models. 
The mean and RMS energies are plotted with the thin and thick lines, respectively, with the colors of Figure~\ref{fig:entropy}. (Neutrino luminosities and RMS energies for all neutrino flavors and for the duration of the simulations are shown in Figures~\ref{fig:lumin} and \ref{fig:ERMS}.)}
\end{figure}

\begin{figure*}
\includegraphics[clip]{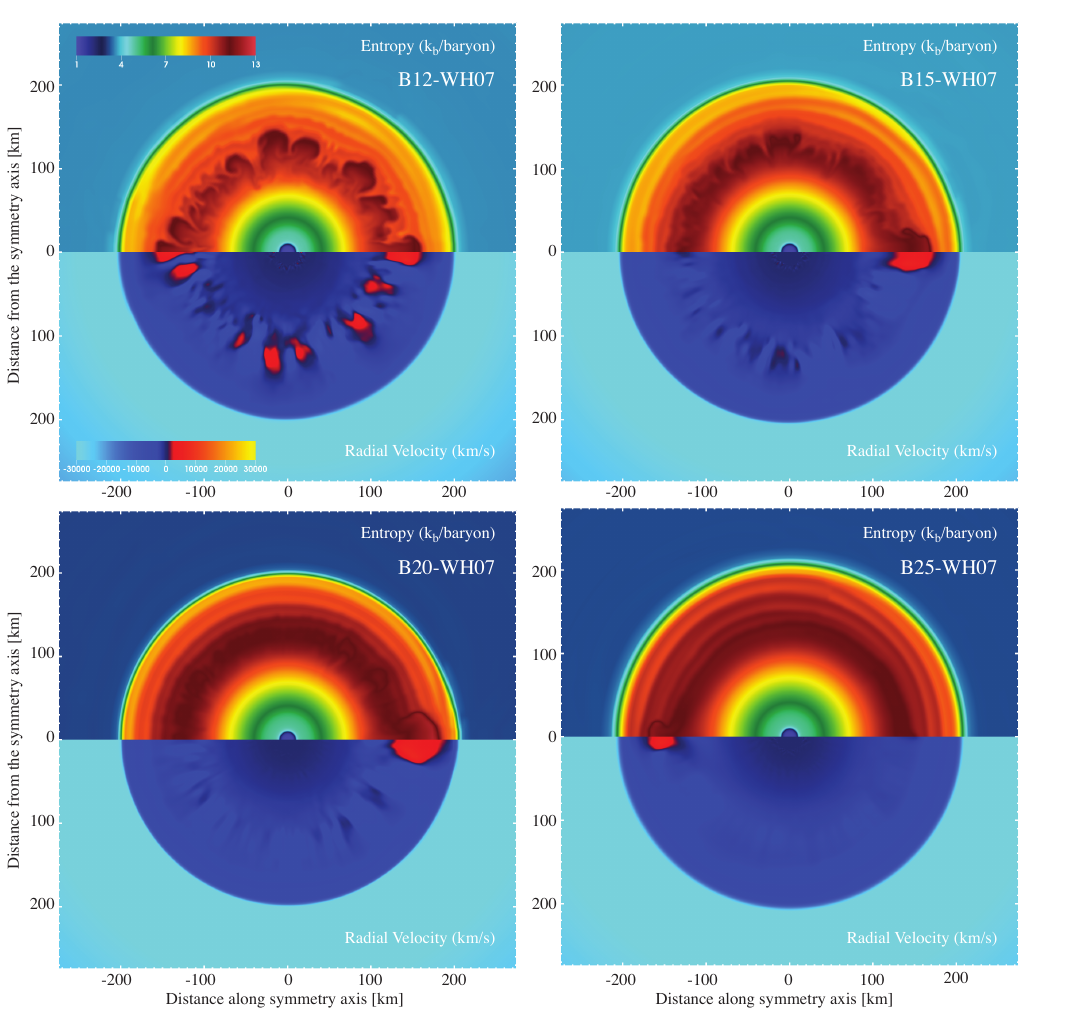}
\caption{\label{fig:entropy}
Specific entropy  (top half of panels) and radial velocity (mirrored bottom half of panels) for each simulation at 85~ms after core bounce. (Animated version of figure available in electronic edition.)}
\end{figure*}

The initial outward propagation of the shock from the core bounce and its subsequent stagnation is followed by a brief period ($\sim$10~ms) of entropy-driven convection in the proto-neutron star (proto-NS) as described above.
After this episode of prompt convection, the region between the proto-NS and the shock enters a period of quiescence during which the competition between neutrino energy emission and deposition establishes a neutrino net heating layer, or `gain region,' below the shock and a net cooling layer between the heating layer and the neutrinosphere \citep{BeWi85}.
These two layers are separated by a gain surface at which net neutrino heating is zero.
Energy exchange between the neutrino radiation and the fluid is dominated by \nue\ and \nuebar\ absorption on free neutrons and protons, respectively, and the inverse processes.

After the quiescent accretion phase two fluid instabilities, neutrino-driven convection and the SASI, emerge and come to dominate the dynamics below the shock.
One or both of these instabilities cause the mean shock trajectory in multi-dimensional core-collapse simulations to diverge from those in 1D simulations, and have proven essential for initiating successful neutrino-driven explosions.

A number of early multi-dimensional hydrodynamics simulations demonstrated that neutrino-driven convection does indeed develop in the heating layer and enhances the revival of the shock \citep{HeBeCo92, HeBeHi94, BuHaFr95, JaMu93, JaMu95}.
These authors and others pointed out that convectively unstable entropy gradients occur in the heating layer for several reasons.
First, neutrino heating is more intense near the gain surface than at larger distances, due to geometric dilution and absorption of neutrinos propagating outward.
Second, material advecting towards the gain surface is continually heated by neutrinos, gaining entropy as it moves inwards.
As has been pointed out by many authors \citep[e.g.,][]{HeBeCo92, HeBeHi94, BuHaFr95, JaMu93, JaMu95, THom00, Jank01, BuRaJa06, FrWa04, BrDiMe06, ThQuBu05, ScJaFo08, MuBu08, MuJaHe12},
convective flows enhance the conditions necessary for the revival of the shock. 
They increase the effective neutrino heating efficiency because rising fluid elements have lower neutrino emissivities after adiabatic expansion, reducing losses due to neutrino cooling, and the nonradial motions increase the dwell time of fluid elements in the gain region thereby increasing the time they are subject to neutrino heating.
In addition, as was pointed out by \citet{BuHaFr95, DoBuMu13, Couc13b} and demonstrated more quantitatively by \citet{MuDoBu13} and \citet{CoOt15}, the rising turbulent convective plumes exerts a dynamical pressure on the shock aiding its expansion. 
We will discuss these effects in the context of our models in more detail below.
After the shock begins to propagate outward, asphericities in the shock surface funnel inflowing matter into accretion streams that reach the proto-NS and maintain the accretion luminosity while delivering material to be heated in the convective, or buoyant, heating cycle.
These effects of multidimensional flows, including the continued accretion power after shock expansion starts, are not possible in 1D simulations and are likely the most important reason multidimensional models achieve explosions where 1D models do not.

The second important fluid instability is the SASI, the tendency of the accretion shock to undergo non-radial low-mode (dipole, quadrapole, etc.) deformations, even in the absence of a negative entropy gradient.
This instability was discovered numerically by \citet{BlMeDe03} and has been investigated further by \citet{GaFo05, BlMe06, OhKoYa06, FoGaSc07, Lami07, YaYa07, ScJaFo08,EnCaBu10, EnCaBu12, FeMuFo14}.
Low mode, non-spherical shock deformations that result from this instability during the period after shock stagnation increase the mean shock radius and deflect post-shock flows laterally, increasing the advection time of matter through the heating layer.
Furthermore, the low-mode core deformations get frozen in when the shock propagates outward and can potentially explain observed asymmetric explosion geometries and neutron star kicks \citep{ScPlJa04, ScKiJa06}.

\subsubsection{Neutrino-driven Convection}
\label{sec:nuconvection}

Development of convection in the gain region, through which matter is continually flowing, requires that a given fluid element spend enough time in the gain region, relative to the convective growth timescale, for a perturbation to be sufficiently amplified to resist being swept through the gain surface into the cooling layer.
\citet{FoScJa06} found that the conditions for the onset of neutrino-driven convection requires that the quantity
\begin{equation}
  \chi = \int_{R_{\rm gain}}^{\Rshock} \max \left[ \omegaBV(r), 0 \right] \frac{dr}{\left| \bar{v}_{\rm r} \right|}
\label{tadv_tconv}
\end{equation}
exceeds a value of about 3, where $R_{\rm gain}$, \Rshock, $\bar{v}_{\rm r}$, and \omegaBV\ are, respectively, the angle averaged gain radius, shock radius, radial velocity, and Brunt-V\"{a}is\"{a}l\"{a} frequency (unstable if positive).
The quantity $\chi$ can be thought of as the ratio of the advection timescale, \tadv, to an average timescale for the growth of convection, \tconv.
The exact value of $\chi$ signaling conditions favorable for the onset of neutrino-driven convection will depend on the magnitude and nature of the fluid perturbations.
We evaluated $\chi$ numerically from Equation~(\ref{tadv_tconv}) for each of our models and plotted the results in Figure~\ref{fig:early}a.
The convective threshold $\chi\approx3$ is reached for each model about 70~ms after bounce.
Between 70 and 100~ms after bounce, $\chi$ rises to between 4 and 5 for the two less massive progenitors, but for the two more massive progenitors $\chi$ stays between 3 and 4, and in the case of B25-WH07 $\chi$ barely rises above 3.
Thus, less massive progenitors should be more prone to the growth of neutrino-driven convection in the heating layer 70--100~ms after bounce than the more massive progenitors.
That this is indeed the case can be seen in the entropy and radial velocity maps of the shocked regions at 85~ms after bounce (Figure~\ref{fig:entropy}).
Convection is well developed in model B12-WH07 and is progressively less developed with increasing progenitor mass through model B25-WH07, where convective activity is barely discernible at 85~ms after bounce.
As is apparent in Figures~\ref{fig:entropy}, \ref{fig:stream}, and~\ref{B12_B25_Entropy_250_tb}, there is a tendency for convective structures and plumes to lie preferentially along the symmetry axis.
This is driven by the inability of convective cells and other fluid motions to cross the polar axis in axisymmetric simulations.
Given our reflecting boundary conditions, motions transverse to the pole are canceled while radial motions remain, resulting in structures elongated along the pole. 
Thus the structures at the pole grow in scale more rapidly than their non-polar counterparts, which favors enhanced neutrino heating and a consequent more rapid growth in entropy.  
The growth in entropy at the pole does not occur in the absence of neutrino heating, and the preference for structure to grow more rapidly along the axis of symmetry goes away when \chimera\ is run in 3D.

The differences in the advection to convection ratios can be understood using quantities plotted in the other panels of Figure~\ref{fig:early} and in Figure~\ref{fig:Lum_Mean_RMS}.
The advection timescales through the gain region (Figure~\ref{fig:early}b; solid lines) generally decrease with increasing progenitor mass between 70 and 130~ms after bounce.
The advection time is a function of the advection velocity through the heating layer and width of the heating layer.
For our simulations, the advection velocity increases with the progenitor mass.
The models for more massive progenitors have higher mass accretion rates (Figure~\ref{fig:early}c; dot-dashed lines), due to denser pre-collapse outer iron cores and silicon shells.
The higher mass accretion rates lead to more massive proto-NSs, stronger gravitational fields, and larger pre-shock and post-shock infall velocities for similar shock radii.
Moreover, the width of the heating layer, the difference between the shock and gain radii (Figure~\ref{fig:early}c; solid and dashed lines, respectively), decreases with increasing progenitor mass as the models with the more massive progenitors have similar shock but larger gain radii.

\begin{figure}
\includegraphics[width=\columnwidth,clip]{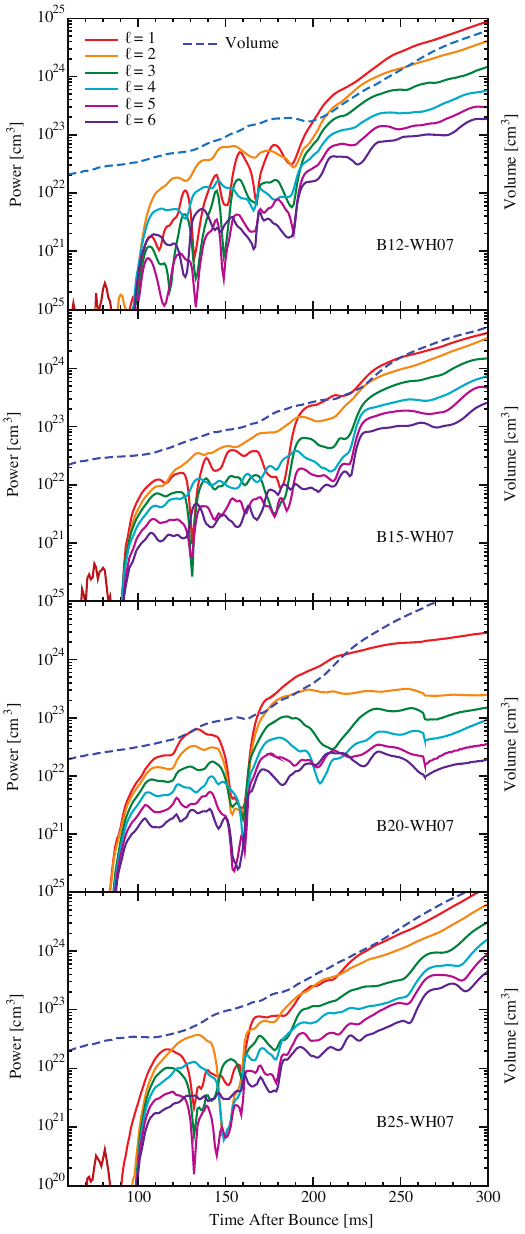}
\caption{\label{fig:mode_power}
Pressure fluctuation power versus time after bounce for all models in modes $\ell=1$ through $\ell=6$ computed using Equation~(\ref{SASI_power1}).
The blue dashed lines show the volume of the heating layer used in the power computation.}
\end{figure}

The convective growth timescale \tconv\ (Figure~\ref{fig:early}c; dashed lines) is a function of the magnitude of the negative entropy gradient established in the heating layer, which increases with both the strength of the neutrino heating and the time a fluid element resides in the heating layer exposed to neutrino radiation.
Both the \nue\ and \nuebar\ luminosities (Figure~\ref{fig:Lum_Mean_RMS}; top) and mean and RMS energies (Figure~\ref{fig:Lum_Mean_RMS}; bottom), defined by
\begin{equation}
  \langle \epsilon_{0} \rangle = \int d\Omega \frac{ \int_{0}^{\infty} d\epsilon_{0} \epsilon_{0}^{3} \psi^{(0)}(\epsilon_{0}) }{ \int_{0}^{\infty} d\epsilon_{0} \epsilon_{0}^{2} \psi^{(0)}(\epsilon_{0}) } ,
\label{mean_rms}
\end{equation}
\begin{equation}
  (\epsilon_{0})_{rms} = \int d\Omega \left[ \frac{ \int_{0}^{\infty} d\epsilon_{0} \epsilon_{0}^{4} \psi^{(0)}(\epsilon_{0}) }{ \int_{0}^{\infty} d\epsilon_{0} \epsilon_{0}^{2} \psi^{(0)}(\epsilon_{0}) } \right]^{1/2} ,
\label{mean_rms}
\end{equation}
where $\epsilon_{0}$ is the fluid frame neutrino energy and $\psi^{(0)}(\epsilon_{0})$ is the zero angular moment of the neutrino occupation function, increase with increasing progenitor mass due to the higher mass accretion rates (Figure~\ref{fig:early}c; dash-dotted lines).
Compensating for this is the smaller accretion timescales for the higher mass progenitors, as discussed above.
The result is that the convective growth timescales are nearly the same for all of the models.
The decreasing advective timescale with increasing progenitor mass, together with the nearly identical convective growth timescales, results in a decreasing $\tadv/\tconv$ with increasing progenitor mass, and the tendency for the onset of convection to be progressively delayed with increasing progenitor mass.

\subsubsection{Standing Accretion Shock Instability}

To provide evidence for the operation of the SASI in our models, which should manifest itself globally, we plot in Figure~\ref{fig:mode_power} the time evolution of the total power in the first six Legendre modes of the pressure fluctuations.
Following \citet{BlMe06} and \citet{MaJa09}, the pressure fluctuation power is computed from the volume integral of the fractional variation of the pressure from its angular mean between the mean neutrinosphere radius \Rnu\ and the mean shock radius \Rshock\,
\begin{equation}
  {\rm Power}(\ell,t) = 2 \pi \int_{\Rnu}^{\Rshock} dr \, r^{2} [G_{\ell}(r,t)]^{2},
\label{SASI_power1}
\end{equation}
where Legendre amplitudes $G_{\ell}(r,t)$ are given by
\begin{equation}
  \frac{ P(r, \theta, t) - \langle P(r, \theta, t) \rangle_{\theta} }{ \langle P(r, \theta, t) \rangle_{\theta} } = \sum_{\ell = 0}^{\infty} G_{\ell}(r,t) P_{\ell}(\cos \theta)
\label{SASI_power2}
\end{equation}
and $P_{\ell}(\cos \theta)$ are the Legendre polynomials.

The power in the lower modes, particularly $\ell=1$ and $\ell=2$, becomes appreciable at $\sim$100~ms after bounce for all models, slightly earlier for B15-WH07 and B20-WH07.
B12-WH07 shows SASI-like oscillations characterized by several sign reversals of all odd modes with intervals of $\sim$20~ms \citep[similar to the period reported by][]{MuJaHe12} and exhibits a persistent growing quadrupole ($\ell=2$) deformation.
The dipole deformation eventually becomes dominant after about 200~ms.
The other models also show growing power in the lower Legendre modes but do not exhibit repeated sign reversals.
B15-WH07 and B25-WH07 both have a sign reversal of the odd modes at about 130~ms after bounce, with the former showing a second and broader collective dip in the odd modes at $\sim$170~ms after bounce.
The suppression of the low-mode SASI oscillations is caused by the buoyancy of large, high-entropy bubbles that form in the expanding lobes of material behind the shock, which inhibit their recontraction \citep{FeMuFo14}. 
After $\sim$200~ms, the modes in all models appear to have become non-oscillatory, and, with the exception of B20-WH07, appear to be growing in proportion to the heating layer volume, while at the same time preserving their relative strengths.
This reflects the fact that the comoving shock and gain region pattern relative to the origin in these models has been frozen in at this time --- i.e., $G_{\ell}(r,t) \rightarrow G_{\ell}(r/\Rshock)$ in Equation~(\ref{SASI_power1}).
For B20-WH07, the mode power after $\sim$200~ms becomes almost constant, rather than increasing with the heating layer volume.
This is a consequence of the off-center and more sphere-like explosion geometry of this model (discussed further in Section~\ref{sec:Morphology}) causing the shock to have a more spherical shape relative to the center of the grid with time. 
The Legendre mode power therefore declines with time relative to the heating region volume.
For all of the models, the oscillatory character of the low Legendre modes switches to one of non-oscillatory growth with thermal runaway in the gain region and shock revival, as we discuss in the following sections.

There is also qualitative evidence of the SASI in our simulations.
A pattern of pole reversals in the radial velocity behind the shock is visible in the animated version of Figure~\ref{fig:entropy} for B12-WH07 with positive radial velocities behind the shock at one pole and negative radial velocities just below the shock at the opposite pole.
This pattern reverses polarity concurrently with the sign changes in the $\ell=1$ coefficient of the pressure fluctuation power (Figure~\ref{fig:mode_power}a) at 150, 170, and 185~ms.

\begin{figure}
\includegraphics[width=\columnwidth,clip]{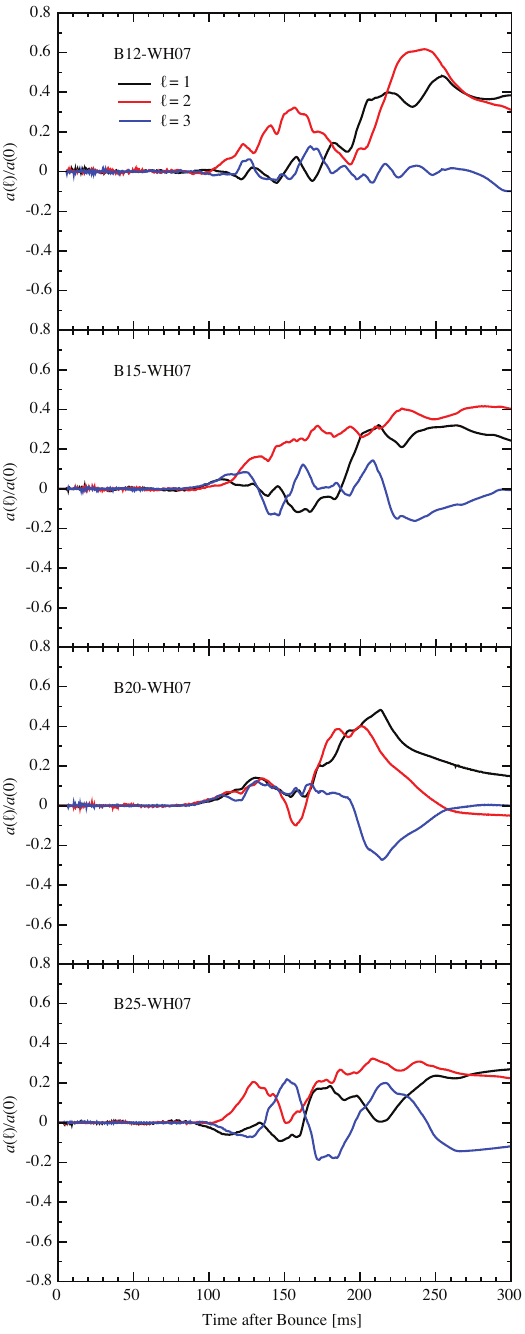}
\caption{\label{fig:SASI_shk} Legendre decomposition of the shock surface as a function of post-bounce time for each of the models. Shown are the three coefficients $a$(1), $a$(2), and $a$(3) normalized by the mean shock radius $a$(0).}
\end{figure}

\begin{figure}
\includegraphics[width=\columnwidth,clip]{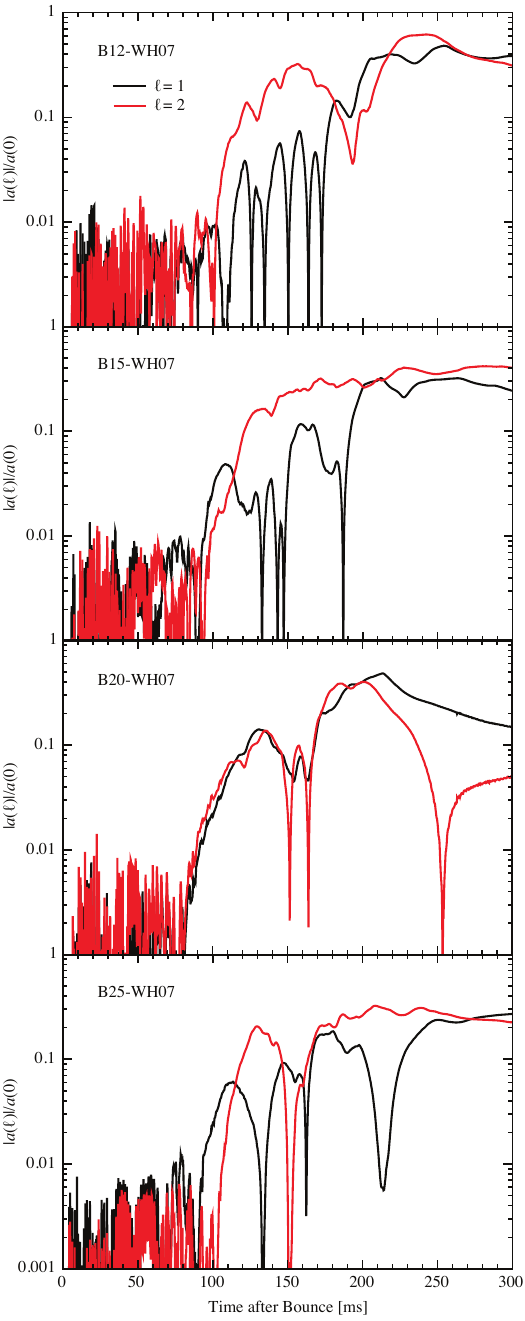}
\caption{\label{fig:SASI_shk_abs} Absolute values of the first two normalized Legendre coefficients of the shock surface plotted on a logarithmic scale to better discern their growth rates.}
\end{figure}

In order to compare our results more directly with those of \citet{MuJaHe12}, we decompose as they do the shock surface into Legendre polynomials
\begin{equation}
  R_{\rm shock}(\theta) = \sum_{\ell = 0}^{\infty} a(\ell) \sqrt{ \frac{2 \ell + 1 }{4\pi} } P_{\ell}(\cos \theta) 
\label{al_1}
\end{equation}
with expansion coefficients given by
\begin{equation}
  a(\ell) = \frac{ 2 \ell + 1 }{2} \int_{0}^{\pi} R_{\rm shock}(\theta) P_{\ell}(\cos \theta) \sin \theta d \theta
\label{al}
\end{equation}
and plot in Figure~\ref{fig:SASI_shk} $a(\ell)/a(0)$, the Legendre coefficients normalized to the mean shock radius for $\ell$ = 1, 2, and 3, and in Figure~\ref{fig:SASI_shk_abs} the absolute values of these normalized coefficients on a logarithmic scale for $\ell = 1$ and $\ell = 2$.

\begin{figure}
\includegraphics[width=\columnwidth,clip]{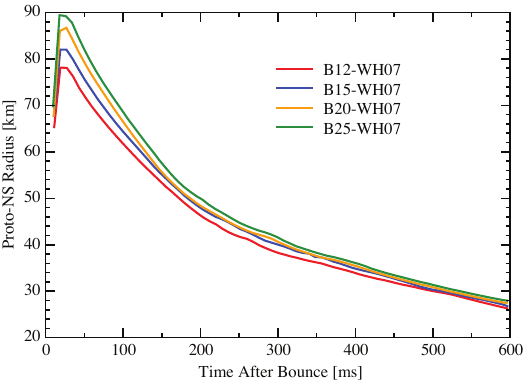}
\caption{\label{fig:NS_Radius}
Proto-NS radii for all models plotted in the colors of Figure~\ref{fig:entropy}. The proto-NS radius is defined as the radius at which the mean density is $10^{11}$~\gcc.}
\end{figure}
Figures~\ref{fig:SASI_shk} and \ref{fig:SASI_shk_abs} present a picture similar to that given by the fractional variation of the pressure given in Figure~\ref{fig:mode_power}, and indicate that each of our models exhibits some low mode oscillations for post-bounce times from $\sim$100~ms to $\sim$200~ms after which time the shock begins to move outward. 
These oscillations are evident for the $\ell = 1$ mode for B12-WH07, B15-WH07, and B25-WH07, with inferred growth rates of roughly 22, 18, and 16~s$^{-1}$, respectively, and possibly the $\ell = 2$ mode for model B20-WH07 .
These SASI growth rates are lower than the value of 45~s$^{-1}$ reported by \citet{MuJaHe12} for their model s27. Our lower growth rates are most likely due to the larger mean radius of our shocks at 100--150 ms post-bounce, as the oscillation frequency and growth rate of the SASI depend on the advection and sound travel times between the shock and the surface of the neutron star \citep{FoGaSc07}.
The low mode oscillations of our models exhibit considerable irregularity indicating that the SASI is being modulated by large scale convection, which is also occurring at this time.

It is tempting to find a parallel between the character of the fluid instabilities that first appear in the gain region after shock stagnation in our models, and the timing of the instabilities found by \citet{MuJaHe12} in their models u8.1 (8.1~\msun) and s27 (27~\msun).
They present detailed arguments to substantiate the fact that convection appears to arise first in the less massive model u8.1, while the SASI precedes convection in the more massive model s27.
The much larger mass accretion rate through the gain surface of their model s27, as compared with their model u8.1, inhibits convection while being conducive to the onset of the SASI.
Our models do indeed exhibit a delay of convective activity that correlates with their mass, as discussed in Section~\ref{sec:nuconvection}, but there is not a substantial period in any of our models during which the SASI is clearly present in the absence of convection.
A comparison of the ratio of advection to convection timescales $\chi$ (Figure~\ref{fig:early}a) for our simulations with $\chi$ for their models \citep[Figure~10 of][]{MuJaHe12} shows that their model s27 exhibits a considerably greater delay before $\chi$ attains the critical value of 3 than is the case for any of our models. 
This provides a greater period of time for the SASI to grow and become evident in their s27 model before convection can begin to grow. 
While the delay in convection in our models increases with their mass, convection does appear in all of our models and becomes large scale soon enough to substantially modulate the SASI.
Had convection been significantly more delayed in our more massive models as was the case in model s27 of \citet{MuJaHe12}, the SASI would likely have exhibited a similar trend to that found in \citet{MuJaHe12} and been manifested more strongly in our more massive models.
We suspect that the differences between the behavior of our models and those of \citet{MuJaHe12} are ultimately the result of the greater neutrino heating rates in our models, discussed in more detail in Section~\ref{sec:Models_CCSNe}.

\citet{ScJaFo08} in their parameter study of convection and the SASI showed that the contraction rate of the proto-NS as it radiates neutrinos can affect both the onset time and the vigor of convection.
A faster contraction leads to a greater pdV heating of the proto-NS and a more rapid release of gravitational energy in the form of neutrino radiation and therefore a more rapid heating of the gain region. 
For the sake of comparison with other work, we show in Figure \ref{fig:NS_Radius} the radii of the proto-NS's of our models as a function of post-bounce times.
We define the radius of a proto-NS as the radius at which the mean density is 10$^{11}$ g cm$^{-3}$.
The reduction in the proto-NS radii of our models between 100~ms and 200~ms varies from 75\% for B12-WH07 to 73\% for B25-WH07 compared with about 72\% for the fast contracting models of \citet{ScJaFo08}. 
(We note here, and further in Section \ref{sec:Models_CCSNe}, that the proto-NS radius evolution of our B15-WH07 is very similar to that of the 15~\msun\ model M15 of \citet{MuJaMa12} evolved from the progenitor S15s7b.)
Our rapidly contracting proto-NSs together with our relatively large heating rates could account for the early onset of convection in all of our models.

\subsubsection{Turbulent Pressure}
\label{sec:turbulentp}

\begin{figure}
\includegraphics[width=\columnwidth,clip]{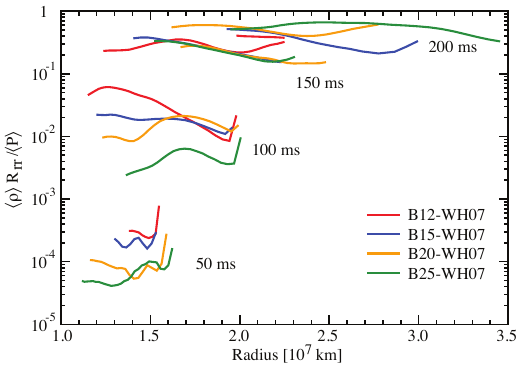}
\caption{\label{fig:B12_B25_Reynolds_Stress}
Ratio of the radial component of the Reynolds stress, $\langle \rho \rangle R_{rr}$, to the thermal pressure averaged over a time window of 5 ms centered at the indicated post-bounce times for our models.}
\end{figure}

\citet{MuDoBu13} and \citet{CoOt15} have identified an important role for the turbulent pressure in multi-D simulations that pushes out the shock and thereby aids its revival. 
To illustrate its importance for our models, we have calculated the radial component of the Reynolds stress, which dominates the radial turbulent force, and is defined by
\begin{equation}
  R_{rr} = \frac{ \langle \rho v'_{r} v'_{r} \rangle }{ \langle \rho \rangle }
\label{Reynolds}
\end{equation}
where $v'_{r}(r, \theta) = \langle v_{r}(r) \rangle - v_{r}(r, \theta)$, and the averages are over all angles for a given radius. 
We show in Figure \ref{fig:B12_B25_Reynolds_Stress} the ratio of the radial component of the Reynolds stress, $\langle \rho \rangle R_{rr}$ to the thermal pressure for our models averaged over a time window of 5 ms centered at the indicated post-bounce times. 
The range in radius for the plots is from the maximum of the gain radius to the minimum of the shock radius. 
Clearly the contribution of the Reynolds stress to the total pressure is unimportant during the first 100~ms after bounce, during which the shock radii of the axisymmetric simulations track those of the 1D simulations. 
However, as the strength of convection rises, the Reynolds stress becomes a significant fraction of the thermal pressure ultimately reaching 30--50\% of the latter at the beginning of shock revival ($\sim$200~ms).

\subsubsection{Accretion Streams}
\label{sec:streams}

\begin{figure}
\includegraphics[width=\columnwidth,clip]{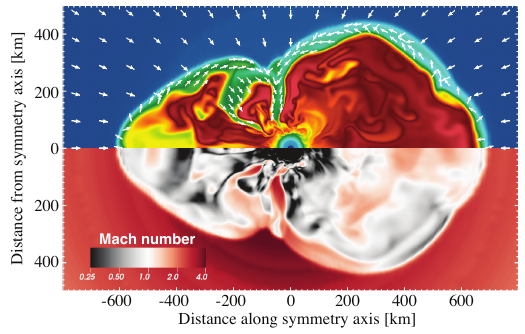}
\caption{\label{fig:stream}
Entropy (top; color scale as in Figure~\ref{fig:entropy}) plotted with selected velocity vectors and mirrored Mach number (bottom) for model B15-WH07 at 225~ms after bounce to illustrate accretion streams. All vectors are plotted with equal length.}
\end{figure}

Asphericity in the shock from the SASI and neutrino-driven convection focus the accretion into streams, which we have illustrated in Figure~\ref{fig:stream} for B15-WH07 at 225~ms after bounce using the entropy with velocity vectors in the accretion flow.
The pre-shock accretion is radial and the entropy is below 4~\kbbar, where \kboltz\ is Boltzmann's constant.
The shock jump conditions require that the passage of the accretion flow through the shock reduces the flow's velocity and increases its density and temperature.
Shock heating is exemplified by the increase in entropy to $\sim$5--10~\kbbar\ in the post-shock layer.
(The rising buoyant neutrino-heated plumes have entropy larger than $\sim$10~\kbbar.)
The shock jump conditions imply that the infall velocity component normal to the shock surface, but not the component(s) tangential to the shock surface, are reduced when passing through the shock.
In the supernova accretion shock, the unaltered tangential component of the incoming velocity results in a post-shock boundary layer flowing from positions with large shock radii toward those with small shock radii.
In this example, the post-shock accretion material flows from both poles toward the shock minimum near the equator, forming a shock triple point.
At a triple point, the shock surface can be very steep and nearly tangent to the radial inflow.
The large component of the infall velocity tangent to the shock near a triple point leads to small decrements in the inflow speed at the shock such that the post-shock flow can be supersonic as illustrated in Figure~\ref{fig:stream} (red; lower panel).

High-velocity and supersonic accretion streams forming at shock triple points have been discussed in 2D and 3D SASI simulations \citep[e.g.,][]{BlSh07,ScJaFo08,IwOhKo09,EnCaBu12}.
Convection can also produce large lobular distortions of the shock and associated local shock minima that focus accretion into streams.
Together they make accretion streams ubiquitous in 2D CCSN simulations.
Because the tangential component of the velocity at the shock is not altered by the jump conditions, less kinetic energy is thermalized at the shock and is therefore available for delivery to the proto-NS, where it can contribute to eventual neutrino heating.
Thus, not only do multi-D simulations allow simultaneous accretion and shock expansion, but the accretion streams themselves frequently are, or become, supersonic. 
These supersonic accretion streams form secondary shocks at the proto-NS surface and thereby deposit considerable thermal energy, much of which adds to the neutrino luminosity.
To the left of the equator in Figure~\ref{fig:stream} we can see a second accretion stream that is being cut off from the direct flow of accreted material from the shock.
During the simulations accretion streams are cut off and new ones form until the radial velocity becomes positive throughout the post-shock layer and no more accreted material is directed into the streams.
(See animated Figure~\ref{fig:entropy} for additional examples.)
The dynamic behavior of the accretion shock has not received the attention afforded the SASI and convection, but previous analyses using tracer particles from \chimera\ simulations has shown that much of the matter passing through the shock is directed toward the proto-NS in accretion streams \citep{ChMeHi12,LeBrHa12}.
We note, that the one or two accretion streams that dominate the accretion in 2D simulations is not characteristic of 3D simulations, which exhibit more numerous but weaker accretion streams \citep{TaHaMu13,TaRaHa14, HaMuWo13, AbOtRa15, TaKoSu14, TaRaHa14, LeBrHi15}.

\subsection{Onset of Explosion}
\label{sec:Expl_onset}

\begin{figure}
\includegraphics[width=\columnwidth]{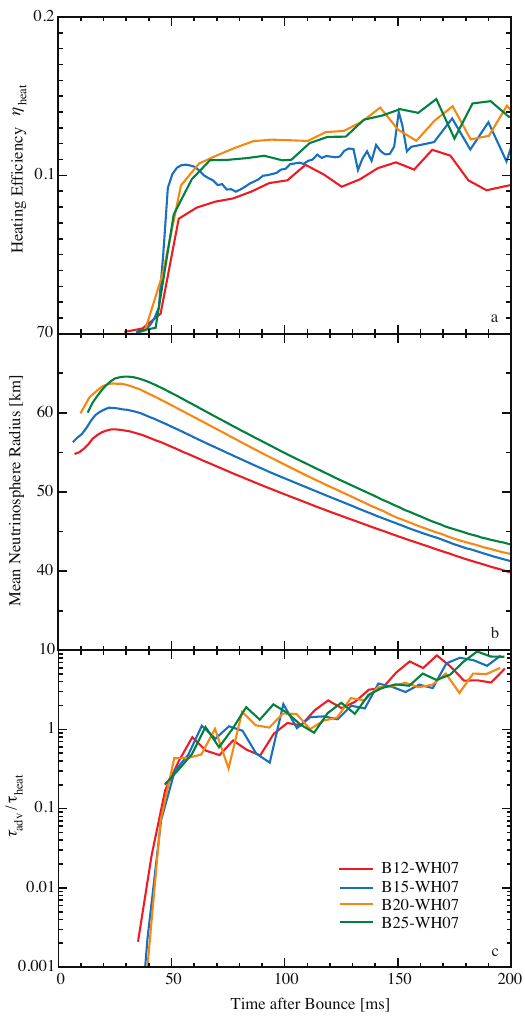}
\caption{\label{B12_B25_Heating}
Heating efficiencies ( $\eta_{\rm heat}$; Panel~a) computed using Equation~(\ref{eta_heat}); mean \nue--\nuebar\ neutrinospheres (Panel~b); and ratio of advection and heating timescales ($\tadv/\theat$; Panel~c) plotted versus time after bounce for all simulations using colors in Figure~\ref{fig:early}.}
\end{figure}

The evolution of the gain layer and the shock following core bounce is critically dependent on the neutrino heating that occurs in the gain layer.
The shock is pushed out to larger radii by the increase in the entropy and pressure behind the shock due to neutrino heating, the low-mode oscillations of the shock induced by the SASI, and by the ram pressure produced by the rising turbulent eddies in the convection driven by neutrino heating behind the shock. 
These processes were discussed individually in Sections~\ref{sec:nuconvection}--\ref{sec:turbulentp}.
Here we discuss the consequence of these processes in the onset of shock revival for our models.

Immediately following core bounce, the shock radius and the gain surface radius are very nearly coincident, forming a narrow heating layer.  
This is because the post-shock matter is very dense and relatively proton-rich at this time, causing energy losses by neutrino radiation accompanying electron capture to be high, \nue\ absorption on neutrons to be suppressed by the dearth of free neutrons, and \nuebar\ absorption on free protons to be low due to the moderately high post-shock electron degeneracy, which suppresses \nuebar\ production.
As the shock continues to move outward into lower density matter, energy gain by \nue\ and \nuebar\ absorption on free nucleons begins to dominate over energy loss by the inverse emission processes, and the radius of the gain surface begins to move inward relative to the shock, causing a heating layer to form and widen (see Figure~\ref{fig:early}c).
This happens as the post-shock densities fall below $\sim$$10^{10}$~\gcc, about 35~ms after bounce for the less massive models and 40~ms after bounce for the more massive models.
The growth of the heating layer from near zero width is reflected in the heating efficiency, $\eta_{\rm heat}$, shown in Figure~\ref{B12_B25_Heating}a and defined by Equation~(\ref{eta_heat}) as the ratio of the neutrino energy deposition rate in the heating layer divided by the sum of the \nue\ and \nuebar\ luminosities at the gain surface
\begin{equation}
  \eta_{\rm heat} = \frac{ 2 \pi \int_{0}^{\pi} d\theta \, \sin \theta \int_{r_{\rm gain}(\theta)}^{\rshock} \rho ( \dot q_{\nue} + \dot q_{\nuebar} ) r^{2} \, dr }{ \Lnue + \Lnuebar },
\label{eta_heat}
\end{equation}
where $\rho$ is the density, $\dot q_{\nue}$ and $\dot q_{\nuebar}$ are the net specific neutrino energy deposition rates by \nue\ and \nuebar, respectively, \Lnue\ and \Lnuebar\ are the neutrino-energy-integrated \nue\ and \nuebar\ luminosities at the gain surface, and $r_{\rm gain}(\theta)$ and \rshock\ are the radii of the gain surface and shock, respectively, as functions of latitude.
Figure~\ref{B12_B25_Heating}a shows that the  heating efficiencies of the models rise from essentially zero to about 10\% at the same time that the heating layer is forming, for the obvious reason that the presence of a heating layer is integral to the definition of $\eta_{\rm heat}$.
Once the heating layer has formed, the subsequent behavior of $\eta_{\rm heat}$ reflects a competition between (1) the decreasing density of this layer, which reduces the absorption opacities and hence the magnitude of $\eta_{\rm heat}$, as the mass accretion rate and pre-shock density decrease with time, and (2) the effect of convection and the SASI in moving the shock out and increasing the volume of this layer, together with the contraction of the neutrinospheres, shown in Figure~\ref{B12_B25_Heating}b, with the consequent rise in the neutrino luminosities and RMS energies (Figure~\ref{fig:early}d).
The competing effects on $\eta_{\rm heat}$ almost balance at this time, and the heating efficiency increases slowly, from about 70~ms to about 160~ms after bounce. 
The jaggedness of these graphs and some others, particularly after 100~ms, is due to the large-scale irregularities that develop as the convective cells grow in the gain region and the down flows become dominated by one or two spatially fluctuating accretion streams.
We note that simulations carried out in 3D have smaller more numerous convective cells and multiple accretion streams, and the corresponding graphs are much smoother.

As pointed out and discussed in \citetalias{BrMeHi13}, the shock radius evolutions of our models are quite similar to each other up to 200~ms after bounce. 
This similarity, at least up to 100~ms, was accounted for by the presence of compensating factors that determine the radius of the shock during its quasi-stationary accretion phase, as given by Equation~1 of \cite{Jank12}.
These factors are graphed for B12-WH07 and B25-WH07 in the top panel of Figure~3 of \citetalias{BrMeHi13} (the neutron star radii plotted there are half their actual values), and, though the separate factors are different for each model, their combination are in fact quite similar, as shown by the solid lines in that figure.

An essential consequence of the presence of convection and the SASI in the heating layer and the outward radial expansion of the shock due to these processes together with the effective turbulent ram pressure is an increase in the residency time of many of the fluid elements there, and therefore an increase in the time that they acquire energy by neutrino energy deposition.
We characterize the residency time of matter in the heating layer by a single parameter, \tadv, the advection time.
Because of the complex flow patterns that develop behind the shock, an alternative approach that better quantifies the complexities of the multidimensional flows is to examine the evolution of the residency time distribution function of the included passive tracer particles \citep{MuBu08,TaKoSu12,HaPlOd14}, which we will examine in a later paper.
We define an advection timescale, as suggested by \citet{BuJaRa06}, to be the interval of time from the time, $t_{\rm shock}(M)$, the shock encloses a given mass shell, $M$, to the time, $t_{\rm gain}(M)$, the gain surface encloses the same mass
\begin{eqnarray}
  \tadv(t(M)) &=& t_{\rm shock}(M) - t_{\rm gain}(M)\\
   t(M) &= &\frac{t_{\rm shock}(M) + t_{\rm gain}(M)}{2}.
\end{eqnarray}
While a fluid element resides in the heating layer, it is heated by neutrino energy deposition.
We characterize this process by a heating timescale \theat\ defined as
\begin{equation}
  \theat = \frac{\Eth}{\dot{Q}_{\nu}},
  \label{eq:tauheat}
\end{equation}
where \Eth\ is the total thermal energy of matter in the gain region and $\dot{Q}_{\nu}$ is the integrated net neutrino energy deposition rate in this gain region (the numerator of Equation~\ref{eta_heat}).
Thus \theat\ is roughly the temperature e-folding time.
We prefer the use of \Eth\ rather than $|E_{\rm tot}|$, which has been frequently used by other groups, where $E_{\rm tot}$ is the sum of the thermal, kinetic, and gravitational energies, as $E_{\rm tot}$ can be positive, or negative, or zero.
In our models, using $E_{\rm tot}$ rather than \Eth\ to compute \theat\ increases by $\sim$30\% the time from bounce for $\theat/ \theat$ to exceed unity.
(See Appendix~\ref{app:thermalE} for a full definition and discussion of the thermal energy.)
A ratio of $\tadv/ \theat >1$ indicates that matter passing from the shock to the gain surface will undergo substantial heating while in the gain region, and conditions will therefore become favorable for a thermal runaway and the revival of the shock \citep{Jank01, ThQuBu05, BuJaRa06}.
As can be seen in Figure~\ref{B12_B25_Heating}c, $\tadv/ \theat$ exceeds unity at $\sim$100~ms after bounce for all models and exceeds three about 50~ms later.

The rate at which $\tadv/ \theat$ rises to unity in our models B12-WH07 and B15-WH07 is somewhat more rapid than that for model u8.1 of \citet{MuJaHe12}, which exceeds unity at 110~ms after bounce and considerably more rapid than their model s27.0 which exceeds unity at 170~ms after bounce.
B15-WH07 briefly exceeds unity at 75~ms and both B12-WH07 and B15-WH07 exceed unity at 97~ms.
Both models B20-WH07 and B25-WH07 exceed unity at 80~ms.
There are a number of factors which could contribute to the differences between our models B20-WH07 and B25-WH07 and model s27.0 of \citet{MuJaHe12}.
They most likely result in greater neutrino heating rates for our models and we explore this further in Section \ref{sec:Models_CCSNe} where we make some detailed comparisons between our models and those of the Garching group.

The rise of $\tadv/ \theat$ above unity at $\sim$100~ms after bounce for all models is likely due to the similar shock radii evolutions of these models at this time, the shorter advection time scales of the more massive models (Figure~\ref{fig:early}b) being compensated by their shorter heating time scales due to their greater \nue\ and \nuebar\ luminosities (Figure~\ref{fig:Lum_Mean_RMS}, upper panel).
We suspect that the similarity in the time of the onset of explosion is due to the fact that our explosions are not marginal.
Had they been marginal, small differences in $\tadv/ \theat$ and other explosion indicators could amplify the differences in the time needed for an explosion to occur, even causing a failure to explode.

\begin{figure}
\includegraphics[width=\columnwidth,clip]{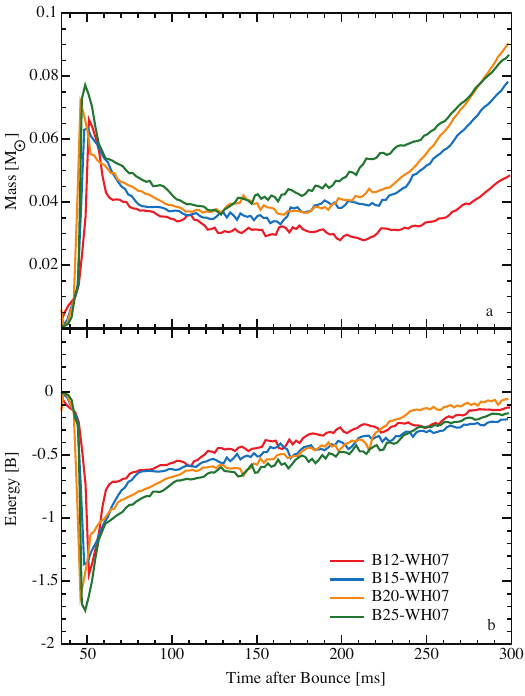}
\caption{\label{B12_B25_Gain}
Mass (a) and total energy (gravitational + thermal + kinetic) (b) of the gain region plotted versus time after bounce for all models using the same colors as Figure~\ref{fig:early}.}
\end{figure}

Another examination of the criteria for shock revival using an accretion shock model with neutrino radiation was performed by \citet{Jank01}. He found that shock revival required both the mass (Figure~\ref{B12_B25_Gain}a) and the total energy (Figure~\ref{B12_B25_Gain}b) of the gain region increase as a function of time.
In Figure~\ref{B12_B25_Gain}a  we can see  that the mass of the gain region for our models initially decreases with time after formation as the matter accreting through the shock becomes less dense.
Ultimately the gain region mass begins to increase as a combination of nonlinear SASI motions and large-scale convection pushing the shock out and increasing the volume and mass of the heating layer.
This turnaround in the mass of the heating layer occurs about 220~ms after bounce for model B12-WH07, and generally decreases with progenitor mass to about 120~ms after bounce for model B25-WH07.
Figure~\ref{B12_B25_Gain}b shows the evolution of total energy in the heating layer $E_{\rm tot}$ in our models.
The initial drop corresponds to the establishment of the gain region (see Figure~\ref{fig:early}c), which is initially gravitationally bound, with a strong return toward zero as the mass in the gain region rapidly drops after 50~ms.
After this initial transient, all models show steadily increasing total energy in the gain region as the neutrino energy deposition becomes more efficient (Figure~\ref{B12_B25_Heating}a).
Heating efficiency increases due to a hardening of the neutrino radiation after $\sim$100~ms, and due to the onset and growth of convective activity in the gain region which circulates cooler material down to the gain layer where it can more readily absorb neutrino energy than the hot material that it displaces.

A comparison of the masses in the gain region of our models with those (G15, M15, S15, and N15) of \citet{MuJaMa12} shows ours to be larger. 
At 100~ms, the gain region masses for their models are a little over 0.02~\msun\ while ours range from 0.035~\msun\ for B12-WH07 to 0.041~\msun\ for B25-WH07. 
We attribute, in part, the increase in gain region mass with progenitor mass exhibited by our models to the increase in the mantle density at a given radius with progenitor mass.
Our greater gain region masses in comparison to those of \citet{MuJaMa12} are likely due, in main part, to our shocks having greater radii ($\sim$200~km at 100~ms) than theirs ($\sim$150~km at 100~ms).

\subsection{Growth of Explosion Energy}
\label{sec:Expl_growth}

Some care is required in quantifying the energy of an explosion that is still developing.
The kinetic energies of the ejecta observed in supernovae and supernova remnants have their origin in the internal and kinetic energies imparted by the central engine.  
The conversion of this internal energy into kinetic energy, and the work to lift the stellar envelope out of the star's gravitational potential, occurs over thousands of seconds as the shockwave propagates toward the stellar surface \citep[see, e.g.,][]{GaGuPl10}.
For this reason, core-collapse supernova simulations, which typically cease after a second or less, have long used the total of internal (or thermal), kinetic, and gravitational potential energies when discussing their explosion energies.  
From the simulations reported herein, we find that a full second or longer is often required to extract a reasonable estimate of the total energy input by the neutrino reheating mechanism.

\subsubsection{Evolution of Unbound Region}
\label{sec:Unbound_region}

\begin{figure}
\includegraphics[width=\columnwidth,clip]{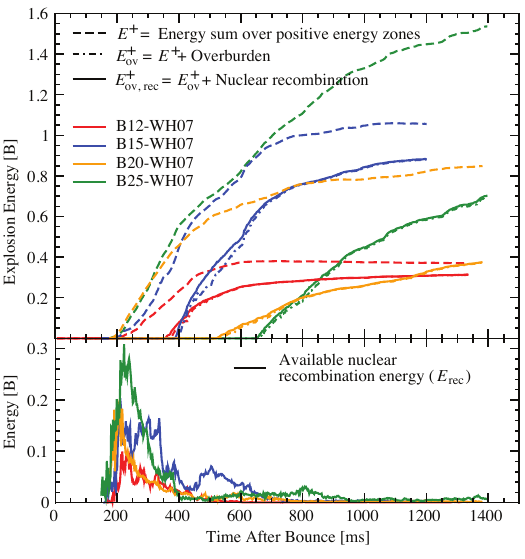}
\caption{\label{Expl_E_vs_t_12M_25M_Comp}
Panel a: Diagnostic energy, \Ediag\ (dashed lines), \Ediagov\ (dash-dotted lines) including binding energy of the overburden on and off the grid, and \Ediagovrec\ (solid lines) including  estimated gain from nuclear recombination, \Erecom, plotted versus time after bounce for all model using colors in Figure~\ref{fig:early}.\\
Panel b: Estimate of potentially recoverable nuclear recombination energy \Erecom\ as described in the text.}
\end{figure}

The measure of the explosion energy most widely used in discussions of CCSN models is referred to as the diagnostic energy, \Ediag, which is the volume integral of the total energy density, $\edtot = \edkin + \edth + \edgrav$, over all zones for which $\edtot > 0$ \citep{BuJaRa06, SuKoTa10, MuJaMa12, BrMeHi13}.
The thermal energy \edth\ is the internal energy minus the rest mass energy of all `conserved' particles.
For example, \edth\ excludes electron rest mass energies but not the rest mass energies of electron--positron pairs.
There are some minor variations in the calculation methods for \Ediag\ between the different groups.
We search for positive energy zones that lie between the shock and the proto-NS and make no restrictions on the velocity of the zone.
We label this unbound region as \Rdiag.
This measure of explosion energy increases monotonically with progenitor mass (with the exception of model B20-WH07, which is considered in Section~\ref{sec:Morphology}) from nearly 0.4~B for model B12-WH07 to nearly 1.6~B for model B25-WH07 at the time of this report on these simulations (see Table \ref{tab:outcomes}). 
Growth of \Ediag\ is depicted in Figure~\ref{Expl_E_vs_t_12M_25M_Comp}a with dashed lines.

\begin{figure}
\includegraphics[width=\columnwidth,clip]{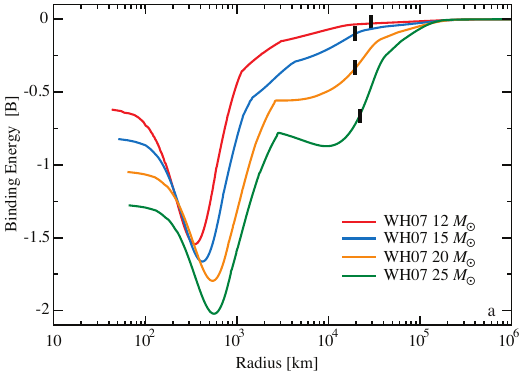}
\includegraphics[width=\columnwidth,clip]{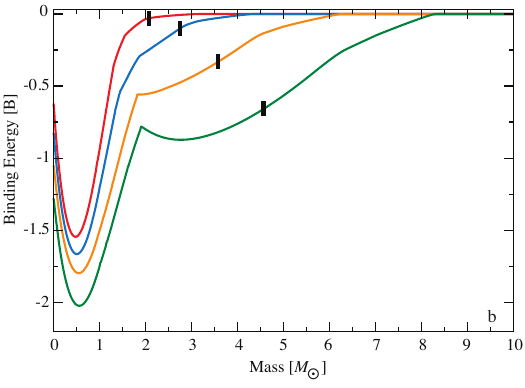}

\caption{\label{Env_Bind} Binding energy of the \citet{WoHe07} progenitor envelopes integrated from the outer edge to a given radius (a) and to a given mass mass (b) immediately prior to collapse.
Tick marks denote the edge of the computational grid for our simulations.}
\end{figure}

An additional estimate of the explosion energy, denoted by \Ediagov, that develops in our completed models takes into account the (negative) total energy of the material above this unbound region, the \emph{overburden}, both on and off the grid.
We refer to this negative contribution as the binding energy of the overburden.
The original progenitor binding energies are plotted in Figure~\ref{Env_Bind}, and the fractions of these progenitors mapped to the grids at the initiation of our simulations are indicated by the region interior to the tick marks. 
The off-grid overburden binding energies for the progenitors used in our simulations (vertical ticks in Figure~\ref{Env_Bind}) are -0.029, -0.100, -0.337, and -0.655~B, respectively, for the 12, 15, 20, and 25~\msun\ progenitors.
These binding energies do not change appreciably during the course of our simulations. 
The overburden energy that we consider is the total energy of all negative energy zones on the grid that lie above the innermost positive energy zones plus the total energy of the off-grid material.
The overburden-corrected diagnostic energy, $\Ediagov\equiv\Ediag+$ overburden energy, is plotted in Figure~\ref{Expl_E_vs_t_12M_25M_Comp}a (dash-dotted lines) and given at the time of this report in Table \ref{tab:outcomes}. 
It is delayed in growth relative to \Ediag, and reaches positivity  at about 350, 380, 530, and 650~ms after bounce, respectively, for B12-WH07, B15-WH07, B20-WH07, and B25-WH07.

Dissociation of heavy nuclei into free nucleons and alpha particles by the shock provides a mechanism to store a significant amount of energy, provided that neutrino heating later lifts that matter into the ejecta.  
This energy is released by neutron--proton and \isotope{He}{4} recombination to form heavier isotopes up to \isotope{Ni}{56} as the matter expands and cools.  
We have attempted to quantify the amount of this instantaneously available potential nuclear recombination energy, \Erecom, in the unbound matter (Figure~\ref{Expl_E_vs_t_12M_25M_Comp}b).
The true recombination energy is determined by the final chemical composition of the matter, thus an estimate of \Erecom\ requires an estimate of the final composition of each parcel of unbound matter.  
This nuclear evolution will occur naturally within \chimera\ as the model evolves (at least to the extent that an \alp-network can compute \alp-rich freezeout), but we wish to make a reasonable prediction at earlier times to gauge the measure of this term in the energy budget.  
We estimate \Erecom\ by assuming that all unbound material or overburden with a density $\rho > 10^{9}$~\gcc\ and a temperature $T > 3 \times 10^{9}$~K  will experience an NSE freeze-out in which all neutron--proton pairs combine into \alp-particles, and all \alp-particles combine to form \isotope{Ni}{56}.
In all other cases, we assume an \alp-rich freeze-out occurs in which neutron--proton pairs combine to \alp-particles, but the buildup of \alp-particles to \isotope{Ni}{56} does not occur.
This is a crude prescription, but sufficient to estimate the amount of energy involved.  
The evolution of the composition and thermodynamic state of the matter over time within \chimera\ will gradually tap this source of potential energy as Figure~\ref{Expl_E_vs_t_12M_25M_Comp}b illustrates, and a more sophisticated approximation of the freeze-out details is not warranted merely for an estimate of the impact on explosion energy.  

Our final estimate of the explosion energy, \Ediagovrec, adds \Erecom\ in the unbound region, that is, $\Ediagovrec\equiv\Ediagov+\Erecom$  (solid lines in Figure~\ref{Expl_E_vs_t_12M_25M_Comp}a and Table~\ref{tab:outcomes}).
At the termination of our simulations \Ediagovrec\ is 0.31, 0.88, 0.38, and 0.70 B, respectively, for B12-WH07, B15-WH07, B20-WH07, and  B25-WH07. 
\Erecom\ is 0.0013, 0.0021, 0.0019, and 0.0074 B, respectively, at this late time as most of the material capable of recombining has already recombined. 
As is evident from these figures, during the epoch when \Ediagovrec\ is still negative, \Erecom\ reaches 0.1--0.3~B, depending on the model, as some of the unbound material has densities exceeding $10^{9}$~\gcc\ and temperatures such that it can expected to undergo an NSE freeze out.
A few 100~ms later, this material has expanded and cooled, and \chimera\ has affected the NSE freeze-out, first through the equilibrium shift of lighter to heavier nuclei, and then through the included \alp-network when conditions no longer support NSE.
Simultaneously, the shock has moved into lower density material while the proto-NS radius and gain surface have retreated inward so that little new material with $\rho > 10^{9}$~\gcc\ becomes unbound.
Therefore, relatively little material remains with the potential to release recombination energy by building \isotope{Ni}{56} from lighter nuclei.
We expect that when the neutrino mechanism has completed its work, when \Ediag\ becomes constant, \Ediag\ and $\Ediagov \approx \Ediagovrec$ will bound the eventual observable explosion energy.
We discuss more sophisticated analyses of energy evolution in Sections~\ref{sec:volumeanalysis} and~\ref{sec:sourcesofediag}.

\begin{figure}
\includegraphics[width=\columnwidth,clip]{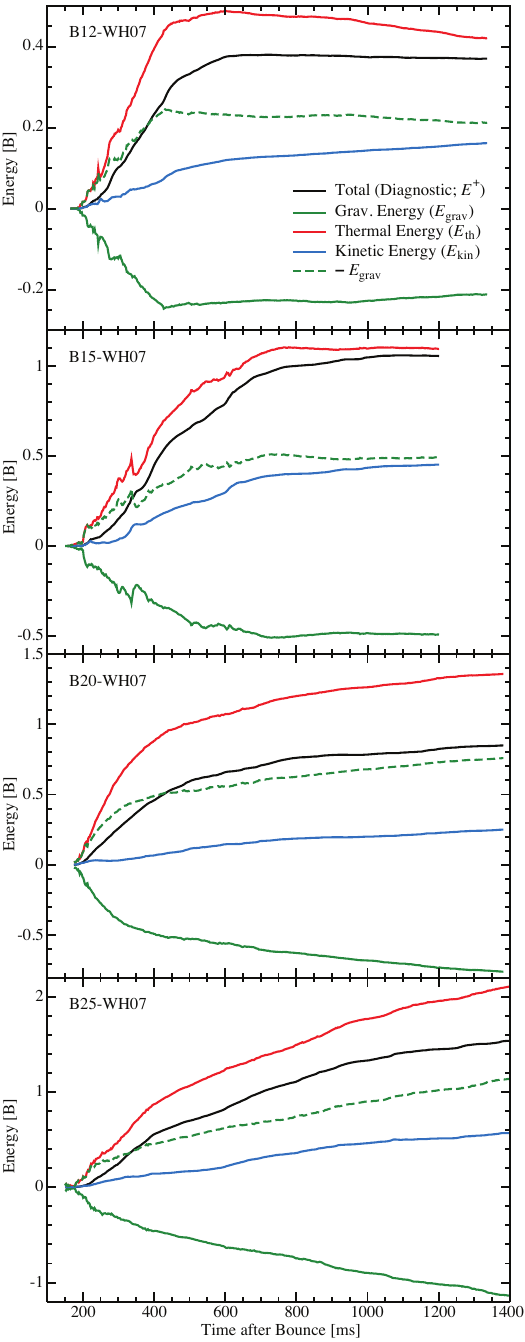}
\caption{\label{B12_B25_Energy_Components} Total diagnostic energy (\Ediag; black lines), and its components kinetic (\Edkin; blue lines),  gravitational (\Edgrav; green lines), and thermal (internal minus particle rest masses, \Edint; red lines) energies  for the unbound material  in all models plotted versus time after bounce. The gravitational energy has also been plotted with the opposite sign ($-\Edgrav$; green dashed lines) for comparisons with the other components.}
\end{figure}

As indicated in Figure~\ref{Expl_E_vs_t_12M_25M_Comp}a, the explosion energy has not completely saturated in any of our simulations even after a full second of shock expansion.
However, the energy \Ediagovrec\ for model B12-WH07 is increasing at a rate of only 0.03~\Bethes.
At this rate, more than three additional seconds of evolution would be required for \Ediagovrec\ for B12-WH07 to increase by an additional 0.1~B.
Thus we feel that the explosion energy of about 0.3~B for this model, achieved at the time of this report, is a representative value of its final explosion energy, though a slight underestimate.
For the other models, the rate of increase in \Ediagovrec\ of 0.15, 0.19, and 0.52~\Bethes\ for B15-WH07, B20-WH07, and B25-WH07, respectively, at the time of this report is non-negligible.
Clearly the final energies of these models will be larger than the energies quoted above.

\begin{figure}
\includegraphics[width=\columnwidth,clip]{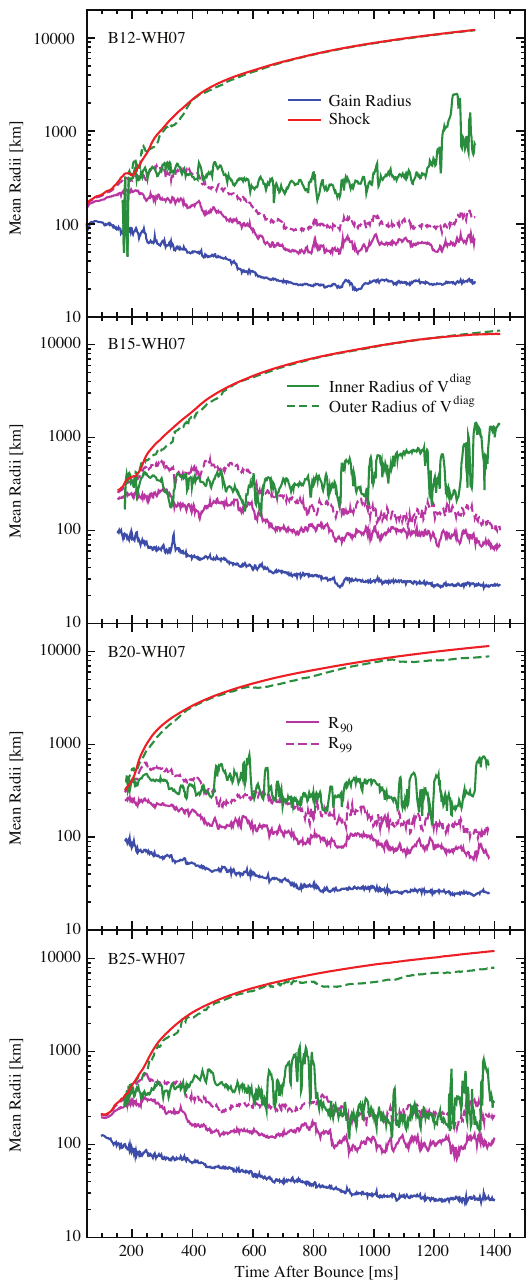}
\caption{\label{B12_B25_Heating_vs_tb}
Mean properties of heating in relation to the unbound region, \Rdiag, plotted for all models versus time after bounce. Solid red lines show the mean shock radius and solid blue lines the mean gain radius, which is also the inner radius of the heating region. The radii enclosing 90\% ($R_{90}$) and 99\% ($R_{99}$) of the neutrino heating are plotted with solid and dashed magenta lines, respectively. Mean radii of the inner and outer boundaries of the unbound region, \Rdiag, are plotted with solid and dashed green lines, respectively. Note that most of the neutrino heating is occurring below the unbound region.}
\end{figure}

Figure~\ref{B12_B25_Energy_Components} shows the diagnostic energy (\Ediag; black lines) for each model together with the components that comprise it: gravitational energy ($\Edgrav$; green solid lines), thermal energy (\Edint; red lines), and kinetic energy (\Edkin; blue lines).
Each energy in the plot is that contained in the diagnostic volume \Rdiag, the volume containing material with positive total energy, therefore all energies are zero up to the starting points of the plot where the diagnostic energy first becomes positive.
The gravitational energy has also been plotted with the opposite sign (green dashed lines) to facilitate comparisons with the other energy components.
As the models evolve and the contributions of the central engine to the explosion energy declines and eventually ceases, we would expect the thermal energy to decrease as adiabatic expansion works to expand the fluid, increasing the kinetic energy and decreasing (in magnitude) the gravitational energy. 
Until the shock breaks through the stellar photosphere, however, the thermal and kinetic energy may undergo several out-of-phase variations caused by variations in the shock speed as it encounters layers with different density profiles \citep[e.g.,][]{GaGuPl10}.
In time, the gravitational energy will tend to zero and the ejecta will become completely unbound, reaching the final explosion state where kinetic energy is the dominant component of the total energy.
Our models show a progression in this process of engine termination and energy conversion from the most massive to the least massive progenitor.
Our least massive model, B12-WH07, is in the most advanced expansion phase where the diagnostic energy is nearly constant.
In the unbound region of B12-WH07 thermal energy decreases from 600~ms after bounce onward, gravitational energy is maximally negative near 400~ms after bounce, and the kinetic energy, while not yet dominant, is increasing.
This occurs as the total mass of the unbound material increases from approximately 0.07~\msun\ to 0.17~\msun\ to 0.39~\msun\ at 400, 600, and 1300~ms after bounce, respectively.
Model B15-WH07 is just beginning this phase, while all energy components in the unbound region \Rdiag\ of models B20-WH07 and B25-WH07 are increasing.
In these latter two models, the mass of material with positive diagnostic energies is still increasing rapidly, causing both the magnitude of the gravitational and the internal energy to increase.

\subsubsection{Localization of Neutrino Heating}
\label{sec:localneutrino}

To understand the development of \Ediag, other measures of explosion energy, and their relation to the neutrino mechanism, it is instructive to first ascertain where most of the neutrino heating is occurring.
Direct neutrino heating of the material in the unbound region is modest, as most of the neutrino heating occurs in still bound material below \Rdiag.
To quantify this we define along each radial ray, $\theta$, the radii of the inner edge, $r_{\rm min}(\theta)$, and outer edge, $r_{\rm max}(\theta)$, of the region along the ray where $\edtot > 0$, that is, of \Rdiag. 
In Figure~\ref{B12_B25_Heating_vs_tb}, we have plotted $R_{\rm min}$ and $R_{\rm max}$ (solid and dashed green lines, respectively), where $R_{\rm min}$ and $R_{\rm max}$ are defined by
\begin{equation}
  R_{\rm min/max} = \frac{ \int_{\Omega, \edtot(\theta) > 0} d\Omega \, r_{\rm min/max}(\theta)}{ \int_{\Omega, \edtot(\theta) > 0} d\Omega },
  \label{eq:Rminmax}
\end{equation}
and are the latitudinal means of $r_{\rm min}(\theta)$ and $r_{\rm max}(\theta)$ over all radial rays containing unbound material, excluding solid angle elements for which there is no unbound material on the corresponding radial ray.
Thus $R_{\rm min}$ and $R_{\rm max}$ are measures of the boundaries of the unbound region \Rdiag\ from which \Ediag\ is computed.
We can compare the above limits of the unbound region to $R_{90}$ and $R_{99}$ (solid and dashed magenta lines, respectively, in Figure~\ref{B12_B25_Heating_vs_tb}), the latitudinal means of the outer radii of volumes containing 90\% and 99\% of the total instantaneous net neutrino heating.
Specifically, we define $R_{x}$ as the latitudinal mean of $r_{x}(\theta)$,  given implicitly by
\begin{equation}
  x = 100 \times \frac{ \int_{\rgain}^{r_x(\theta)}  \rho(r,\theta) \qdotnu(r,\theta) r^2 dr}{ \int_{\rgain}^{\rshock}  \rho(r,\theta) \qdotnu(r,\theta) r^2 dr },
  \label{eq:Rx}
\end{equation}
where $r_{x}(\theta)$ is the outer radius of the volume containing x\% of the net neutrino heating along the radial ray $\theta$.
In Equation~(\ref{eq:Rx}) $\qdotnu(r,\theta)$ is the net neutrino energy deposition rate per unit mass, \rgain\ is the gain radius, and $\rshock$ is the radius of the shock.
We also show in Figure~\ref{B12_B25_Heating_vs_tb} the mean radius of the lower boundary of the heating layer (the mean gain radius) with blue lines.
As can be seen from Figure~\ref{B12_B25_Heating_vs_tb}, there is only a small overlap between the unbound region \Rdiag\ and the region where the bulk of the neutrino heating is occurring, indicating that most of the neutrino heating is occurring below the zones of positive total energy.
It should be noted that while Figure~\ref{B12_B25_Heating_vs_tb} is strongly indicative of the small overlap between \Rdiag\ and the region where the bulk of the neutrino heating is occurring, the actual situation is somewhat complicated by the fluctuations of $r_{x}(\theta)$ about its angular mean value.

\begin{figure}
\includegraphics[width=\columnwidth,clip]{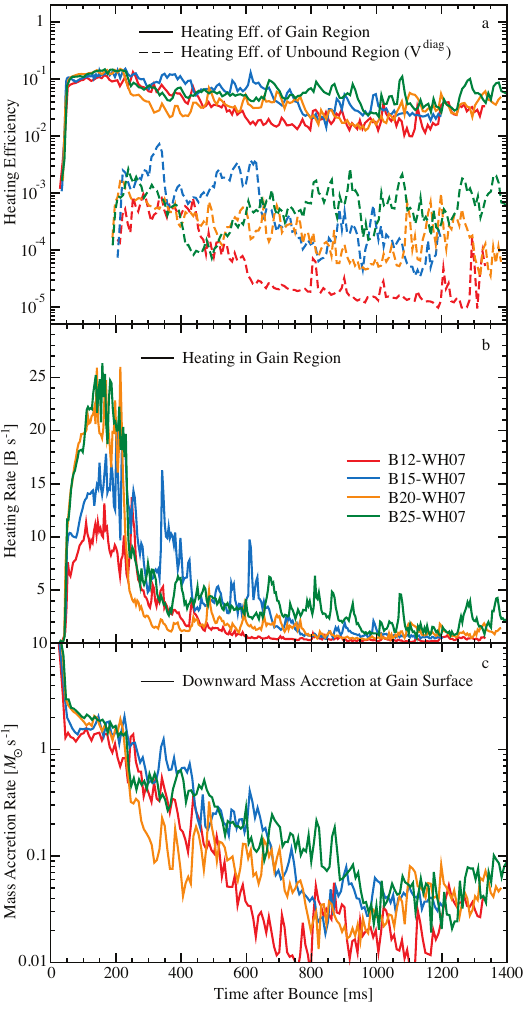}
\caption{\label{fig:heat}
Heating and accretion measures versus time after bounce for all models plotted with the colors used in Figure~\ref{fig:early}.
Panel~a: Heating efficiencies (ratio of the net neutrino energy deposition in region to $\Lnue + \Lnuebar$ at the gain surface) of the heating layer (solid lines) and unbound regions \Rdiag\ (dashed lines). 
Panel~b: Net total neutrino energy deposition rate in the heating averaged with a 4~ms time window.
Panel~c: Total inward ($v_r < 0$) mass accretion rate through the gain surface.
Note that the net accretion rate through the gain surface, the difference between the inward flow and the outward flow, is much smaller at late times than the inward flow shown here.}
\end{figure}

\begin{figure}
\includegraphics[width=\columnwidth,clip]{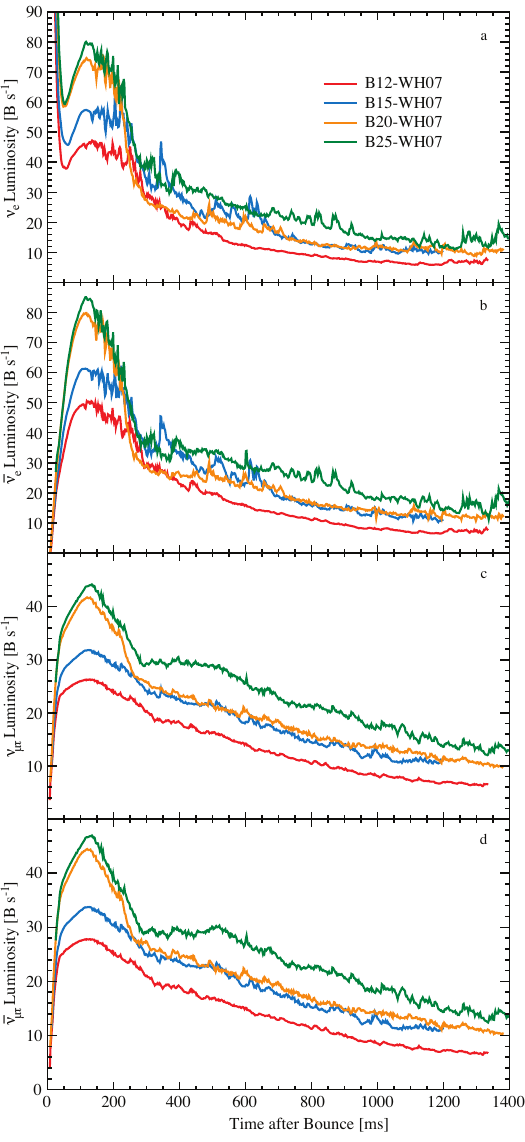}
\caption{\label{fig:lumin} Total neutrino luminosities, $L_{\nu}$, through the 1000-km shell for (a) \nue, (b) \nuebar, (c) \numt, and (d) \numtbar\ versus time after bounce for all models plotted in the colors of Figure~\ref{fig:early}.}
\end{figure}

\begin{figure}
\includegraphics[width=\columnwidth,clip]{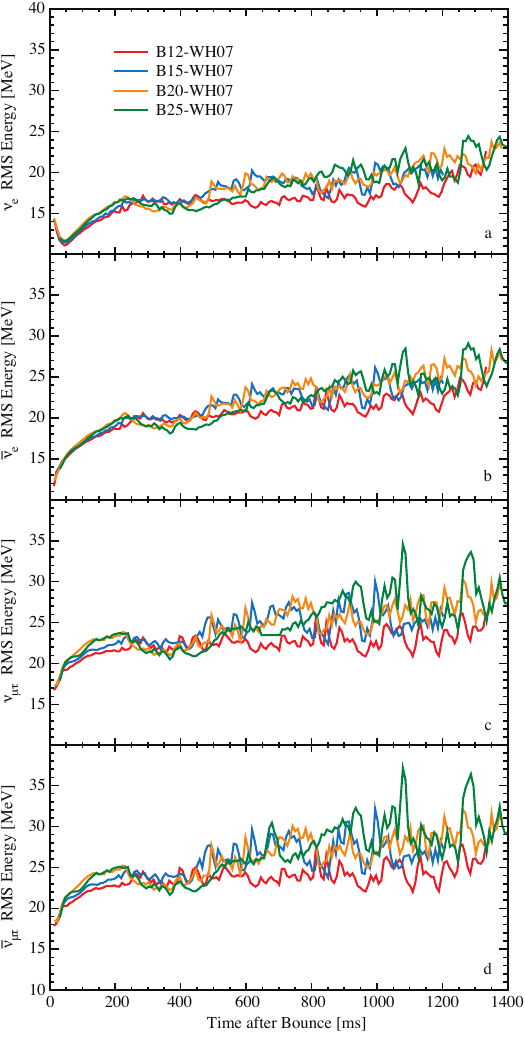}
\caption{\label{fig:ERMS} Neutrino RMS energies, $\epsilon_{\nu}$, integrated over all directions at the 1000-km shell for (a) \nue, (b) \nuebar, (c) \numt, and (d) \numtbar\ versus time after bounce for all models plotted in the colors of Figure~\ref{fig:early}.}
\end{figure}

Another indicator of the relatively small impact of direct heating in the unbound region \Rdiag\ are the orders-of-magnitude differences between overall heating efficiency (Equation~\ref{eta_heat}) of the heating layer (Figure~\ref{fig:heat}a; solid lines) and the heating efficiency, $\eta^{\rm diag}$, of the unbound region (dashed lines).
The heating efficiencies of both the heating layer and the unbound region rise from zero as the volumes of the relevant regions grow from zero. 
Following the initial rise of $\eta^{\rm diag}$, there is, with the exception of model B25-WH07, a slow decline (amid large fluctuations) as the neutrinospheres, the gain surface, and the region of strong neutrino heating retreat inward from  \Rdiag, the results of the contracting proto-NS and the growth of the negative density gradient above the proto-NS. 
For model B25-WH07, $\eta^{\rm diag}$ is approximately constant aside from a significant dip near 400~ms after bounce.
Another feature of B25-WH07 is the overlap of $R_{\rm min}$ and $R_{99}$ after $\sim$800~ms after bounce.
Both of these features are plausibly linked to the still rapid growth of \Ediag\ in this model.

The total neutrino heating rates in the heating layer are shown in Figure~\ref{fig:heat}b. 
Like the overall heating efficiencies, they grow rapidly from zero with the width of the heating layers. 
The neutrino heating is a function of the luminosities (\Lnue\ and \Lnuebar, Figure~\ref{fig:lumin}a and b) and the square of the RMS energies (\Enue\ and \Enuebar, Figure~\ref{fig:ERMS}a and b).
Because the RMS energies are slowly increasing, due to the compression of the proto-NS and the mean neutrinospheres, but are otherwise fairly smooth, the shape of the neutrino heating curves tend to follow the shapes of the neutrino luminosity curves, each of which exhibits a steep rise followed by a decline (Figure~\ref{fig:lumin}). 
\Lnue\ and \Lnuebar\ (as well as $L_{\numt}$ and $L_{\numtbar}$) reach a post-breakout peak for all models between 100 and 150~ms after bounce, and the heating rates do as well.
There is a trend of increasing heating rates with progenitor mass for all models, which reflects increasing \Lnue\ and \Lnuebar\ with progenitor mass.
Following the peak in \Lnue, \Lnuebar, and the heating rates, these quantities decline with the mass accretion rate (Figure~\ref{fig:heat}c), as the latter provides the considerable accretion component of  \Lnue\ and \Lnuebar.
There are several other features to note.
The accretion rate just outside the shock for B12-WH07 and B15-WH07 is smoothly declining during shock stagnation and revival, but for B20-WH07 at about 200~ms and B25-WH07 at 220~ms after bounce, this accretion rate drops several-fold in a few milliseconds corresponding to a density decrement in the progenitors structure.
This is reflected in large drops in the heating rates for these models at these times.
Interestingly, the mass accretion rates through the gain surface, the neutrino luminosities, and the neutrino heating rates all show a particularly sharp decline for B20-WH07 starting at about 200~ms, although there is no corresponding drop in the mass accretion rate through the shock. 
In Section~\ref{sec:Morphology} we will examine this behavior of model B20-WH07 relative to the other models relating to the particular morphology of the explosion of B20-WH07.
In all four models, even the well developed explosion of B12-WH07, accretion continues at a noteworthy rate ($>0.01$~\msuns) at the time of this report.
We finally note that following the post-bounce time of $\sim$450~ms there is a decline in the neutrino heating efficiencies of model B12-WH07 relative to those of the other models as the RMS energies, \Enue\ and \Enuebar, for this model fail to rise as fast as those of the other models during this time.

Comparing our heating rates and heating efficiencies of our model B15-WH07 with those of the 15~\msun\ models M15 and G15 of \citet{MuJaMa12}, evolved from the quite different progenitor S15s7b2 of \citet{WoWe95}, we note that ours are larger.
Our heating efficiencies peak at about 10\% at 150--200~ms while theirs peak at 8.5\% and 7.5\% for their models G15 and M15, respectively, and occur a little earlier.
The total heating rate in the gain region for B15-WH07 peaks at about 16~\Bethes\ while those for their G15 and M15 peak at 9 and 7~\Bethes, respectively.
These differences are likely due to the larger gain layer mass of model B15-WH07 and differences in the neutrino rates.
The latter is discussed further in Section~\ref{sec:Models_CCSNe}.

\subsubsection{Fixed Volume Energy Analysis}
\label{sec:volumeanalysis}

\begin{figure}
\includegraphics[width=\columnwidth,clip]{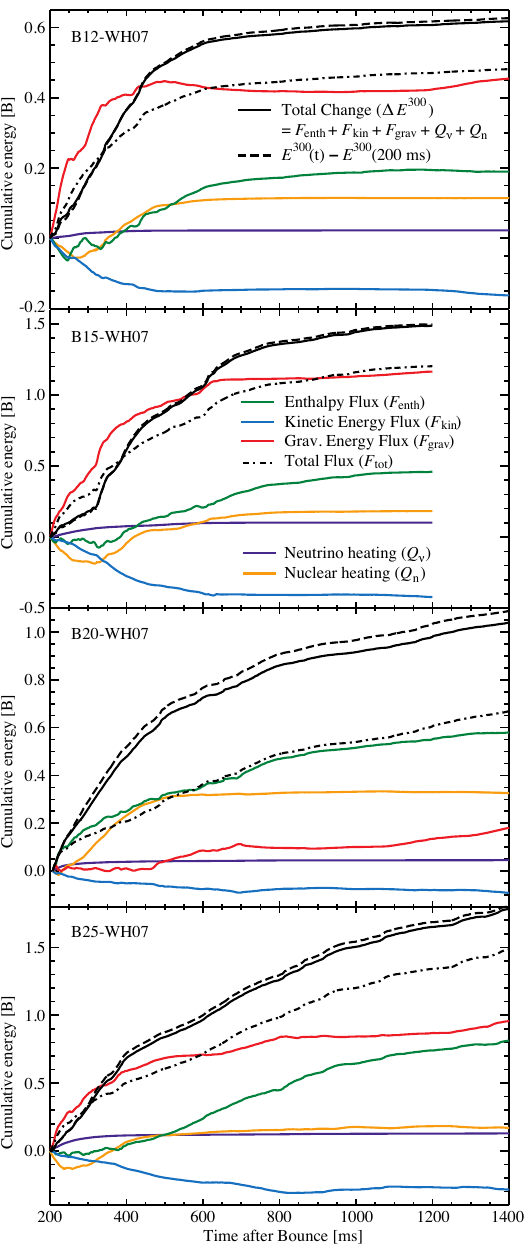}
\caption{\label{fig:cumeng300} Cumulative energy inputs beginning at 200~ms after bounce to the volume, \Rsph{300}, which approximates \Rdiag, and extends from 300~km to the outer edge of the simulation. The cumulative change, \dEvol{300}(t), in the total energy of \Rsph{300} evolves with contributions from the surface fluxes of enthalpy (\FluxCum{enth}{300}; green lines), kinetic energy (\FluxCum{kin}{300}; blue lines), and gravitational energy (\FluxCum{grav}{300}; red lines), and from the volume contributions of neutrino energy deposition (\Qnu; purple lines) and nuclear energy release (\Qnuc; orange lines). The contribution to \dEvol{300}(t) from all the surface fluxes is shown by the dot-dashed black limes. The total cumulative contribution from all volume sources and surface fluxes, which  is \dEvol{300}(t) by definition, is shown by the solid black lines. The change in the energy of \Rsph{300} computed by the volume integral of total energy, $\Evol{300}(t)-\Evol{300}$(200~ms), is plotted with black dashed lines.}
\end{figure}

\begin{deluxetable*}{lccccc}
\tabletypesize{\scriptsize}
\tablecaption{Summary of cumulative energy input\label{tab:sources}}
\tablecolumns{6}
\tablewidth{0pt}
\tablehead{
\colhead{} & \multicolumn{5}{c}{Models} \\
\cline{2-6} \\
\colhead{Cumulative energy input [B]} & \colhead{B12-WH07} & \colhead{B15-WH07} & \colhead{B20-WH07} & \colhead{B25-WH07} 
}
\startdata
$\nu$-heating (\Qnu) & 0.021 & 0.099 & 0.046 & 0.124 \\
Nuclear heating (\Qnuc) & 0.116 & 0.190 & 0.327 & 0.174 \\
Enthalpy flux (\FluxCum{enth}{300}) & 0.190 & 0.461 & 0.580 & 0.812 \\
Kinetic energy flux (\FluxCum{kin}{300}) & -0.163 & -0.422 & -0.093 & -0.285 \\
Gravitational energy flux (\FluxCum{grav}{300}) & 0.455 & 1.163 & 0.184 & 0.956 \\
Total flux (\FluxCum{tot}{300}) &  0.481 & 1.202 & 0.670 & 1.484 \\
Total energy change in \Rsph{300} (\dEvol{300}) & 0.618 & 1.495 & 1.043 & 1.782 \\
\cutinhead{Heating and influx approximation to \Ediag}
Gravitational energy influx (\FluxCum{grav}{in,300}) & -0.452 & -1.008 & -0.757 & -1.437 \\
Kinetic energy influx (\FluxCum{kin}{in,300}) & 0.106 & 0.289 & 0.226 & 0.468 \\
Enthalpy influx (\FluxCum{enth}{in,300}) &  0.691 & 1.664 & 1.232 & 2.230 \\
Total influx (\FluxCum{tot}{in,300}) &  0.346 & 0.945 & 0.701 & 1.261 \\
\Ediagapprox\ = \FluxCum{tot}{in,300} + \Qnu + \Qnuc &  0.484 & 1.233 & 1.074 & 1.558 \\
Diagnostic energy (\Ediag - \Ediag(200 ms)) &  0.367 & 1.054 & 0.828 & 1.521
\enddata
\end{deluxetable*}

To develop a quantitative measure of the energy inputs responsible for the growth of the diagnostic energy and related measures of the explosion strength, we look now at the sources of the diagnostic energy \Ediag.
Ideally we would consider the energy fluxes into and out of the unbound region \Rdiag, but this is complicated by its irregular shape and the sudden changes to its volume as additional zones become unbound or bound.
Instead, we select a fixed ``analysis volume'' \Rsph{300}, extending from a radius 300~km to the outer simulation boundary, and obtain approximate values for the energy fluxes into and out \Rdiag\, by examining the energy fluxes into and out of \Rsph{300}.
We use the same volume for all simulations as it roughly corresponds the unbound regions, \Rdiag, for each of these simulations as depicted in Figure~\ref{B12_B25_Heating_vs_tb}.
We briefly explore below the consequences of choosing other lower boundaries for the analysis volume.

In Appendix~\ref{app:energy}, we derive the energy rate of change equation for a fixed volume (Equation~\ref{eq:dvolenergydt}).
Integrating this equation from an initial time of 200~ms (roughly when \Ediag\ becomes noticeably positive as shown in Figure~\ref{Expl_E_vs_t_12M_25M_Comp}a) gives Equation~(\ref{eq:volenergy}).
Applying this equation to the analysis volume \Rsph{300} yields the cumulative change to the total energy \Evol{300} as
\begin{equation}
\dEvol{300}(t) = \Qnu(t) + \Qnuc(t) + \FluxCum{enth}{300} + \FluxCum{kin}{300} + \FluxCum{grav}{300}.\label{eq:dE300}
\end{equation}
First, we ask if the cumulative change in the total energy, $\dEvol{300}(t)$, calculated by the time integral of all sources and net fluxes, is equal to the change in the volume integrated total energy since the analysis start time, $\Evol{300}(t)-\Evol{300}\mbox{(200~ms)}$.
A comparison is shown in Figure~\ref{fig:cumeng300} in which we plot $\dEvol{300}(t)$ from Equation~(\ref{eq:dE300}) as solid black lines and the cumulative change in the volume integrated total energy $\Evol{300}(t)-\Evol{300}\mbox{(200~ms)}$ as dashed black lines for each model.
The similarity of these two lines indicates the suitability of the approximations made in Appendix~\ref{app:energy} to define \dEvol{300}, the accuracy of the post-processed source integrations, and the energy conservation maintained by \chimera.

The cumulative contributions to $\dEvol{300}(t)$ (Equation~\ref{eq:dE300}) are shown in Figure~\ref{fig:cumeng300} and are the cumulative neutrino energy deposition in  \Rsph{300} (\Qnu; purple lines), the nuclear energy release in \Rsph{300} (\Qnuc; orange lines), and the cumulative net enthalpy flux (\FluxCum{enth}{300}; green lines), kinetic energy flux (\FluxCum{kin}{300}; blue lines), and gravitational energy flux (\FluxCum{grav}{300}; red lines) across the surface of  \Rsph{300}.
These quantities are defined in Appendix~\ref{app:energy}. 
The values at 1400~ms after bounce (1200~ms for B15-WH07) are tabulated for all models in the top block of Table~\ref{tab:sources}.

\subsubsection{Sources of Diagnostic Energy \Ediag}
\label{sec:sourcesofediag}

\begin{figure}
\includegraphics[width=\columnwidth,clip]{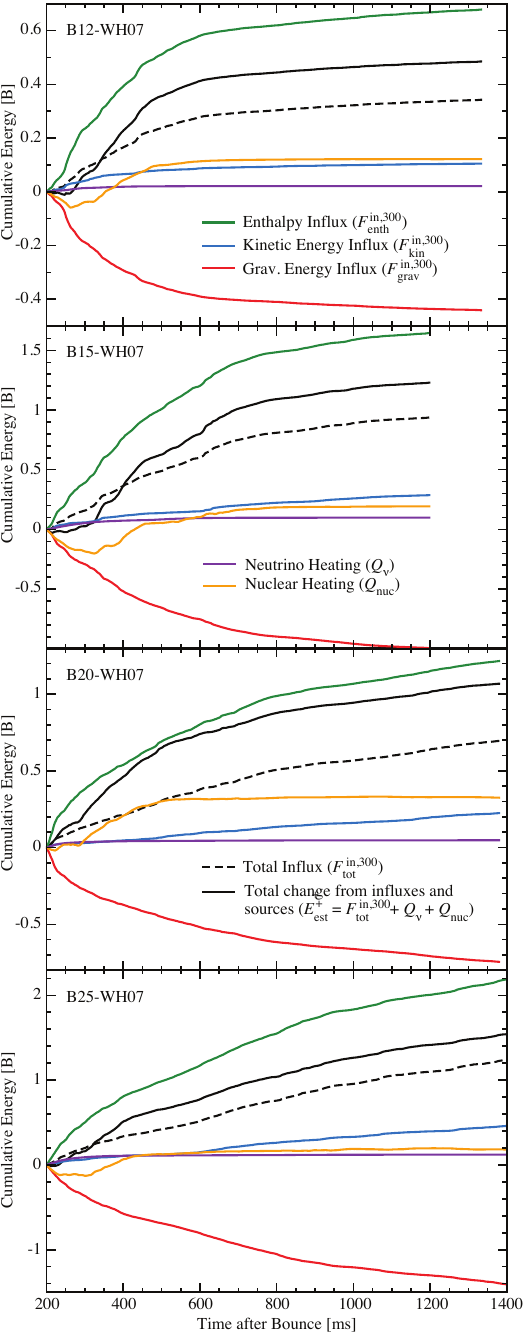}
\caption{\label{fig:influx300} 
Estimated inputs to the diagnostic energy (black lines), $\Ediagapprox \equiv \FluxCum{tot}{in,300} + \Qnu(t) + \Qnuc(t)$ from the cumulative energy influx and the volume sources to \Rsph{300} integrated from 200~ms after bounce. 
Contributing terms are the cumulative influxes of enthalpy (\FluxCum{enth}{in,300}, green lines), kinetic energy (\FluxCum{enth}{in,300}, blue lines), gravitational energy (\FluxCum{grav}{in,300}, red lines), which sum to the total energy influx (\FluxCum{tot}{in,300}; black dashed lines), and volume energy sources from direct neutrino heating (\Qnu, purple lines) and nuclear heating (\Qnuc, orange lines).}
\end{figure}

We now refine our analysis to better approximate the energy flows that are responsible for the evolution of \Ediag.  This consists largely of distinguishing between those energy flows that are apart of \Rdiag and those that are not.
Guided by animations of our simulations (e.g., animated Figure~\ref{fig:entropy}), we find that, from 200~ms after bounce and beyond, down flows through the lower boundary, $R_{\rm lower}=300$~km, of \Rsph{300} consist largely of accretion streams of bound material that, because they are bound, were never part of \Rdiag, while upflows across $R_{\rm lower}$ consist mainly of rising plumes of high-entropy material.
These buoyant plumes rising through $R_{\rm lower}$ are mostly unbound, or soon to become unbound, and therefore contribute to \Ediag.
To first approximation, then, we consider the advection of enthalpy, kinetic energy, and gravitational energy \emph{into} \Rsph{300} through $R_{\rm lower}$ as indicative of the \emph{net} advection of these quantities into \Rdiag, and ignore the outflow of material from \Rsph{300} at $R_{\rm lower}$, as this material is largely bound and is therefore not a part of \Rdiag.
Since the upper boundary of \Rdiag\ is contained within the \Rsph{300}, the small upper boundary fluxes are not needed in this approximation.

In Figure~\ref{fig:influx300} the various time integrated energy influxes to \Rsph{300} for each model are plotted  with colored lines, the total time integrated energy influxes, $\FluxCum{tot}{in,300} = \FluxCum{enth}{in,300}+\FluxCum{kin}{in,300}+\FluxCum{grav}{in,300}$, are plotted with dashed black lines, and the volume energy sources due to neutrino energy deposition ($\Qnu(t)$) and nuclear recombination energy ($\Qnuc(t)$) are plotted with purple and orange lines, respectively.
Summing these gives the estimated diagnostic energy, $\Ediagapprox \equiv \FluxCum{tot}{in,300} + \Qnu(t) + \Qnuc(t)$ (solid black lines).
The bottom section of Table~\ref{tab:sources} summarizes the contributions to our estimate of the diagnostic energy growth, \Ediagapprox, from the cumulative influxes and volume sources at the time our simulations were terminated.
Each of these contributions is discussed briefly below.

The negative contributions to \Ediagapprox\ from the cumulative gravitational energy influx \FluxCum{grav}{in,300}, (Figure~\ref{fig:influx300}, red lines) are proportional to the net inflow of matter into \Rsph{300} through $R_{\rm lower}$.
The gradual flattening of the slope of these lines reflects the declining rate of mass influx into \Rsph{300}  with time.

The cumulative influx of kinetic energy \FluxCum{kin}{in,300} (Figure~\ref{fig:influx300}, blue lines) contributes from $\sim$0.1~B (B12-WH07) to $\sim$0.46~B (B25-WH07) to \Ediagapprox. 
This is modest compared with the cumulative influx of enthalpy, \FluxCum{enth}{300} (Figure~\ref{fig:influx300}, green lines) which contributes from $\sim$0.68~B (B12-WH07) to  $\sim$2.2~B (B25-WH07) to \Ediagapprox, providing by far the largest positive contribution to \Ediagapprox.
The influx of kinetic energy and enthalpy continues at late times, unlike the nuclear and direct neutrino heating, and would likely provide most of the additional explosion energy to our models if the simulations were continued.
The influx of kinetic energy and enthalpy occurs simultaneously with the inflowing negative gravitational energy of that material. 
As can be seen in Figure~\ref{fig:influx300} (red lines) these negative gravitational energy influxes are non-negligible.
The total cumulative influx \FluxCum{tot}{in,300} (gravitational energy, enthalpy, plus kinetic energy) is plotted with black dot-dashed lines in Figure~\ref{fig:influx300} and the final values are given in the bottom section of Table~\ref{tab:sources}.
This cumulative total influx of energy dominates over the direct volumetric sources of energy provided by nuclear recombination and direct neutrino heating.
Interestingly, the relative importance of \Qnuc\ over \Qnu\ grows in weaker explosions, a factor that may influence prior analysis of the importance of nuclear energy generation in core-collapse supernovae \citep[see, e.g.,][]{NaTaKo14}.

The final values of \Ediagapprox\ in Table~\ref{tab:sources} can be compared to the diagnostic energies $\Ediag - \Ediag$(200~ms) at the end of the simulations, also listed in Table~\ref{tab:sources}.
(We list the change in \Ediag\ from its value at 200~ms to be consistent with the definition of \Ediagapprox). 
We see that \Ediagapprox\ makes a reasonable approximation to $\Ediag - \Ediag$(200~ms). 
The similarity of \Ediag\ and \Ediagapprox\ suggests that considering the influxes (\FluxCum{tot}{in,300}) at a fixed boundary and the heating terms within that volume (\Qnu\ and \Qnuc) captures the major features of the growth of the diagnostic energy \Ediag.

\begin{figure}
\includegraphics[width=\columnwidth,clip]{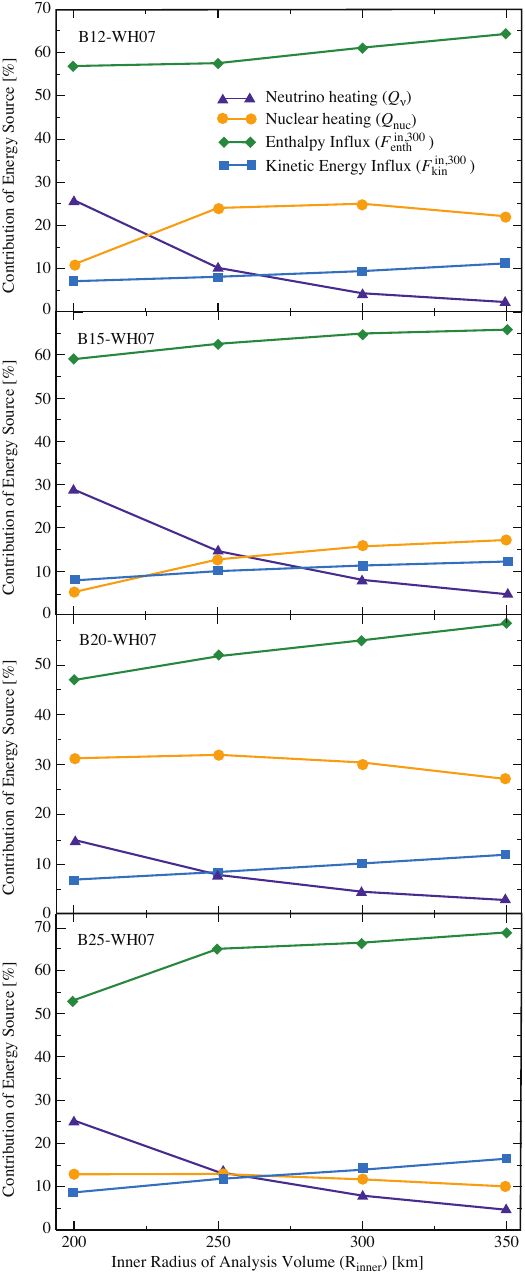}
\caption{\label{fig:B12_B25_Source_Summary} 
Relative contributions of direct neutrino heating (purple), nuclear heating (orange), kinetic energy influx (blue) and enthalpy influx (green) to the final energy of the analysis volume (\Ediagapprox) as a function of the radius of its lower boundary ($R_{\rm inner}$). The relative contribution of the kinetic energy and enthalpy influxes were computed by maintaining their relative magnitudes while scaling them so that their sum is the total influx (\FluxCum{tot}{in,300}). Lines are plotted to connect symbols.}
\end{figure}

To summarize the contribution of the various energy sources to \Ediagapprox, and to examine the consequences of our choice of 300~km for the lower boundary, $R_{\rm lower}$, of \Rsph{300}, we show in Figure~\ref{fig:B12_B25_Source_Summary} the relative contributions of direct neutrino heating, nuclear heating, kinetic energy influx, and enthalpy influx for various $R_{\rm lower}$ to the estimated explosion energy \Ediagapprox.
We account for the negative gravitational energy influxes in these percentages by maintaining the ratio of the kinetic energy and enthalpy influxes while scaling them so that their sum was that of the total energy influx ($\FluxCum{tot}{in,300} = \FluxCum{enth}{in,300}+\FluxCum{kin}{in,300}+\FluxCum{grav}{in,300}$).

As is evident from Figure~\ref{fig:B12_B25_Source_Summary}, for all models enthalpy influx is the dominant source of energy into the analysis volume regardless of the choice of lower boundary.
The relative contribution of direct neutrino heating increases for smaller $R_{\rm lower}$ as would be expected.
The relative contribution of the other two sources, kinetic energy influx and nuclear heating, are only mildly dependent on $R_{\rm lower}$. Both exhibit a modest decrease with decreasing $R_{\rm lower}$, except for the nuclear heating in model B12-WH07, where the decrease is large.
The primary conclusion of this section is that we find that the major source of energy input into volumes \Rsph{200} through \Rsph{350} excluding outflows, and by implication into \Rdiag, is the influx of enthalpy, contributing typically $\sim$60\% of the total energy. 
The contributions of the other three energy sources are each less than half that of the enthalpy flux, but their relative contributions will depend on the shape and, particularly, the radius of the lower boundary of \Rdiag.
After 500 to 800~ms the influx of enthalpy dominates the continuing rise \Ediagapprox, and therefore of the explosion energy \Ediag.

\subsubsection{Nuclear Transmutations}
\label{sec:Nuclear}

\begin{figure}
\includegraphics[width=\columnwidth,clip]{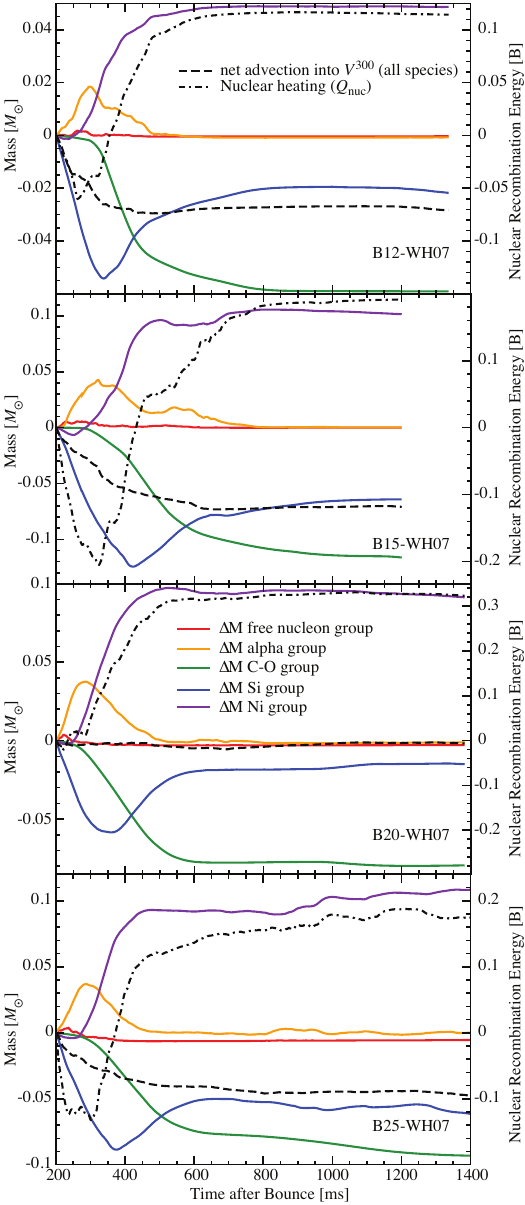}
\caption{\label{B12_B25_M_Nuclear_R300_Sum_vs_tb200}
``Flow'' of nuclear binding energy and nuclear masses relative to 200~ms after bounce for all simulations. 
The red, orange, green, blue and violet lines show the changes in the masses of free nucleon, alpha, C--O, Si, and Ni groups in \Rsph{300}, respectively.
The black dashed line shows the net mass of nuclei advected into \Rsph{300}.
The dot-dashed black line shows the net change of the nuclear binding energy in \Rsph{300} due to nuclear burning (\Qnuc).}
\end{figure}

As noted in Section \ref{sec:methods}, \chimera\ implements a 14-species \alp-network (\alp, \isotope{C}{12}-\isotope{Zn}{60}) to track nuclear transmutations among \alp-nuclei in zones that are not in NSE. 
We give a brief account here of the nuclear transmutations that occur in our simulations, focusing as before on the volume from 300 km out to the edge of our grid. 

The cumulative inputs to \Evolc{300}{th} from nuclear reactions (\Qnuc; Figure~\ref{fig:cumeng300}, orange lines), including dissociation/recombination and shock burning, provide a net input to the thermal energy inside \Rsph{300} of 0.116, 0.190, 0.327, and 0.174~B, respectively, for models B12-WH07, B15-WH07, B20-WH07, and B25-WH07 from 200 ms post-bounce to the end of the simulations.
To follow the sequence of nuclear transmutations that take place in \Rsph{300}, Figure~\ref{B12_B25_M_Nuclear_R300_Sum_vs_tb200} shows the evolution of the mass of key groups of nuclei within \Rsph{300}, together with the net advection of nuclei into and out of \Rsph{300}.
For convenience, the 17 nuclear species evolved in these simulations are grouped into 5 key groups, the `free nucleons' group consisting of free neutrons and protons; the `alpha' group, consisting only of \isotope{He}{4}; the C--O group consisting of \isotope{C}{12}, \isotope{O}{16}, and \isotope{Ne}{20}; the Si group consisting of \isotope{Mg}{24}, \isotope{Si}{28}, \isotope{S}{32}, \isotope{Ar}{36}, \isotope{Ca}{40} and \isotope{Ti}{44}; and the Ni group consisting of \isotope{Cr}{48}, \isotope{Fe}{52}, \isotope{Ni}{56}, \isotope{Zn}{60}, and the auxiliary heavy nucleus.

The evolution of the masses of these groups is a function of nuclear processes within the volume and advection into and out of the volume are chronicled in Figure~\ref{B12_B25_M_Nuclear_R300_Sum_vs_tb200}.
For B12-WH07 (Figure~\ref{B12_B25_M_Nuclear_R300_Sum_vs_tb200}a), the transmutation of the Si group by shock burning into $\alpha$-particles and a small quantity of free nucleons is the dominant nuclear process from 200 to 300~ms after bounce.  
Such dissociations produce the net decline in \Qnuc\ seen in this epoch in B12-WH07, as well as in B15-WH07 and B25-WH07.  
Much of this dissociated matter is advected downward, leading to a net loss in mass in \Rsph{300}\ for these models. 
For B20-WH07, this dip in \Qnuc\ is missing as the shock dissociation is occurring just inside 300~km.
(This feature would be visible in B20-WH07 if 250~km was used as the inner radius for this analysis.)  
Furthermore, in B20-WH07, the flux of the Si group downward across 300~km is well matched by the flow of $\alpha$-particles and free nucleons upward across 300~km, resulting in a near zero total mass flux.
Between 250 and 300~ms after bounce, recombination of $\alpha$-particles and free nucleons in the rising plumes behind the shock, which is by then well above 300~km, dominates over shock dissociation and drives an increase in \Qnuc.
As a result, the abundance of $\alpha$-particles in the volume peaks near 300~ms after bounce for all of the models.
Ni group production begins at this time, both as a result of this recombination of $\alpha$-particles behind the shock and from silicon burning in the shock. 
The total Ni group mass in \Rsph{300}\ reaches near final values by 400--500~ms after bounce, but, in the more massive models, this mass continues to show some evolution until the time of the reporting of these simulations. 
By this time, accretion through the 300~km surface consists primarily of material rich in $\alpha$-particles, while the rising plumes are rich in free nucleons.

Beyond this point in time, oxygen burning is the dominant nuclear process, continuing beyond 600~ms after bounce.  
Oxygen burning starts as early as 220~ms after bounce (in B12-WH07) as the most advanced parts of the shock progress into the outer portion of the silicon shell where \isotope{O}{16} represents a significant admixture in the composition dominated by \isotope{Si}{28} and \isotope{S}{32}.
In the 12 \msun\ progenitor, both the extent of this O-enriched silicon shell ($\sim$1900--2800~km) and its oxygen concentration ($\sim$4\% by mass) are small. 
In the 15 \msun\ progenitor, the extent remains small ($\sim$2800--3900~km), but oxygen represents a quarter of the mass.  
The more massive progenitors combine inner, small and mildly O-enriched regions ($\sim$5\% by mass from 2300--2700~km for 20 \msun\ or 2600--2900~km for 25 \msun) with much larger, more heavily enriched outer O-enriched silicon shells.  
In the 20 \msun\ progenitor, the oxygen mass fraction rises gradually from 30\% at 2700~km to 47\% at 8600~km, where it jumps sharply to 70\% as the actual oxygen shell is reached.
In the 25 \msun\ progenitor, the outer edge of the silicon shell exhibits a gradual decline in silicon group mass fraction with increasing radius, thus the oxygen mass fraction rises gradually from 25\% at 2900~km, leveling off near 73\% by 12000~km in radius.

Oxygen burning accelerates in all of the models as more of the shock reaches the O-enriched outer silicon layer, and as the admixture of oxygen increases with radius.  
For B12-WH07, the sharp increase in the rate of C--O group destruction visible at 290~ms in Figure~\ref{B12_B25_M_Nuclear_R300_Sum_vs_tb200}a corresponds with the shock reaching the oxygen layer, where the \isotope{O}{16} mass fraction jumps to 78\%.
Oxygen burning progressively decelerates in B12-WH07 between 450 and 800~ms after bounce as the post-shock temperature declines due to the expansion of the shock, causing oxygen burning to become less efficient.
For B15-WH07, oxygen burning is delayed until 270~ms after bounce by the larger initial radius of the O-enriched silicon layer, proceeds initially more rapidly than in B12-WH07 because of the higher \isotope{O}{16} enrichment in this layer and accelerates after the shock reaches the oxygen shell at 380~ms after bounce.
The deceleration of oxygen burning is delayed in this more strongly exploding model until $\sim$550~ms after bounce and some oxygen burning continues until 1 second after bounce.
For B20-WH07, the onset of oxygen burning occurs near 240~ms after bounce, when the shock first reaches the lightly enriched, inner part of the O-enriched silicon layer and accelerates near 280~ms after bounce as the shock reaches the more heavily \isotope{O}{16} enriched, outermost portion of the silicon shell.  
The shock does not reach the oxygen shell in B20-WH07 until 790~ms after bounce, well after oxygen burning has completed (at $\sim$600~ms after bounce), indicating that oxygen burning in this relatively underpowered model occurs exclusively in the silicon shell.
B25-WH07 exhibits early behavior similar to B20-WH07, with oxygen burning commencing about 250~ms after bounce and accelerating near 290~ms after bounce, however oxygen burning in this stronger explosion continues for more than 1~second after bounce.  
This is long after the shock reaches regions where \isotope{O}{16} is the dominant constituent ($\sim$650~ms after bounce), and even after the shock enters the silicon-depleted region where the abundance of oxygen (and other species) levels off ($\sim$900~ms after bounce).
Thus, in this more strongly exploding model, oxygen burning does progress into the oxygen shell.
Visible in Figure~\ref{B12_B25_M_Nuclear_R300_Sum_vs_tb200}d, the continued rise in \Qnuc\ within \Rsph{300} for this model beyond 500~ms after bounce, when the nickel and $\alpha$-particles masses have leveled off, indicates that oxygen burning is contributing appreciably to the total nuclear energy generation.
Similar behavior is exhibited in B15-WH07, but not for the more weakly exploding B12-WH07 and B20-WH07 models.

\begin{figure}
\includegraphics[width=\columnwidth,clip]{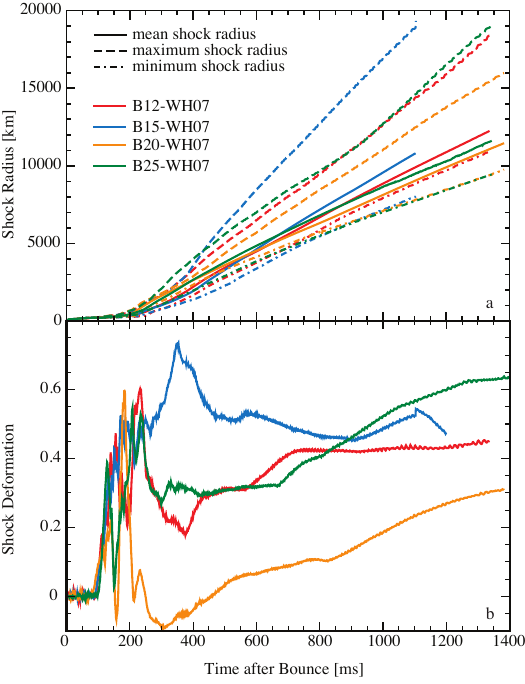}
\caption{\label{fig:shape}
Shock shape quantities plotted versus time after bounce for all models using colors of Figure~\ref{fig:early}. Panel a: Mean (solid lines), minimum (dot-dashed lines), and maximum (dashed lines) shock radii. Panel b: Shock deformation \dshock\ defined in Equation~(\ref{Shock_Deformation}).}
\end{figure}

\begin{figure*}
\includegraphics[clip]{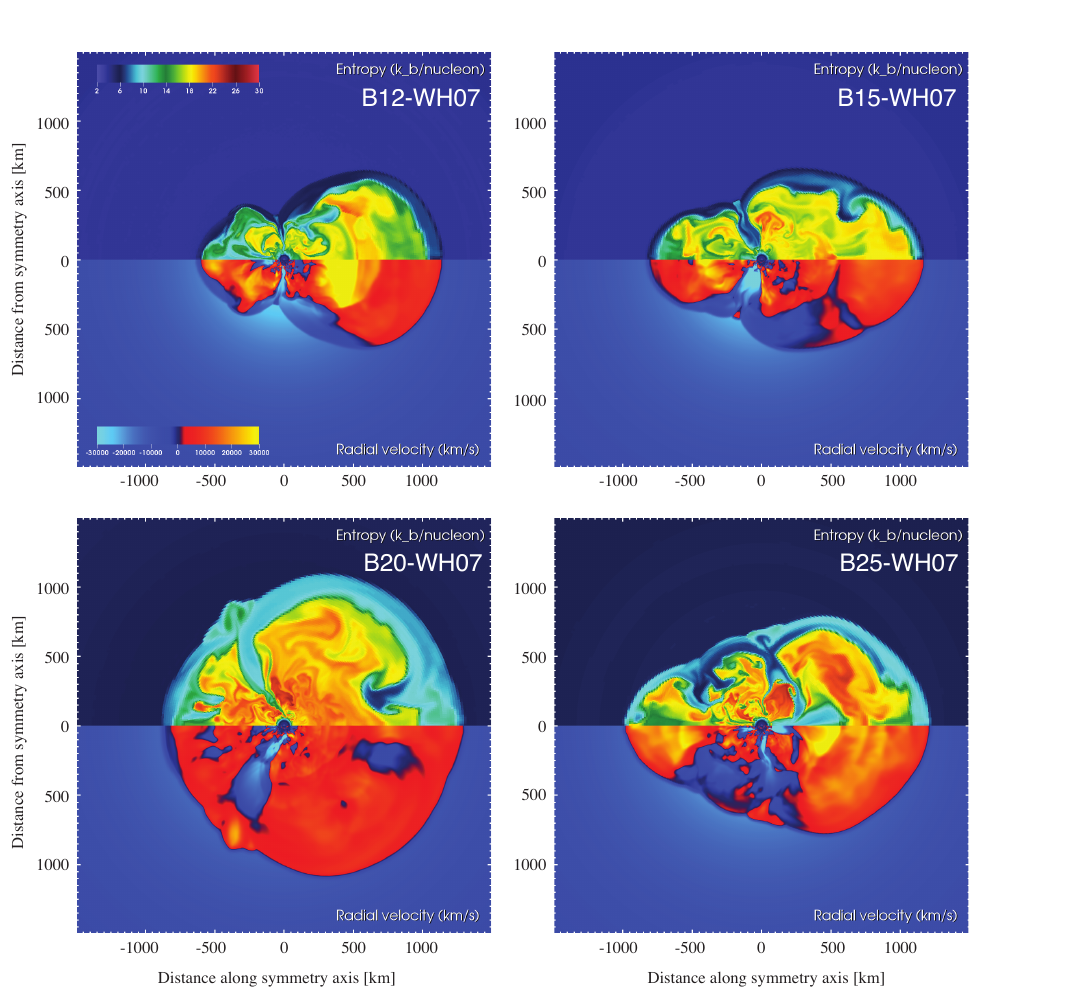}
\caption{\label{B12_B25_Entropy_250_tb}
Profiles of entropy (upper portion of frames) and radial velocity (lower portion of frames) for all four models at 250~ms after bounce. Plotted as in Figure~\ref{fig:entropy} with the entropy scale extended.}
\end{figure*}

\subsection{Explosion Morphology}
\label{sec:Morphology} 

The overall morphology of the shock has an impact on the development of the explosion and, in particular, we believe that the difference in morphology for B20-WH07 relative to the other models helps explain its lower explosion energy.

From the mean, maximum, and minimum shock radii of the models (Figure~\ref{fig:shape}a), it is clear that the shock has been revived and explosions are developing for all models by $\sim$200~ms after bounce as was discussed in Section~\ref{sec:Expl_onset} and \citetalias{BrMeHi13}.
By $\sim$300~ms, the shocks in all of the models are expanding at a fairly steady pace and mean shock radii larger than 10000~km are eventually achieved in all of these simulations.

A simple measure of the shock morphology is the shock deformation parameter \dshock\ defined by \citet{ScKiJa06} as
\begin{equation} 
\dshock= \frac{\max[\rshock\cos\theta]-\min[\rshock\cos\theta]}{2\times \max[\rshock\sin\theta]}-1.
\label{Shock_Deformation}
\end{equation}
Irrespective of the location of the origin, prolate, oblate, and spherical shock geometries are characterized by positive, negative, and vanishing values of \dshock, respectively.
It is evident from the history of \dshock\ plotted in Figure~\ref{fig:shape}b that the shock geometries in all models, with the exception of model B20-WH07, develop a distinctly prolate structure at the onset of the explosion that is maintained throughout expansion.
Model B20-WH07, by contrast, remains roughly spherical in shape for hundreds of milliseconds after the explosion develops, and only starting at $\sim$800~ms after bounce does a prolate shock geometry ($\dshock > 0$) distinctly develop.
We suspect that the tendency of the shocks in our models to assume a prolate shape (eventually, in the case of B20-WH07) and to preferentially explode along the axis of symmetry is, at least in part, a consequence of the imposition of axisymmetry with an impenetrable polar axis (reflecting boundary condition) that forces flows converging at the pole to be directed either radially inwards or outwards along the poles.

The trend of increasing diagnostic energy with progenitor mass evident in our models (Figure~\ref{Expl_E_vs_t_12M_25M_Comp}a) is broken by B20-WH07, which has a relatively low diagnostic explosion energy \Ediag, between those of models B12-WH07 and B15-WH07, rather than between B15-WH07 and B25-WH07 as the trend would otherwise imply.
The slope of the diagnostic energy versus time for B20-WH07 (Figure~\ref{Expl_E_vs_t_12M_25M_Comp}a) initially matches the slopes of its neighbors in mass, B15-WH07 and B25-WH07, but the slope declines relative to those models about 300~ms after bounce, while the \Ediag\ growth rate for B15-WH07 declines only after 600~ms and there is no abrupt flattening of the \Ediag\ curve for B25-WH07 through 1400~ms after bounce. 
This relatively early decrease in \Ediag\ growth rate for B20-WH07 seems tied to the morphology at the shock, particularly the nature of the accretion streams delivering material directly from the shock to the proto-NS.

Accretion streams can be seen in the low-entropy, negative-radial-velocity streams inward from the shock, shown in Figure~\ref{B12_B25_Entropy_250_tb} at 250~ms after bounce, just after shock revival. 
The three models with the systematic trend in diagnostic energy, B12-WH07, B15-WH07, and B25-WH07, show a pronounced prolate shock structure with well-defined accretion streams inward from the shock.
The shock shape for model B20-WH07 is more spherical, though off-center, than prolate. 
None of the down flows are being fed directly from newly shocked material, as a shell of material behind the shock, more than 100~km thick and moving outward, cuts off direct flow into the accretion streams by 240~ms after bounce.
A similar transition to outward velocities behind the entire shock occurs more than 100~ms later for the other models, at about 370, 500, and 410~ms after bounce, respectively, for B12-WH07, B15-WH07, and B25-WH07. 

The mass accretion rates through the gain surface (Figure~\ref{fig:heat}c) also confirm the mass accretion suggested visually by Figure~\ref{B12_B25_Entropy_250_tb}. 
The mass accretion rate through the gain surface of model B20-WH07 falls to less than half that of the other models between 250 and 450~ms after bounce. 
Because a significant fraction of the \nue\ and \nuebar\ luminosities at this time are generated by the gravitational energy released by the accretion of mass through the gain surface and neutrinospheres, it would be expected that the luminosities for model B20-WH07 show a significant decline at this time, relative to those of the other models.
This is reflected in the luminosities of all neutrino species (Figure~\ref{fig:lumin}).
The \nue\ and \nuebar\ luminosities for model B20-WH07 do indeed show a significant decline during the 250--450~ms interval, falling below those of B15-WH07 and almost to the level of B12-WH07. 

We thus attribute the relatively low diagnostic energy of model B20-WH07 to a post-shock envelope structure that has been frozen at the time of shock revival and has inhibited the accretion streams and mass accretion through the gain surface (Figure~\ref{fig:heat}c) and neutrinospheres relative to the other models at this critical time.
These morphological differences in model B20-WH07 and their impact on the development of the explosion seem to be initiated by the growth of a large plume on the north, or right, pole of the simulation that spreads into the equatorial region just before the revival of the shock. 
This can be seen in the animation of Figure~\ref{fig:entropy}.
This event seems to be the result of a stochastic variation in the growth of the buoyant plumes rather than related to any feature of the progenitor.
Confirming this supposition will require further examination as the available number of 2D simulations grows.

Note also in Figure~\ref{fig:heat}c the very significant decline, by almost an order of magnitude, in the mass accretion rate of model B12-WH07 relative to the other models, beginning at $\sim$600~ms after bounce. 
This B12-WH07 mass accretion rate trough is the result of both the original lower density oxygen-rich layer being advected to the gain surface and the temporary choking off of the down flows.
This accretion rate trough is correlated with the very slow growth of the explosion energies for this model beginning at this time (Figure~\ref{Expl_E_vs_t_12M_25M_Comp}a). 

These examples of the connection between mass accretion and the growth of the explosion energy are illustrative of several important aspects of the multidimensional dynamics of the neutrino-driven supernova mechanism precluded in spherical symmetry.
The first is that mass accretion in streams can continue to deliver thermal energy to the proto-NS and enhanced neutrino emission by the release of the gravitational potential energy as the matter accretes on the proto-NS, thereby continuing to pump energy into the developing explosion.
Any inhibition of these accretion streams will be reflected in a slower growth of the explosion energy, as seen in B20-WH07.
The second is that the morphology of the explosion and the topology of these accretion streams, and therefore the final explosion energy, can be affected randomly from model to model by the stochastic nature of the fluid instabilities that develop prior to the explosion. 
Similar to the point made by \citet{ScKiJa06} in connection with the bimodality of neutron star velocities, the final explosion energy of a given 2D model can be subject to random variations due to the stochastic nature of the fluid flow in the gain region immediately prior to the onset of the explosion.
The extent to which this potential variability carries over to 3D simulations remains to be determined.

\subsection{Proto-Neutron Star}
\label{sec:PNS}

\begin{figure}
\includegraphics[width=\columnwidth,clip]{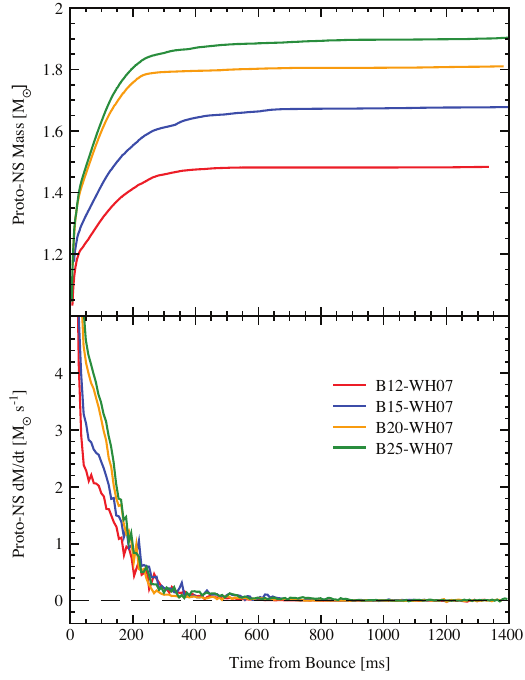}
\caption{\label{B12_B25_NS_mass_vs_tb}
Proto-NS rest mass (upper panel) and the rate of mass accretion onto the proto-NS (lower panel) for all models plotted versus time after bounce in the colors used in Figure~\ref{fig:early}.}
\end{figure}

\begin{figure}
\includegraphics[width=\columnwidth,clip]{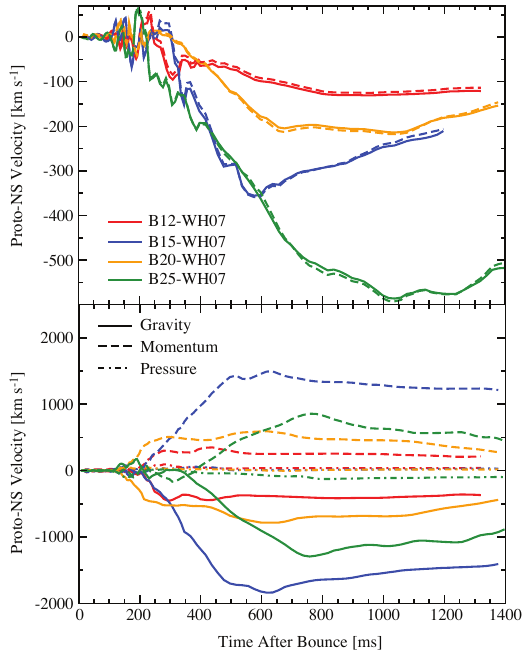}
\caption{\label{B12_B25_NS_v_vs_tb}
Velocity of the proto-NS (upper panel) and inputs to the proto-NS velocity (lower panel) for all models plotted versus time after bounce in the colors used in Figure~\ref{fig:early}.  The velocities plotted in the upper panel are calculated by assuming momentum conservation (Equation~\ref{V_ns_mom_cons}; solid lines) or by integrating the external forces acting on the proto-NS (Equations~(\ref{Accel_sources}) and (\ref{V_ns_accel}); dashed lines). Contributions to the velocity of the proto-NS by each term in the right-hand side of Equation~(\ref{Accel_sources}) plotted in the lower panel are: the gravitational force on the proto-NS by the matter exterior to it (solid lines); transport of momentum to and from the proto-NS (dashed lines); and pressure on the surface of the proto-NS (dot-dashed lines).}
\end{figure}

The proto-NS baryonic rest masses, $M_{\rm bary}$, are plotted as a function of time in Figure~\ref{B12_B25_NS_mass_vs_tb}, upper panel, and are 1.461, 1.676, 1.806, and 1.898 \msun\ (Table \ref{tab:outcomes}), respectively, for B12-WH07, B15-WH07, B20-WH07, and B25-WH07 at the time of this report, where we have defined the proto-NS as the matter with densities above $10^{11}$~\gcc.
The baryonic masses can be translated to gravitational masses, $M_{\rm grav}$, with the relation
\begin{equation}
M_{\rm grav} = M_{\rm bary} - 0.075 \, \msun \left(\frac{M_{\rm grav}}{\msun}\right)^2,
\end{equation}
with the constant obtained by fitting results using a wide range of EoSs by \citet{TiWoWe96}.
This gives $M_{\rm grav}$ of 1.345, 1.506, 1.611, and 1.685~\msun\ for B12-WH07, B15-WH07, B20-WH07, and B25-WH07, respectively.

The proto-NS growth rate is plotted in Figure~\ref{B12_B25_NS_mass_vs_tb} (lower panel).
Proto-NS growth is strong in the pre-explosive phase, but the growth rate drops sharply as the accretion rate from the infalling core decreases and matter passing through the shock increasingly becomes included in the outward flow, as is shown by the increasing mass of the gain region in Figure~\ref{B12_B25_Gain}a.
From 300~ms, when the shock is already launched, to 1200~ms after bounce, the proto-NS rest mass grows by 0.0226, 0.0644, 0.0160, and 0.0443~\msun\ for models B12-WH07, B15-WH07, B20-WH07, and B25-WH07, respectively.
The low growth rate of the B20-WH07 proto-NS mass also corresponds to the reduced accretion and reduced build up of explosion. (See Section~\ref{sec:Morphology}.)

Another feature of the proto-NSs in our models, evident from the gain surface accretion rates (Figure~\ref{fig:heat}c), is the lack of the so-called neutrino-driven wind phase, where mass accretion onto the proto-NS is replaced by a low-density, outward flowing wind from the proto-NS surface, driven by neutrino energy deposition.
Mass accretion inward through the gain surface at 0.01--0.1~\msuns\ is still occurring at the time of this report, though the net accretion through the gain surface (inward minus outward) at late times has become quite small ($\sim$$10^{-3}$~\msuns).
There is some evidence of the increased entropy symptomatic of the neutrino-driven wind, but only for some latitudes, with accretion dominating overall.

It has been established \citep{ScPlJa04, ScKiJa06, NoBrBu10, NoBrBu12, WoJaMu10, WoJaMu13} that neutron star velocities in the range of observations can be generated by the anisotropic mass ejection that arises naturally from fluid instabilities in simulations of neutrino-driven CCSNe.
Since the proto-NSs in \chimera\ models are tied to the origin and cannot move when exchanging momentum with the surrounding gas, we compute the proto-NS velocity by two previously used methods.

In the first method, we assume conservation of linear momentum and compute the proto-NS velocity from the negative of the total linear momentum of the ejected gas \citep{ScKiJa06, WoJaMu10, WoJaMu13},
\begin{equation}
  {\bf v}_{\rm ns}(t) = - {\bf P}_{\rm gas}(t)/M_{\rm ns}(t),
\label{V_ns_mom_cons}
\end{equation}
where ${\bf v}_{\rm ns}$ is the velocity of the proto-NS and  
\begin{equation}
  {\bf P}_{\rm gas}(t) = \int_{R_{0} < r < R_{\rm s}} \rho {\bf v}\, dV
\end{equation}
is the gas momentum integrated over the computational grid exterior to the proto-NS, where $M_{\rm ns}$ is the baryonic mass of the proto-NS star; $R_{0}$ is a fiducial radius (chosen to be 100~km) inside which  essentially all of the momentum transfer between neutrinos and matter occurs; and $R_{\rm s}$ is the surface of the computational grid.
There is a neutrino contribution to the momentum given by
\begin{equation}
  {\bf P}_{\rm \nu}(t) = \int_{R_{0} < r < R_{\rm s}} p_{\nu} \hat{\bf r} \, dV + \int_{t_{0}}^{t} \oint_{r = R_{s}} p_{\nu} c \, \hat{\bf r}\, dS,
\end{equation}
where $p_{\nu}$ is the neutrino momentum integrated over the computational grid exterior to the proto-NS plus the neutrino momentum radiated out of the grid.
This contribution is small and we do not include it further.
From Equation~(\ref{V_ns_mom_cons}) we find the evolution of the proto-NS velocities plotted with solid lines in Figure~\ref{B12_B25_NS_v_vs_tb}, upper panel.

The second method involves computing the acceleration of the proto-NS by summing up all the forces acting upon it, and integrating with respect to time to compute the velocity \citep{ScKiJa06, NoBrBu10, WoJaMu10, WoJaMu13}.
We compute
\begin{eqnarray}
& {\displaystyle \dot{\bf P}_{\rm ns}(t) = G \int_{r>R_{0}} M_{\rm ns}\, \hat{\bf r} \frac{dm}{r^{2}} - \oint_{r = R_{0}} \rho {\bf v}( {\bf v} \cdot \hat{\bf r} ) \, dS
} & \nonumber \\
& {\displaystyle - \oint_{r = R_{0}} p\,\hat{\bf r}\, dS,
} &
\label{Accel_sources} 
\end{eqnarray}
and then
\begin{equation}
  {\bf v}_{\rm ns}(t) = \int_{t_{0}}^{t} \dot{\bf P}_{\rm ns}(t')/M_{\rm ns}(t')\, dt'.
\label{V_ns_accel}
\end{equation}
The proto-NS star velocities as a function of time for our models, computed from Equations~(\ref{Accel_sources}) and~(\ref{V_ns_accel}) using a fiducial radius $R_{0} =$100~km for evaluation of the forces, are plotted in Figure~\ref{B12_B25_NS_v_vs_tb}, upper panel, with dashed lines.

For three of the models (all but B12-WH07) the proto-NS velocity magnitude declines towards the end of the simulation.
The velocity contributions from the three forces in Equation~(\ref{Accel_sources}) are plotted in Figure~\ref{B12_B25_NS_v_vs_tb}, lower panel.
The dominant contribution to the proto-NS velocity is the gravity term, which is partially canceled by the momentum transport term, indicating that the dominant process giving rise to the proto-NS velocity is anisotropic mass loss with similar velocities.
The cause of the decline in the magnitude of the proto-NS velocity is the continuing accretion of matter onto the proto-NS (Figure~\ref{fig:heat}c) and the onset of mass ejection from the opposite pole from that which occurred at earlier times.

It is gratifying to note that both methods of computing the proto-NS kick velocities yield similar velocities as a function of time, as it indicates that \chimera\ conserves linear momentum to good accuracy, despite the gravitational potential not being constructed in \chimera\ to explicitly conserve linear momentum.
The final proto-NS velocity has only been attained for model B12-WH07, which has a final computed velocity of $\sim$100~\kmps.
For the remaining three models we can only suggest that models B15-WH07 and B20-WH07 will have final velocities of $\sim$100--200~\kmps\ and model B25-WH07 will have a final velocity of $\sim$500~\kmps.
Their velocities at the time of this report are listed in Table \ref{tab:outcomes}.

\section{Comparison with Observed Supernovae}
\label{sec:Obs_CCSNe}

In this section, we summarize attempts to extract the initial progenitor masses, explosion energies, and the synthesized \isotope{Ni}{56} masses from observations of CCSNe. 
There are large error bars associated with attempts to extract these quantities, particularly the initial progenitor masses and explosion energies.
Different approaches, such as extracting the progenitor mass of a CCSN from a comparison of stellar modeling with pre-explosion images versus hydrodynamic modeling of the light curve, can give very different results. 
Nevertheless, crude as the current determination may be, these observations are more nuanced than the 1~B explosion energy and 0.1~\msun\ of \isotope{Ni}{56} that modelers usually consider.

We have carried out our simulations to the point where the explosion energies can be reasonably extrapolated.
 We formally compare the results of our simulations with observations of CCSN explosion energies and synthesized \isotope{Ni}{56} mass, \Mni.
We stress at outset that these comparisons are far from definitive, as they result from simulations which involve numerous approximations, e.g., 2D versus 3D, ray-by-ray MGFLD transport versus fully multi-dimensional transport, etc. 
Furthermore, though our radial and latitudinal resolutions are higher than most simulations by other groups, we have not carried out a numerical resolution convergence study and are therefore unable to say how our results might change with a more refined grid.
(A resolution study of CCSNe simulations is potentially complicated by stochastic changes in observables like the total explosion energy. 
An ensemble of models for each grid resolution may be required to draw firm conclusions.)
Future numerical improvements in \chimera\ will likely change our computed values of explosion energies and synthesized \Mni. 
However, if numerical simulations are to have any usefulness, they must converge to the observed values over time, as all of the relevant physics is incorporated.
We therefore log our current results against some observational standards, despite the large (essentially unknown) error bars in both models and observations, in order to track futures changes in results with an improved \chimera\ code and the results of other groups.

Finally, the placement of our results on the ZAMS axis represents the ZAMS masses reported by \citet{WoHe07} from which they initiated their stellar evolutionary calculations to produce CCSNe progenitors used herein. 
Due to potential differences between these progenitor models and the structures of real stars of various masses,  we are hesitant to suggest any trends with ZAMS mass in the observables from our results.

\subsection{Background}

Numerous caveats must be kept in mind when quoting observed values for the explosion energies, progenitor masses, and \isotope{Ni}{56} masses of particular events.
The best determined of these are the \isotope{Ni}{56} masses.
The newly synthesized \isotope{Ni}{56} decays to \isotope{Co}{56} and then to \isotope{Fe}{56}, with half lives of 6.1~days and 77.3~days, respectively.
When possible, measuring the gamma-ray luminosity when the supernova envelope becomes transparent to gamma-rays is preferable; however, this requires a nearby supernova.
The nebular phase of a CCSN is powered by these radioactive decays, and the early nebular-phase bolometric luminosity, when the envelope is optically thick to gamma-rays and all the gamma-ray luminosity can be assumed to be thermalized and re-emitted, provides a minimum of the mass of \isotope{Ni}{56} synthesized \citep{WeWo80}, as a non-negligible portion of the thermalized energy is expended by the work it does on the expanding material \citep{Utro07}.
We will refer to this method of using the bolometric luminosity of a CCSN during its nebular phase as the ``BL method.''

A complementary approach is to use the correlation between the maximum gradient at the transition phase in the V band and the photometric estimate of the \isotope{Ni}{56} mass \citep{ElChDa03}.
We will refer to this method as the ``S method." 
The principal uncertainties in such analyses are the distance to the observed supernova, the extinction of light as it passes through the interstellar medium (in the host galaxy and the Milky Way), and the date of the explosion.

Correlating explosion parameters with the mass of the progenitor adds additional sources of uncertainty. 
The progenitor mass of an observed CCSN can best be ascertained if its image can be resolved in archival images of the host galaxy, preferably in images taken with a range of photometric filters.
The publicly accessible HST archive has been particularly fruitful in this regard.
To date, almost all of the progenitors revealed in pre-explosion images of the location of CCSNe have been red supergiants (RSGs).
Determining the zero-age main sequence (ZAMS) progenitor masses, \Mzams, from these images then depends on matching the inferred luminosities of these immediate pre-collapse RSGs with stellar evolutionary models leading up to core collapse.
These comparisons depend on all the attendant uncertainties in stellar evolution modeling including those from convective and rotational mixing and reaction rates for advanced nuclear burning stages.
Additional uncertainties include the host galaxy distance, progenitor metallicity, extinction corrections for circumstellar dust surrounding the progenitor, which can potentially be destroyed by the X-ray and UV radiation of the supernova, and background subtraction of the light from stars in the progenitor's vicinity.
Consistency checks are possible if the progenitor can be determined to be a member of a stellar cluster with a determined age.
This also can work as an alternative method of ascertaining the progenitor mass if the progenitor is in a compact cluster and cannot be resolved (e.g., SN 2004dj).

Another method of determining supernova progenitor \Mzams\ is to infer the ejecta masses through semi-analytical, or numerical radiation-hydrodynamics, modeling of the bolometric light curves and photospheric velocities.
This requires corrections for estimated mass loss during progenitor evolution and a compact remnant mass estimate. 
Since this method requires high-quality photometric and spectroscopic data, it has only been applied to a handful of events where such data are available \citep[e.g.,][]{ZaPaTu03, BaBlPa05, Utro07, UtCh08, UtCh09, PaVaZa09, DaBoPu14}.
Typically, these models of explosions in RSGs predict significantly higher ZAMS masses than those obtained by direct imaging.
The reasons for this difference are unknown, but the most likely candidates are the assumption of spherical symmetry in the models \citep{UtCh13}, unaccounted inhomogeneity particularly in the outer layers \citep{ChUt14}, and possibly the lack of non-LTE and spectral transport  \citep{UtCh09}.
Alternately, the fault could lie in the stellar evolutionary models on which the analysis of direct images relies. 
We include results for progenitor masses from these recent radiation-hydrodynamics simulations, in part, to emphasize the uncertainty that persists in deriving ZAMS and ejecta masses.

Finally, the explosion energy of a CCSN can only be estimated by comparing light curves and ejecta velocities with radiation-hydrodynamics simulations.
To this end \citet{LiNa83, LiNa85} computed a grid of 27 radiation-hydrodynamics simulations.
They parameterized the stellar envelope and simulated the explosion by setting up an outwardly-directed matter velocity in approximately one-third of the stellar mass.
They found an approximate expression for the ejected mass, $M_{\rm eject}$, the explosion energy, $E_{\rm expl}$, and the radius of the immediate progenitor, $R_{\rm env}$, in terms of the duration, $\Delta t$, of the light-curve plateau, the absolute V magnitude at the midpoint of the plateau, and the velocity, $v_{\rm ph, mid}$, of material at the photosphere at the mid-plateau epoch.
Neglect of the energy input from the $\isotope{Ni}{56}\rightarrow \isotope{Co}{56}\rightarrow\isotope{Fe}{56}$ decay chain by \citet{LiNa83, LiNa85} caused their results for the plateau length  $\Delta t$ to be too short \citep[cf.][]{JeFrMa12}.
Since the \citet{LiNa83, LiNa85} expressions for $M_{\rm eject}$ and $E_{\rm expl}$ depend strongly on $\Delta t$, values of $M_{\rm eject}$ and $E_{\rm expl}$ obtained by use of these expressions will tend to be overestimated.
However, correlations obtained through the use of these expressions should remain valid.
\citet{Hamu03} applied the fitting formulae of \citet{LiNa83, LiNa85} to a sample of 13 SNe IIP that had sufficient data for the needed observational parameters to be derived.
He found explosion energies to vary between 0.6~B to 5.5~B, with most of the energies clustering between 1~B and 2~B.
The \isotope{Ni}{56} masses varied between 0.0016 and 0.26~\msun.
While there were large error bars, several correlations emerged from this analysis --- more massive progenitors produced more energetic explosions and CCSNe with greater energies produced more \isotope{Ni}{56}.

Recently, \citet{DeLiWa10} have provided an extensive  grid of radiation-hydrodynamics simulations of artificial 0.1--3~B explosions driven by a piston at the base of non-rotating \citep{WoHeWe02} and rotating \citep{HeLaWo00} RSG progenitor stars  with \Mzams\ between 11 and 30~\msun.
Their results suggest that the velocities of the material at the outer edge of the oxygen-rich shell $(v_{\rm e,O})$, or of the photosphere at 15 $(v_{\rm p, 15d})$ or 50 $(v_{\rm p, 50d})$ days after shock breakout, are correlated with \Mzams\ and $E_{\rm expl}$.

More recently, \citet{Pozn13} compiled a list of 23 SN IIP events whose progenitors had been determined from their presence in archival images of the host galaxy, or through upper limits to \Mzams\ implied by non-detection in archival progenitor images.
For 17 of these supernovae, \citet{Pozn13} determined expansion velocities at the photosphere using the minimum of the 5169~\AA\ \ion{Fe}{2} spectral feature \citep{PoBuFi09} and propagated to day 50, $v_{\rm p, 50d}$, on the plateau using Equation~(2) of \citet{NuSuEl06}.
\citet{Pozn13} found an approximately linear relation (albeit with substantial error bars) between $v_{\rm p, 50d}$ and the progenitor ZAMS masses, implying a strong dependence of the explosion energy on \Mzams, such that $E_{\rm expl} \propto v_{\rm p, 50d}^{3}$.

To compare our explosion energy and synthesized \isotope{Ni}{56} results with estimates of these quantities from observations, we will consider those events for which the progenitor can be observed in archival images of the host galaxy, allowing a restricted range of possible ZAMS progenitor masses to be inferred, excluding objects with only \Mzams\ upper limits.
Also, we only consider events in which the inferred progenitor mass lies within the 12 to 25 \msun\ range covered in our simulations.
To estimate the explosion energies, we will use the values of $v_{\rm p, 50d}$ compiled by \citet{Pozn13}, together with the masses derived from archival images, to interpolate the explosion energy in the radiation-hydrodynamics grid of \citet{DeLiWa10}.
The uncertainties we quote for the explosion energies are based on the uncertainties in the ZAMS mass of the progenitor and uncertainties in the $v_{\rm p, 50d}$ velocities taken from \citet{Pozn13}, and uncertainties in the ZAMS structure of the progenitors used by \citet{DeLiWa10} radiation-hydrodyamical simulations.
Rigorous accounting for the latter is beyond the scope of our analysis, so we adopt the expedient of computing the explosion energy using the \citet{DeLiWa10} results for progenitor masses 10\% above and below the progenitor mass in question.

\subsection{Observational Sample}
\label{sec:sample}
Here we are dealing with very-small-number statistics.
Two of the nearest events, SN~1987A and SN~1993J, which have fairly massive and well-observed progenitors, are both peculiar.
Aside from these two outliers, the rest of the supernovae that have thus far had their progenitors observed on archival images of the host galaxy, or upper limits placed on their luminosity, are SN Type IIP whose progenitors are mostly red supergiants with inferred masses clustering around the low end of the progenitor ZAMS mass range for which core-collapse is a possible evolutionary outcome, $\Mzams\approx 8 \pm 1$~\msun\ \citep{Smar09, FrErEl11}.
However, there are several events with progenitors having inferred $\Mzams \approx 12$--25~\msun\ within the range of progenitors used in our simulations.
We describe them briefly in the following, along with our estimates of their explosion energies.

\paragraph{SN 1987A}
The challenges of modeling the progenitor of SN~1987A, discovered 1987 February~24 in the Large Magellanic Cloud, have included accounting for the blue supergiant structure of the progenitor, and the triple-ring nebula with its axisymmetric but nonspherical structure and high nitrogen abundance. 
The difficulties are reflected in the number of attempts made to model the progenitor.
Some, but not necessarily all, of the anomalies have been explained by models of low-metallicity single-star evolution \citep{Ar87, HiHoTr87, TrWe87}, models of single-star evolution with mass loss \citep{SaKaNo88, ArBaKi89, Ma87}, and models of single-star evolution with abnormal convection \citep{LaElBa89, We89}, rapid rotation or binarity \citep{FaReHe87, JoPoHs88, PoFaSt91}, and accretion or mergers \citep{PoJo89, deLoVa92, PoJoRa90, PoMoIv07}.
A common denominator of these investigations was a progenitor  \Mzams\ of 15--20~\msun.
Models of the light curve and inferred ejecta velocities provide estimates of the explosion energy of SN 1987A of 1--2~B \citep{ArBaKi89, BePi90, ShNo90, MaHiHo92}.
The mass of ejected \isotope{Ni}{56} is estimated to be 0.075~\msun\ (in the range 0.055--0.090~\msun) based on the bolometric luminosity of the remnant after day 126 \citep[][and references therein]{SuBo90, SuPhDe91}.

\paragraph{SN 1993J}
The other peculiar supernova with a well observed progenitor is SN 1993J, discovered 1993 March~28 in the nearby galaxy M81. 
The very sharp initial peak in the early light curve; the transition from hydrogen-rich Type II to hydrogen-poor and helium-rich Type Ib spectral features within a few weeks; and a spectral energy distribution of the observed progenitor that is inconsistent with a single star; are strong indications that at the time of the explosion the progenitor was stripped of most of its hydrogen envelope, most likely by mass transfer and loss due to a binary companion. 
The detection of a massive binary companion \citep{SmMaKu05} supports this conclusion. 
Radiation-hydrodynamics models of the light curve suggest that the ejected \Mni\ is 0.06--0.09~\msun\ \citep{WoEaWe94, ShSuKu94}, based on a distance to M81 of 3.3~Mpc.
With the M81 distance revised to 3.63~Mpc \citep{FrHuMa94}, this estimate becomes 0.07--0.11~\msun. Luminosity constraints and estimates of the ejected mass based on light curve modeling indicate $\Mzams \approx 12$--17~\msun\ \citep{WoEaWe94, ShSuKu94, YoBaBr95}. An explosion energy of 1--2~B is consistent with the results of radiation-hydrodynamics light-curve models \citep{WoEaWe94, BaBlPa94}.

\paragraph{SN 2004A}
Archival images of the SN~2004A host galaxy NGC~7247 were analyzed by \citet{HeSmCr06} and more recently  by \citet{MaReMa13} using late-time imaging and a better accounting of circumstellar dust to find $\Mzams = 12.0 \pm 2.1$~\msun.
Using $v_{\rm p, 50d} = 3410 \pm 180$~\kms\ \citep{Pozn13}, we infer an explosion energy of 0.76--1.3~B using the \citet{DeLiWa10} grid.
Ejected \Mni\ was estimated by various methods as $0.046^{+0.031}_{-0.017}$~\msun\ \citep{HeSmCr06}.

\paragraph{SN 2004dj}
Discovered on 2004 July~31, SN~2004dj was coincident with the compact star cluster Sandage~96  in the SC galaxy NGC~2403.
At a distance of $3.3 \pm 0.1$~Mpc, it is the closest normal SN~IIP observed.
The progenitor was not resolved on archival images, but constraints on its mass were obtained by estimates of the cluster's age, which were made from archival images by \citet{MaBoSi04} and  \citet{WaYaZh05}, and after the SN faded by \citet{ViSaBa09}.
The most probable progenitor mass was determined to be 12--15~\msun, although a mass as high as 20~\msun\ could not be excluded.
The \isotope{Ni}{56} synthesized in this event has been estimated to be $0.020 \pm 0.002$~\msun\ \citep{ChFaSh05, ZhWaLi06}.
Using the expansion velocity given by \citet{Pozn13}, $v_{\rm p, 50d} = 3080 \pm 130$~\kms\, we infer an explosion energy of 0.7--0.9~B from the \citet{DeLiWa10} grid assuming a ZAMS mass of 12--15 \msun\ for the progenitor.

\begin{figure}
\includegraphics[width=\columnwidth,clip]{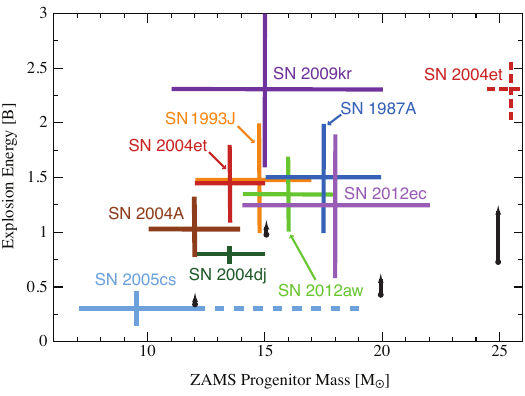}
\caption{\label{Explosion_Energy_Comparisons}
Inferred explosion energies for Type IIP supernovae whose progenitors have been observed on archival images described in Section~\ref{sec:sample}. \Mzams\ with estimated errors inferred from archival images are indicated by the solid horizontal lines and \Mzams\ inferred from hydrodynamic modeling by dashed horizontal lines.
The error estimates of the ZAMS progenitor masses and explosion energies are probably minimum estimates given the number of uncertainties in their determination.
Simulation explosion energies, \Ediagovrec\ as described in Section~\ref{sec:Expl_growth} and shown in Figure~\ref{Expl_E_vs_t_12M_25M_Comp}a, are plotted for the end of the simulations as filled circles with the vertical arrows using the ZAMS masses of the \citet{WoHe07} progenitors.
The length of each vertical arrow indicates the additional \Ediagovrec\  that would accrue if the rate of \Ediagovrec\  increase at the end of our simulations were to remain constant for an additional second. 
\Ediagovrec\  for models B15-WH07, B20-WH07, and B25-WH07 may substantially increase with further evolution.
The error bars on our simulation results are unknown.
}
\end{figure}

\paragraph{SN 2004et}
Discovered 2004 September~27 in the nearby ($D=5.5$~Mpc) galaxy NGC 6964, SN~2004et received extensive coverage of its photometric and spectroscopic evolution.
The ejected \isotope{Ni}{56} mass was estimated to be $0.060 \pm 0.02$~\msun\ (BL~method) and $0.062 \pm 0.02$ \msun\ (S~method) \citep{SaAnSr06}, $0.068 \pm 0.009$ \msun\ (BL method) \citep{UtCh09}, $0.06 \pm 0.03$ \msun\ (BL~method) and $0.056 \pm 0.016$~\msun\ (S~method) \citep{MiPoCh07}, $0.057 \pm 0.03$~\msun\ (BL~method) and $0.057 \pm 0.02$~\msun\ (S~method) \citep{MaDiSm10}, and $0.062 \pm 0.02$~\msun\ (BL~method) \citep{JeFrMa12}.
\citet{LiVaFi05} identified  a yellow supergiant with $\Mzams = 15^{+5}_{-2}$~\msun\ at the site of SN~2004et.
\citet{CrSmPa11} used higher resolution pre- and post-explosion images to show that there were at least three stars at the site of SN~2004et.
They concluded that the progenitor was most likely a K to M RSG with a ZAMS mass between 7 and 15~\msun\ (two assessments of the contributions to the observed components yielded $8^{+5}_{-1}$ and $10^{+5}_{-1}$~\msun).
Detailed late-time spectral modeling of the supernova by \citet{JeFrMa12} finds that the nebular-phase optical and near-infrared spectra are well reproduced by a 15~\msun\ ZAMS progenitor.
Radiation hydrodynamics modeling of the bolometric luminosity and spectral evolution of this event yields a considerably higher \Mzams\ of $27 \pm 2$~\msun\ and an ejecta mass of $22.9 \pm 1$~\msun\ \citep{UtCh09}.
Using $v_{\rm p, 50d} = 3940 \pm 110$~\kms\ \citep{Pozn13}, we infer an explosion energy of 1.1--1.8~B from the \citet{DeLiWa10} grid assuming $\Mzams =12$--15~\msun. 
The somewhat higher explosion energy of $2.3 \pm 0.3$~B is inferred from the radiation-hydrodynamics model of \citet{UtCh09}, to go with the larger progenitor mass.
Given the discordant results for the explosion energy and progenitor mass from \citet{UtCh09} we have marked it separately in Figure~\ref{Explosion_Energy_Comparisons} using dashed lines.

\paragraph{SN 2005cs}
The sub-luminous Type II SN 2005cs was discovered on 2005 June 28 in the `Whirlpool Galaxy' M51. 
The progenitor was identified on archival images by \citet{LiDyFi06} and by \citet{MaSmDa05} with an inferred \Mzams\ of 7--9~\msun\ and $9_{-2}^{+3}$~\msun, respectively, for an 8.4~Mpc distance to M51. 
\citet{TaVi06} revised the distance to M51 to 7.1~Mpc, which would reduce the above progenitor mass estimates to $\Mzams=9.6 \pm 5.2$~\msun\ using the formula given by \citet{Nady03}. 
Similar results were obtained by \citet{TsVoSh06} from their observations of the supernova and their use of the IIP-analytical model of \citet{Popo93} and the simulations of \citet{LiNa85}. 
\citet{PaVaZa09} analyzed an extensive data set to estimate $\Mzams\approx10$--15~\msun\ by a semi-analytic fit of the data to the model of \citet{ZaPaTu03}. 
Finally, \citet{UtCh08} with radiation-hydrodynamics modeling obtain a larger \Mzams\ of 17.2--19.2~\msun. 
Interestingly, if this latter estimate is correct there would be a strong implication that subluminous Type~II SNe arise from two distinct progenitor populations, a low mass ($\sim$7--10~\msun) population and a relatively high mass ($\gtrsim$20~\msun) population. 
Three studies have estimated the ejected \isotope{Ni}{56} mass for SN~2005cs using the BL~method.
\citet{TsVoSh06} compared the SN~2005cs light curve tail with that of SN~1987A and estimated \Mni\ as 0.017--0.018~\msun. 
\citet{TaVi06} obtained a somewhat smaller estimate: $\Mni = 0.009 \pm 0.003$~\msun.
\citet{PaVaZa09} also find a small \Mni\ $\approx 0.003$~\msun.
\citet{UtCh08} obtained  $\Mni=0.0082 \pm 0.0016$~\msun\ from radiation-hydrodynamics modeling.  
Taking the lowest progenitor mass of 11~\msun\ from the \citet{DeLiWa10} grid, and  $v_{\rm p, 50d} = 2160 \pm 130$~\kms\ from \citet{Pozn13}, we infer an explosion energy of 0.27--0.39~B. 
Using the IIP-analytical model of \citet{Popo93} and the simulations of \citet{LiNa85}, \citet{TsVoSh06} obtain an explosion energy of 0.17--0.39~B. \citet{TaVi06}, from the \citet{Nady03} formulae, derive an explosion energy of $0.09_{-0.10}^{+0.17}$~B.
Using the semi-analytical model of \citet{ZaPaTu03}, \citet{PaVaZa09} derive an explosion energy of $\sim$0.26--0.3~B.
The radiation-hydrodynamics modeling of \citet{UtCh08} gives an explosion energy of $0.41 \pm 0.03$~B for SN~2005cs.

\paragraph{SN 2009kr}
SN 2009kr was discovered 2009 November~6 in the spiral galaxy NGC 1832, and identified as either a Type~IIL \citep{ElDyLi10} or a transitional event between the Type~IIL and the Type~IIP \citep{FrTaPa10}.
Archival imaging of the supernova site reveals a yellow supergiant progenitor with estimated \Mzams\ of 18--24~\msun\  \citep{ElDyLi10} and $15_{-4}^{+5}$~\msun\ \citep{FrTaPa10}.
The latter authors criticized the former for not comparing the measured luminosity of the progenitor with models at the end of core helium burning, as recommended by \citet{Smar09}.
Taking the latter estimate of the progenitor \Mzams\ and  $v_{\rm p, 50d} = 4960 \pm 280$~\kms\  \citep{Pozn13}, we infer an explosion energy of 1.6--3.0~B using the \citet{DeLiWa10} grid.
No estimates of \Mni\ have been published.

\paragraph{SN 2012aw}
Analysis of pre-explosion archival HST images of M95 indicate that the progenitor star of this Type~II SN was a red supergiant.
\citet{FrMaSm12} estimated the ZAMS mass as 14--26~\msun.
Smaller photometric uncertainties allowed \citet{VaCePo12} to narrow that range to 15--20~\msun.
\citet{KoKhDa12} found the \Mzams\ of SN~2012aw to be less than 15~\msun\ by utilizing an improved circumstellar dust model.
\citet{JeSmFr14} obtained \Mzams\ of 14--18~\msun\ on the basis of nucleosynthesis models and optical and near-infrared spectroscopy of the nebular phase. 
Radiation-hydrodynamics modeling of this event by \citet{DaBoPu14} gives an envelope mass $\sim$20~\msun.
With a value of $v_{\rm p, 50d} = 4040 \pm 90$~\kms\ given by \citet{Pozn13}, the inferred explosion energy from the grid of  \citet{DeLiWa10} is 1.0--1.7~B, using a ZAMS mass of 14--18~\msun.
An independent estimate of the explosion energy of 1--2~B was made by \citet{BoKuSy13} by comparing the photospheric ejecta velocity at 15 days and the velocity of the outer edge of the oxygen-rich shell, with the \citet{DeLiWa10} grid, and 1.5~B by the radiation-hydrodynamics model of \citet{DaBoPu14}.
Using the BL method, \citet{BoKuSy13} estimated an ejected \isotope{Ni}{56} mass of $0.06 \pm 0.01$~\msun. Using the same method, \citet{DaBoPu14} estimated the mass to be 
0.05--0.06~\msun.

\paragraph{SN 2012ec}
The Type IIP SN 2012ec was discovered 2012 August~11 in the galaxy NGC 1084.
The progenitor was identified by \citet{MaFrSm13}, from which they infer a ZAMS mass of 14--22~\msun.
We infer an explosion energy of 0.6--1.9~B from the \citet{DeLiWa10} grid and the velocity $v_{\rm p, 50d} = 3890 \pm 410$~\kms\ reported by \citet{Pozn13}.
No estimates of the ejected \isotope{Ni}{56} mass have been published.

\begin{figure}
\includegraphics[width=\columnwidth,clip]{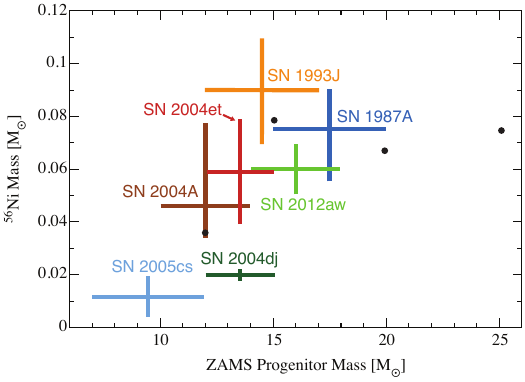}
\caption{\label{Ni56_comparisons}
Inferred ejected \isotope{Ni}{56} mass, \Mni, for Type~IIP supernovae with \Mni\ estimates described in Section~\ref{sec:sample} whose progenitors have been observed on archival images.
Progenitor \Mzams\ estimates are those used in Figure~\ref{Explosion_Energy_Comparisons}.
Ejected \Mni\ of our models are shown by the filled red circles.
As in the preceding graph, the error bars on our simulation results are unknown.
}
\end{figure}

\subsection{Simulations compared to observed sample}
In Figure~\ref{Explosion_Energy_Comparisons} we plot a comparison of the inferred explosion energies for the supernovae described above with the explosion energies \Ediagovrec\ of our four models.
The intent of this comparison is to place the results of our simulations in the context of the observations rather than to suggest that these models, with their many approximations, are a complete representation of the exploding star. 
Our ZAMS masses in Figure~\ref{Explosion_Energy_Comparisons} and \ref{Ni56_comparisons}  pertain to the ZAMS of the \citet{WoHe07} progenitors and the structure associated with these may not reflect the pre-explosion structure of actual stars.
\citet{Smar15} has recently compared progenitor masses determined from the \citet{WoHe07} models to those determined from the commonly used models of \citet{ElTo04}, and found a shift to values 1--2~\msun\ larger over the mass range considered here when using the \citet{WoHe07} models. 

The explosion energy of B12-WH07 is somewhat weaker than those inferred from the six observed supernovae.
This is also true of B20-WH07 and B25-WH07, but the explosion energies of these latter two models are still increasing at a substantial rate and the current \Ediagovrec\ should be regarded as a low estimate.
The explosion energy of B15-WH07 is already close to the canonical 1~B and is still increasing at a rate of 0.15~\Bethes.
It is certainly encouraging that our \emph{ab initio} models are exhibiting explosion energies that are within the range of those inferred from observed supernovae.

In Figure~\ref{Ni56_comparisons} we compare the ejected \isotope{Ni}{56} masses, \Mni, of 0.035, 0.077, 0.065, and 0.074~\msun, respectively, for B12-WH07, B15-WH07, B20-WH07, and B25-WH07 with the seven observed \Mni\ estimates for the supernovae described above.
The \isotope{Ni}{56} masses produced by our models agree remarkably well with the masses inferred from observed supernovae.
Ejected \Mni\ in our models were obtained by summing the \isotope{Ni}{56} mass in all unbound zones, neglecting the possibility that some of this matter will become bound (and fall back) as the result of subsequent interactions in the stellar envelope. 
The \isotope{Ni}{56} mass fractions for zones that were never in NSE were obtained by following their nuclear evolution with \chimera's included \alp-network, as described in Section~\ref{sec:methods}. 
For zones that were once in NSE, but then cooled to temperatures below the criterion for NSE, these mass fractions were obtained by `deflashing' these zones, computing NSE abundances compatible with \chimera's nuclear network, and following their subsequent nuclear evolution with the network. 
\alp-networks are somewhat limited in their ability to accurately follow the recombination of \isotope{Ni}{56}\ \citep{TiHoWo00}, a shortcoming which we will partially address in a forthcoming post-processing analysis of tracer particles evolved with these models (J.~A. Harris et al., in prep.). 
A better solution, to use a larger, more realistic nuclear network, is also being pursued with \chimera\ \citep{ChMeHi12}.  
Such a larger network is particularly important for following nucleosynthesis processing during \alp-rich freezeout and when the electron fraction differs significantly from 0.50, the value assumed within an \alp-network.

\section{Discussion}
\label{sec:discussion}

There have been a wide range of CCSN simulations computed in the last decade, using a wide variety of parameters, methods, input physics, resolutions, approximations, and codes.
We discuss here the simulations that are the most similar in methodology to the \chimera\ models, their similarities and differences compared to our results, and some of the issues in making such comparisons.

\subsection{Other Axisymmetric Simulations}
\label{sec:Models_CCSNe}

In this section we compare the results of CCSNe simulations from different groups.
At this stage these comparisons are between simulations with multiple significant differences in the included physics (e.g., Newtonian versus general relativistic gravity, single-flavor versus three-flavor neutrino transport, etc.), numerical methods, or computational approach (e.g., ray-by-ray versus multidimensional neutrino transport).
It is hoped that the compendium of results from different groups, present and future, will ultimately point us to the essential ingredients, physical and numerical, needed to accurately model core collapse supernovae. Once a convergence of results from different groups is achieved, it should be clear whether or not the neutrino reheating mechanism is a viable explosion mechanism, at least from a theoretical standpoint. 
The ultimate test will be to compare the outcomes of CCSN models to the observables of the next galactic CCSN.

The most similar simulations to ours are those using the \vertex\ codes, both of which include a complete set of neutrino opacities and velocity-dependent transport with energy-group coupling in the ray-by-ray approximation. 
A significant difference between \chimera\ and the \vertex\ codes is that \chimera\ uses ray-by-ray MGFLD the while the Garching codes uses ray-by-ray two-moment plus variable Eddington factor, which is formally better.
Another difference is our use of the revised Cooperstein EoS in the region between the neutrinosphere and the shock.
The \prometheusvertex\ code \citep{RaJa02,MaDiJa06}, with Newtonian hydrodynamics and the same psuedo-Newtonian (spherical GR corrections to Newtonian) gravity treatment implemented in \chimera, was used by \citet{BuRaJa06, BuJaRa06} and \citet{MaJa09} to evolve, in axisymmetry, the 15~\msun\ progenitor s15s7b2 of \citet{WoWe95}. 
\citet{MuJaMa12} also evolved s15s7b2 using the \coconutvertex\ code \citep{MuJaDi10}, which employes the conformal flatness GR approximation rather than a pseudo-Newtonian approximation to the hydrodynamics, gravity, and ray-by-ray transport.
These simulations found that explosions of s15s7b2 were marginal in that some simulations led to explosions, while others with slightly different input physics or numerics did not.

The simulations of this progenitor reported by \citet{BuRaJa06,BuJaRa06} were of rather low resolution (32 or 64 angular zones) and did not incorporate a full 180-degree grid in latitude, which excludes the development of the low order ($\ell = 1, 2$) convective and SASI modes, and did not lead to explosions. 
\citet{MaJa09} employed a full 180-degree grid and finer angular resolution (128 and 192 angular zones).
Their simulations produced an explosion rather late ($\sim$570~ms after bounce) for their rotating case, but did not explode by 670~ms after bounce when restarted at 420~ms after bounce with finer radial zoning. 
The two non-rotating models evolved by \citet{MaJa09} with different \eos s did not explode before termination at $\sim$400~ms. 
On the other hand, a vigorous explosion was obtained by $\sim$400~ms after bounce for this progenitor by \citet{MuJaMa12} with \coconutvertex. 
A diagnostic energy of 0.13~B was achieved by 770~ms after bounce increasing at 0.7~\Bethes.
They attributed the more robust explosion of s15s7b2 to a more realistic GR model, which gave rise to more luminous neutrino emission with larger RMS energies and more vigorous SASI activity.

The 11.2 \msun\ progenitor of \citet{WoHeWe02} has also been investigated with \vertex\ in axisymmetry.
While no explosion was obtained in simulations with a restricted angular grid \citep{BuRaJa03}, an explosion commenced at 225~ms after bounce when a full 180-degree angular grid was employed \citep{BuJaRa06}, and when continued to 300~ms after bounce reached a diagnostic energy of 0.025~B, increasing at the rate of 1~\Bethes\ \citep{MaJa09}.

To better understand the differences in the results of the \chimera\ simulations reported here and \vertex\ simulations, we compare in more detail some of the radiation transport results of our model B15-WH07 and the model M15 described in \citet{MaJa09} and \citet{MuJaMa12} and evolved with the \prometheusvertex\ code and initiated from the S15s7b2 of \citet{WoWe95}.
As a function of radius, S15s7b2 is denser near the center by almost a factor of 2 and less dense by a factor of a few beginning near the S15s7b2 Si/O+Si shell at about $1.8 \times 10^{3}$~km and extending to the 15~\msun\ \citet{WoHe07} O+Si/O shell at about $4 \times 10^{3}$~km.
Beyond that the S15s7b2 density once again becomes substantially less than that of the \citet{WoHe07} 15~\msun\ progenitor used for B15-WH07.
These differences will, in particular, cause differences in accretion rates as a function of time between the two models.
Notwithstanding these progenitor differences, we will compare some of the important radiation-hydrodynamics results between M15 and B15-WH07 in an attempt to elucidate some of the factors that could account for our different outcomes.

Comparing Figure~2 of \citet{MuJaMa12} with our Figures~\ref{fig:early}c and~\ref{fig:NS_Radius} shows significant differences in the time dependence of the shock radii, but strong similarities in the time dependence of the proto-NS radii. 
The proto-NS radii decline from a maximum soon after bounce of about 75~km (\prometheusvertex) and 82~km (\chimera) to about 30~km at 500~ms after bounce (both codes).
The shock radius of the \citet{MuJaMa12} M15 plateaus at $\sim$150~km from 80--150~ms. then peaks at $\sim$200~km from 150 to 230~ms, retreats, and is finally revived at $\sim$430~ms. 
The shock radius of B15-WH07 plateaus briefly at 200~km from 90 to 110~ms then begins to slowly increase in radius after that, accelerating outward from 200 to 300~ms. 
Part, but certainly not all of the differences between the initial shock trajectories may be attributed to our use of the revised Cooperstein EoS between $10^{11}$~\gcc\ and the NSE--nonNSE boundary, as described in Section~\ref{sec:methods}.
Numerical experiments have shown that using the revised BCK EoS in that region instead of the Lattimer-Swesty EoS causes the shock to initially plateau $\sim$10~km farther out in radius.

Comparing neutrino luminosities and mean energies as plotted in our Figure~\ref{fig:Lum_Mean_RMS} with Figure~14 of \citet{MaJa09} for their 15~\msun\ rotating model (they do not provide such plots for their non-rotating model M15) we choose 200~ms at which time the total luminosity of their 15~\msun\ rotating model, also plotted in their Figure~14, becomes nearly equal to that of their model M15.
At this time the \chimera\ \nue\ and \nuebar\ mean energies are both 14.5~MeV, and nearly equal to those of \prometheusvertex. The \chimera\ luminosities are considerably higher, however, 54~\Bethes\ for both \Lnue\ and \Lnuebar\ versus 34~\Bethes\ for the \prometheusvertex.
A small downward correction to the \chimera\ luminosities and mean energies of approximately 6\% and 3\%, respectively, are required as the \chimera\ luminosities and mean energies are relative to an observer comoving with the fluid which, at 1000~km where these quantities are evaluated, the fluid velocity is radially inward at $\sim$0.03~c.

The above differences in the neutrino luminosities together with \chimera's use of the revised Cooperstein EoS between $10^{11}$~\gcc\ and the NSE--nonNSE boundary might account for the fact that the \prometheusvertex\ explosions are marginal while ours are not.
This could also account for the differences in the heating rates, advection time to heating time ratios, gain region masses and energies, and other radiation-hydrodynamic quantities discussed earlier that are more favorable for producing explosions in the \chimera\ simulations than in those of \prometheusvertex.

\citet{DoBuZh15} have evolved the same four \citet{WoHe07} progenitors examined in this paper to $\sim$600~ms after bounce in axisymmetry using the \castro\ code \citep{AlBeBe10, ZhHoAl11, ZhHoAl13} and did not obtain any explosions.
\castro\ is quite distinct in that it has multidimensional, rather than ray-by-ray, MGFLD, but neglects GR and the energy coupling between neutrino energy groups (i.e., inelastic scattering), both of which improve the prospects for an explosion \citep{MuJaMa12, LeMeMe12b}.
\castro\ is also the only code using a Cartesian mesh, which requires that they implement adaptive mesh refinement to adequately resolve the proto-NS.
\citet{DoBuZh15} argue that the large variations in the neutrino field that can arise along different rays in a ray-by-ray approach are coupled to hydrodynamic structures along each ray and can enhance the prospect of explosions (see Section~\ref{sec:rbr}).
In their simulations, \citet{DoBuZh15} used the STOS \citep{ShToOy98, ShToOy98b} \eos, which \citet{Couc13a} and \citet{SuTaKo13} have found makes explosions more difficult to achieve.

A fourth group has used a version of the Zeus-MP code \citep{HaNoFi06,IwKoOh08,IwKoOh09} for hydrodynamics and the IDSA scheme for neutrino transport \citep{LiWhFi09}, which we will refer to as `\zidsa.'
\zidsa\ is fully Newtonian and the transport omits inelastic (energy exchanging) scattering in favor of the elastic versions.
Early versions neglected \numt\ and \numtbar\ completely, but some recent works \citep{TaKoSu14,NaTaKu14} have included a leakage scheme for these species.
\citet{SuKoTa10} obtained a $\sim$0.1~B explosion, growing at $\sim$0.2--0.3~\Bethes, for a rotating 13~\msun\  \citep{NoHa88} progenitor, and a weaker explosion for a non-rotating simulation using the same progenitor.
\citet{SuTaKo13} computed simulations of two progenitors comparing the LS \citep{LaSw91} and STOS \eos s with both progenitors.
Through 600~ms of post-bounce evolution with the 15~\msun\ \citet{WoWe95} progenitor, they found no explosion with the LS \eos\ and characterized the explosion with the STOS \eos\ as weak, but did not quote an explosion energy.
Explosions of about 0.1~B were obtained from the 11.2~\msun\ \citet{WoHeWe02} progenitor using both \eos s, but the explosion using the STOS \eos\ was weaker and slightly more delayed.
\citet{TaKoSu14} obtained explosions, while examining stochastic variation and resolution effects at angular resolutions of 64, 128, and 256, from the 11.2~\msun\ \citet{WoHeWe02} progenitor with a \numt-\numtbar\ leakage scheme added.
The diagnostic energies attained by their models when the simulations were terminated were low (0.05--0.13~B for the axisymmetric simulations), but the simulations were terminated  300--570~ms after bounce, before the explosions were fully developed.

Two recent studies using the \zidsa\ code have modeled a large number of progenitors with 128 latitudinal zones and 384 radial zones covering the inner 5000~km of the progenitor structure.
\citet{SuYaTa14} computed simulations for nine progenitors from \citet{WoHe07} including the 12, 15, and 20~\msun\ progenitors used herein and several additional 15~\msun\ progenitors from other authors.
Among the progenitors matching our set, they found explosions for only the 12~\msun progenitor, which had a diagnostic energy of 0.1~B at 600~ms after bounce and little sign of growth.
They also obtained explosions by $\sim$400--500~ms after bounce for the 55 and 80~\msun\ progenitors, which had energies of 0.15~B at 800~ms after bounce that grew $\sim$0.2--0.3~\Bethes, as well as a late explosion for the 40~\msun\ progenitor more than 1 second after bounce.
\citet{NaTaKu14} have simulated all 101 solar metallicity progenitors (10.8--75~\msun) from \citet{WoHeWe02} for up to 1.5 seconds after bounce, or until the shock reached the outer boundary, and obtained explosions for 88 of 101 progenitors.
Of the remaining simulations, one shock did not reach the boundary and 12 models experienced failures when the simulation left the range of the \eos.
They found explosions with energies of $\sim$0.2--0.6~B, correlated with increasing values of the compactness parameter introduced by \citet{OcOt11} to measure the compactness of the stellar core at or near bounce.
While this compactness parameter roughly correlates with progenitor mass, there are non-monotonic deviations in progenitor internal structure that have been seen to affect the outcomes of parameterized 1D simulations of the same 101 progenitors \citep{UgJaMa12}, and core compactness is not an observable parameter that can be translated into observational comparisons like those in Figure~\ref{Explosion_Energy_Comparisons}.

Many of the exploding simulations from other groups have not met the observed values depicted in Figure~\ref{Explosion_Energy_Comparisons} --- often several times less energetic than our simulations.
\textit{For most of these `underpowered' explosion models a better characterization would be that they are `incomplete' or `unfinished,' as several of these models had \Ediag\ growth rates that would have reached observed energies if maintained for another half to full second.}
Our simulations shed light on the requirement to complete the neutrino-driven phase in order to more precisely determine the explosion energy.

If we consider a flat \Ediag\ curve as an indicator of the completeness of simulations, we can see that B12-WH07 is quite flat from 600~ms after bounce and may be considered `done' by this measure.
\Ediag\ for B15-WH07 and B20-WH07 may be sufficiently slowly growing beyond about 800~ms to be considered `done,' but B25-WH07 has not met this criterion yet.
However, if energy were to stop being injected into the unbound region, we should expect \Ediag\ to decline as it works to unbind the overlying material.
Only B12-WH07 unambiguously exhibits this feature.

The energy measure \Ediagovrec\ includes the binding energy of the overlying material.
\Ediagovrec\ should represent the final energy state of the explosion when \Ediagovrec\ stops growing.
For B12-WH07, and perhaps B15-WH07, \Ediagovrec\ is slowly converging with \Ediag, but for B20-WH07 and B25-WH07 this is clearly not yet the case.
From the continuing growth of \Evol{300} (total energy in the simulation above 300~km radius, see Section~\ref{sec:volumeanalysis}) it would also appear that neither B15-WH07, nor B20-WH07, nor B25-WH07, have completed the development of explosion energy.
However, it does appear that 1400~ms of post-bounce evolution (about 1200~ms post-shock-revival) is adequate for B12-WH07 to saturate this energy measure as well.

The continuing growth of \Ediag\ for B25-WH07 and the growth of \Ediagovrec\ and \Evol{300} for B15-WH07, B20-WH07 and B25-WH07 indicate that these explosions are not yet complete.
For B15-WH07, it might have completed by $\sim$1500~ms after bounce if the outer boundary had contained more of the progenitor, but for B20-WH07 and B25-WH07 it is not unreasonable to expect that 2 or more seconds of post-bounce evolution may be required to settle on a final explosion energy.
In all cases, the time to fully develop the explosion is significantly longer than the time required to revive the shock after core bounce, and we should withhold judgement on the `understrength' of simulated explosions that have not reached a final explosion energy.

\subsection{Early Characteristics of Explosions}

We discussed in \citetalias{BrMeHi13}, and in Section~\ref{sec:Expl_onset}, the striking feature of our models that they all have nearly the same early shock radius evolutions.
To account for the similarity in the shock trajectories up to about 100~ms after bounce we modeled in \citetalias{BrMeHi13} the shock radius during the shock stagnation phase using the shock stagnation model of \citet{Jank12} and found that the modeled radii were very similar for all of our simulations despite much larger variations in the input quantities (proto-NS radius, mass accretion rate, etc.), reflecting the self-regulation operative in accretion.
It also demonstrated a self-consistency in our results, as the similarity found in the modeled shock radii matches the similarity found in the shock radii extracted directly from the simulations at the same epoch.
The radius in this shock stagnation model peaks at about 100~ms after bounce when evaluated for all four of our simulations and then begins to decline.
This is also the epoch where 2D simulations diverge from their 1D counterparts, as convection and SASI activity drive expansion of the shock in 2D simulations while 1D simulations contract toward failure.
The flattening of our simulation shock trajectories at about 100~ms can be seen in Figure~\ref{fig:early}c, where it is also evident the shock expansion due to multidimensional effects remains similar in all simulations through at least 150~ms after bounce.
We suspect that the similarity in the shock trajectories after 100~ms after bounce is most likely due to the fact that our explosions are not marginal.
Once these multidimensional effects set in, they rapidly establish the conditions for an explosion, which are similar for all models.
Had our explosions been marginal, small differences in the models would have led to large differences in the evolution, some exploding rather quickly, others taking many 100~ms, and still other not exploding at all.

As we noted in \citetalias{BrMeHi13}, the mean shock radius exceeds 500~km (one commonly used explosion indicator) at $\sim$210~ms after bounce for B20-WH07 and B25-WH07, and at $\sim$235~ms after bounce for B12-WH07 and B15-WH07.
Most of the other indicators and potential drivers of explosion are moving smoothly toward a more explosion-favorable state.
The ratio of the advective and heating timescales (Figure~\ref{B12_B25_Heating}c) exceeds unity for all simulations, and rises in a similar manner toward conditions favorable for shock revival.
The mass in the gain region (Figure~\ref{B12_B25_Gain}a) rises slowly, leading up to explosion, and then rises rapidly in response to the engulfment of mass by the rapidly expanding shock.
The total energy in the gain region (Figure~\ref{B12_B25_Gain}b) rises steadily through the initiation of explosion and, like the preceeding indicators, shows no abrupt changes that would indicate an explosion had been suddenly ``triggered.''

A common pattern for the trajectory of the mean shock in axisymmetric models, including ours, is a modest flattening of the mean shock trajectory, followed by a slow expansion driven by multidimensional effects, sometimes with strong oscillations in the shock radius, that then accelerates as the shock is relaunched.
The successful explosions in \citet{BuJaRa06,MaJa09,SuKoTa10,MuJaMa12,MuJaHe12,SuKoTa10,SuTaKo13,SuYaTa14}; and \citet{NaTaKu14} all exhibit this pattern.
Some axisymmetric simulations \citep[][F.~Hanke, in prep.]{HaMuWo13} with \prometheusvertex\ have shown contraction of the mean shock on a 1D-like trajectory before later revival.
In these `retreating shock' simulations and some others \citep[e.g.,][]{MaJa09,SuYaTa14} the eventual shock revival has been attributed to a decrease in ram pressure at the shock from a sudden drop in the accretion rate due to features in the stellar structure of the progenitor.
In contrast, the relatively smooth and continuous expansion of the shock radius in our simulations does not appear to require a sudden decrease in accretion ram pressure to trigger an explosion.
That the rapid increase in shock radius due to the sudden decrease in accretion rates near certain composition boundaries, namely the Si/O+Si interface, ``trigger'' the explosions in some simulations by other groups may be due to the marginality of their explosions and the large density decrements in the structure of their progenitors. 
These \chimera\ models do not exhibit this trigger because their explosions are more robust and because drops in the accretion rates occur less abruptly, and at later times, in these simulations due to the structure of the \citet{WoHe07} progenitors. 
Unlike progenitor progenitor model S15s7b2 of \citet{WoWe95} or S11.2 of \citet{WoHeWe02}, the density drops at the Si/O+Si interface of our progenitors are smaller and occur at larger radii.
The shock has already been revived by the time they encounter these drops in accretion rates.  

Through 200~ms after bounce, our simulations have a smooth decline in the accretion rate at the shock.
For the two lighter progenitor models, this smooth decline continues through shock revival, but the two more-massive progenitor models experience a two-fold decrease in accretion rate ahead of the shock over a period of 20~ms, which is about five-fold faster than the previous halving of the accretion rate seen in the two less-massive progenitor models.
For B20-WH07 this accretion rate drop begins at 200~ms and for B25-WH07 at 220~ms after bounce.
Though this rapid decline seems related to the onset of explosion at 210~ms, examination of the mean shock radius (see Figure~2 in \citetalias{BrMeHi13}) shows no corresponding change in the already rapid shock expansion.
It thus appears that the sudden, rapid drop in accretion rate is not the driver of explosion for any of our simulations, though the smooth decline in accretion that proceeds the rapid drop may be an important aspect in allowing the explosions to develop.

\subsection{The Ray-by-Ray Approximation}
\label{sec:rbr}

\begin{figure}
\includegraphics[width=\columnwidth,clip]{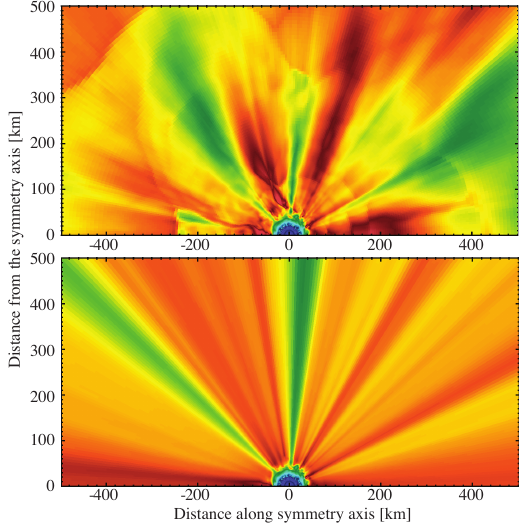}
\caption{\label{B12-lum-image}
Comparison of \nue\ `luminosity,' $4\pi r^2 F(r,\theta)$, spatial distribution for model B12-WH07 at 262~ms after bounce (upper panel) and a stationary solution using the 262~ms hydrodynamic profile (lower panel). The color scale ranges from 0~\Bethes\ (blue) to 60~\Bethes\ (red) for both images. }
\end{figure}

Three of the four codes discussed above use the ray-by-ray approximation and have achieved explosions in at least some 2D simulations.
The fourth code (\castro) has multi-dimensional transport and does not use the ray-by-ray approximation.
\citet{DoBuZh15} concluded, from the lack of explosions in their simulations of the same \citet{WoHe07} progenitors used herein, that the ray-by-ray approximation may be enhancing the explosions in the simulations of the other groups by artificially correlating neutrino emission to the overall shock and hot bubble features.
\citet{SuKoTa10}, in contrast, speculated the opposite --- that the ray-by-ray approximation might be inhibiting strong explosions.
Here we examine certain aspects of the ray-by-ray approximation, noting, however, that a full examination will have to wait until a single code can compute the full dynamic evolution of the supernova engine with the ray-by-ray approximation and with otherwise identical neutrino transport that does not make the ray-by-ray approximation.

Recently, \citet{SuTaMa14} have reported on the development of a multi-angle, multifrequency, 3D neutrino transport solver.
Using this solver, they computed the stationary-state neutrino radiation field given a background taken from a 3D supernova simulation and compared it to a stationary state radiation field computed on the same background using a ray-based version of their solver. 
\citet{DoBuZh15} performed a similar study with \castro.
\citet{SuTaMa14} and \citet{DoBuZh15} express concern that the ray-by-ray approximation could alter the triggering of explosions by overheating in regions above the most intense neutrino emissivity.
\citet{SuTaMa14} found that deviations in the heating rate were somewhat smaller than the luminosity deviations, and the largest deviations were found above the impact points of the accretions streams onto the proto-NS.
As this excess heating in the ray-by-ray approximation occurs in strong (often supersonic) downflows, the impact on the overall explosion trigger and strength is likely reduced.
We have computed the stationary state neutrino radiation field in the stellar background taken from our B12-WH07 model at 262~ms after bounce.
Our stationary state (Figure~\ref{B12-lum-image}; lower panel) resembles those reported by the other authors, with radially oriented `streaks' of neutrinos emerging from the proto-NS, where they decouple at about 50~km.
We can contrast this `streaky' field pattern with the actual \nue-luminosity field in our simulation at 262~ms after bounce (Figure~\ref{B12-lum-image}; upper panel).\footnote{This snapshot contains features from the velocity-dependent observer corrections missing from the stationary radiation field in the lower panel. In \chimera\ the velocity used in those terms is computed from the motion of the grid during the Lagrangian hydrodynamic step that is bypassed in the static transport test and the effect washes out as the radiation field converges.}
We can clearly see that the angular variation varies strongly with radius, which in turn indicates there is rapid temporal variability in the neutrinos emerging from the decoupling region, and \textit{a priori} rapid variability in the hydrodynamic and thermodynamic conditions in this region.
A computation of the stationary state radiation field for a \textit{given} background will assume that local hydrodynamic and thermodynamic conditions are frozen in, particularly extremal conditions such as hot spots at the bases of accretion plumes.
This will in turn exaggerate the relative angular variations of the neutrino radiation field determined by ray-by-ray and non--ray-by-ray approaches.

The ray-by-ray approach has permitted simulation groups, including ours, to build simulations that include much of the neutrino physics, such as observer corrections in the radial direction and non-iso-energetic neutrino scattering, in 2D and 3D models, but at a price in fidelity that is not yet fully known.
To connect ray-by-ray and non--ray-by-ray simulations, and to use both effectively toward understanding the supernova mechanism and its observational consequences, full comparison simulations with each scheme must be performed \emph{with the same physics}, so that the ray-by-ray approximation can be better assessed.
If gauged to be a good approximation, the ray-by-ray approximation can serve us well, as the community moves toward simulations with fully multidimensional transport.

\section{Summary}
\label{sec:summary}

In this paper, we present results for four non-rotating axisymmetric core-collapse supernova simulations, designated B12-WH07, B15-WH07, B20-WH07, B25-WH07, initialized from progenitors with zero-age main sequence masses of 12, 15, 20, and 25~\msun. 
These models, evolved from the main-sequence to the onset of core-collapse by \citet{WoHe07}, were continued by us through core collapse, bounce, and shock revival to explosion with our \chimera\ code. 
Our simulations were carried out to 1400~ms after bounce, a long time for such simulations, allowing the explosions to more fully develop.
(Model B15-WH07 had to be terminated at 1200~ms after bounce when the shock reached the edge of the grid at 20,000~km.)
Some of the important observable outcomes of the simulations are summarized in Table \ref{tab:outcomes}.
Because of the length of the simulations, we were able to follow the development of the explosion energies of our models to the point where the explosion energy had nearly saturated for model B12-WH07, and is increasing at a much reduced rate relative to the first few hundred ms of post-bounce growth for models B15-WH07 and B20-WH07.
Model B25-WH07 clearly requires longer simulation time to finalize its explosion energy and other observable measures.

By $\sim$70~ms after bounce, distinct heating and cooling layers have been established behind the stagnated shock in all models, and conditions become favorable for entropy-driven convection in the heating layer. 
In models from more massive progenitors, greater neutrino heating, favoring convective growth, is more than compensated for by the reduced advection time of material through the heating layer, which suppresses convective growth, thus convection takes longer to become fully developed. 
By $\sim$100~ms after bounce, convection and the SASI push the shock outward relative to 1D counterparts.
These instabilities cause the shock to undergo dipole and quadrupole distortions, the heating layer to expand, and the residency time for material in the heating layer to increase. 
Increased \nue\ and \nuebar\ RMS energies from the shrinking neutrinospheres and expansion of the gain region more than compensate for the density decline in the gain region, therefore the neutrino heating efficiencies slowly increase during the build up to explosion. 
Most importantly, the ratio of the advection time scale to the heating time scale in the gain region increases, exceeding unity by $\sim$100~ms and 3 by $\sim$150~ms after bounce in each of our models, signaling the potential for thermal runaway.
The mass of the heating layer, which initially decreases after formation of the gain region, grows as convection and the SASI push the shock out, providing another indication that thermal runaway is imminent.
The mean, maximum, and minimum shock radii for all models grow rapidly starting at $\sim$200~ms after bounce, indicating that explosions are developing.
With the exception of model B20-WH07, all models develop a distinctly prolate shock shape as their explosions develop.
Model B20-WH07 develops an off-center spherical shock shape as the explosion develops, and only later, at $\sim$800~ms after bounce, does the shock become significantly prolate.

\begin{deluxetable*}{lccccc}
\tabletypesize{\scriptsize}
\tablecaption{Summary of Simulation Outcomes\label{tab:outcomes}}
\tablecolumns{6}
\tablewidth{0pt}
\tablehead{
\colhead{} & \multicolumn{5}{c}{Models} \\
\cline{2-6} \\
\colhead{Outcome} & \colhead{B12-WH07} & \colhead{B15-WH07} & \colhead{B20-WH07} & \colhead{B25-WH07} 
}
\startdata
Diagnostic Energy (\Ediag) [B] &  0.370 & 1.056 & 0.848 & 1.536 \\
Diagnostic + Overburden Energy (\Ediagov) [B] & 0.312 & 0.880 & 0.373 & 0.696 \\
Diagnostic + Overburden + Recom Energy (\Ediagovrec) [B] & 0.314 & 0.883 & 0.375 & 0.703 \\
Ejected \isotope{Ni}{56} Masses (\Mni) [\msun] & 0.035 & 0.077 & 0.065 & 0.074 \\
Proto-NS Baryonic Rest Mass ($M_{\rm bary}$) [\msun] & 1.481 & 1.676 & 1.806 & 1.898 \\
Proto-NS Gravitational Mass ($M_{\rm grav}$) [\msun] & 1.345 & 1.506 & 1.611 & 1.685 \\
Proto-NS Velocity Magnitude [\kms] &  124 & 205 & 150 & 516
\enddata
\end{deluxetable*}

About 150--175~ms after bounce, the first unbound zones, those with positive total energy, appear.
The diagnostic energy \Ediag, the total energy of the unbound region, remains small until $\sim$200~ms.
The total energy \Ediagovrec, which includes the binding energy of matter overlying the unbound region, becomes positive at 370, 400, 530, and 660~ms after bounce for models B12-WH07, B15-WH07, B20-WH07, and B25-WH07, respectively.
At the time of this report, the best estimate of the explosion energy, \Ediagovrec, is 0.34, 0.88, 0.38, and 0.70~B, and increasing at 0.03, 0.15, 0.19, and 0.52~\Bethes, respectively, for models B12-WH07, B15-WH07, B20-WH07, and B25-WH07.

We have analyzed energy flows in a region \Rsph{300} from 300~km to the outer simulation boundary that approximately contains the unbound material and lies above the central neutrino engine.
We find that much ($\sim$60\%) of the growth of the explosion energy arises from the enthalpy influx upward through the lower boundary --- the advection of neutrino heated and dissociated material into \Rsph{300} from below --- and that this dominance of the enthalpy influx is maintained when the lower boundary of  \Rsph{300} is varied between 200 and 350~km.
We find only a relatively modest amount of the explosion energy comes from direct neutrino heating of already unbound material, or by the influx of kinetic energy.
The nuclear energy release is also relatively modest, and occurs by nucleon and \alp-particle recombination, silicon burning, and oxygen burning, in sequence, as the shock propagates first through the silicon layers and then through the oxygen-rich layers of the progenitor.

The differences in model B20-WH07 relative to the others help illustrate the effects of variations in the morphology and the importance of streamed accretion from the shock to the proto-NS on the growth of the explosion energy.
In B20-WH07, outward radial velocities behind the entire shock at revival cut off the accretion stream to the proto-NS, resulting in a decline in accretion powered \nue\ and \nuebar\ luminosities and a corresponding reduction in the heating at the bottom of the gain region.
As a consequence, the diagnostic energy of B20-WH07 increases more slowly during the 250--450~ms post-bounce epoch than our other simulations, and the trend of increasing diagnostic energy with progenitor mass is broken.

We have compiled a set of observed CCSNe with estimated explosion energies and \isotope{Ni}{56} masses for which a progenitor mass have been estimated within the 12--25~\msun\ range simulated herein.
The explosion energies of our models B15-WH07 and B25-WH07 are comparable, within the large observational errors, to the inferred energies of these observed supernovae, while the explosion energies of the other two models are somewhat smaller and their energies are still growing. 
The ejected \isotope{Ni}{56} masses from our models compare quite well with those inferred from the observed CCSN sample.

The proto-NS baryonic rest masses at the end of our simulations are 1.48, 1.68, 1.81, and 1.90~\msun, respectively, for models B12-WH07, B15-WH07, B20-WH07, and B25-WH07.
With the exception of model B12-WH07, the proto-NSs are still accelerating non-negligibly at the end of our simulations.
We infer a final velocity of the proto-NS for model B12-WH07 of $\sim$100~\kms, and estimate the final proto-NS velocities for B15-WH07 and B20-WH07 of $\sim$100--200~\kms\ and B25-WH07 approximately $\sim$500~\kms.

These axisymmetric simulations represent significant progress toward understanding the CCSN explosion mechanism in several important respects:
\begin{enumerate}
\item We obtain robust explosions across a broad range of progenitor masses (12--25~\msun), not just for lower-mass massive stars.
\item An analysis of the energy sources and fluxes powering the CCSNe elucidates the contribution of direct neutrino and nuclear heating, kinetic energy influx, and enthalpy influx, and demonstrated the dominance of the latter. 
\item Our explosion energies, \isotope{Ni}{56} ejecta masses, proto-NS masses, and NS kick velocities, summarized in Table~\ref{tab:outcomes}, are all within range of observations.
\item We demonstrate that assessing observable quantities requires the end of significant accretion onto the proto-NS and thus lengthly simulations that increase in simulated post-bounce time with progenitor mass.
\end{enumerate}

Further analyses from these simulations of detailed ejecta nucleosynthesis (J.~A. Harris et al., in prep.) and of multi-messenger signals from gravitational waves \citep{YaMeMa15} and neutrinos (O.~E.~B. Messer et al., in prep.) are forthcoming.

Finally, it must be kept in mind that these simulations are not definitive, but are limited by the approximations inherent in the \chimera\ code used for these simulations. 
Further important improvements in numerical realism include removing the restriction of axial symmetry and moving to 3D, as has been initiated by \citet{HaMuWo13} using \vertex, \citet{LeBrHi15} using \chimera, and \citet{TaKoSu12,TaKoSu14} using \zidsa; performing these simulations in full general relativity rather than hybrid approximations; employing multi-dimensional transport instead of ray-by-ray; and better approximations to Boltzmann transport.
Resolution studies need to be performed in order to assure numerical convergence.

Future simulations that capture fully developed explosions must be of similar duration and the computed domain must be of adequate size to contain the shock throughout, as illustrated by the premature termination of our B15-WH07 simulation.
We can estimate that the time for the explosion to fully develop, which generally increases with progenitor mass, ranges from 1 second to 2 or more seconds after shock revival.
The dearth of simulations carried out this long restricts comparisons across groups and accurate conclusions regarding the robustness of explosions obtained. 

\acknowledgements

We thank Christian Cardall for a careful reading of the manuscript.
We also wish to thank members of the CCSN modeling community for valuable discussions about these models over the past 2 years, including Ernazar Abdikamalov, Sean Couch, Josh Dolence, Thomas Janka, Bernhard M\"uller, Evan O'Connor, Christian Ott, Kohsuke Sumiyoshi, and Yudai Suwa.
This research was supported by the U.S. Department of Energy Office of Nuclear Physics; the NASA Astrophysics Theory and Fundamental Physics Program (grants NNH08AH71I and NNH11AQ72I); and the National Science Foundation PetaApps Program (grants  OCI-0749242, OCI-0749204, and OCI-0749248). PM is supported by the National Science Foundation through its employee IR/D program. The opinions and conclusions expressed herein are those of the authors and do not represent the National Science Foundation.
The simulations here were performed via NSF TeraGrid resources provided by the National Institute for Computational Sciences under grant number TG-MCA08X010; resources of the National Energy Research Scientific Computing Center, supported by the U.S. DOE Office of Science under Contract No. DE-AC02-05CH11231; and an award of computer time from the Innovative and Novel Computational Impact on Theory and Experiment (INCITE) program at the Oak Ridge Leadership Computing Facility, supported by the  U.S. DOE Office of Science under Contract No. DE-AC05-00OR22725.

\begin{appendix}

\section{Thermal Energy}
\label{app:thermalE}

In classical thermodynamics, binding energies are not included in internal energy.
Supernova simulations have somewhat different requirements, in that the compression or decompression of the fluid can trigger a rearrangement of the nucleons into different nuclei and nuclear matter states.
The nuclear equations of state used for supernova simulations therefore include the binding energy of nuclei self-consistently in the `internal' energy and this internal energy is used when evolving the hydrodynamic equations or including neutrino heating and cooling.
For analysis purposes we need a more traditional, or thermal, form.
This avoids the possibility of negative internal energy in a fluid element, which would complicate the computation of heating time scales.
It is the thermal component that is available to drive buoyancy and expansion and keeping the nuclear binding energy out of the thermal energy permits the separate analysis of the contributions of nuclear burning, dissociation, and recombination on the thermodynamic state of the developing explosion.

The specific thermal energy \etherm\ includes an ideal nuclear gas, $\frac{3}{2} kT/\bar{A}$, a trapped photon gas, $a T^4/\rho$, and the internal energy of the electron--positron gas $\varepsilon_{e^+e^-}$ with the rest mass of the electrons from net charge excluded while including the mass--energy of the thermal electron--positron pairs
\begin{equation}
\etherm = \frac{3}{2} kT/\bar{A} + \frac{a T^4}{\rho} + \left( \varepsilon_{e^+e^-} - Y_e m_e\right),
\end{equation}
where $\varepsilon_{e^+e^-}$ is the total specific degenerate electron gas energy returned by the electron--positron EoS, $a$ is the radiation constant, $\bar{A}$ is the mean nuclear mass, and $m_e$ is the electron mass.
This `thermal' energy includes the degeneracy energy of the electron--positron gas, which while not technically thermal is available to do the same type of work, and the rest-mass energy of $e^+e^-$ pairs.

\section{Fixed Volume Energy Diagnostics}
\label{app:energy}

To consider the evolution of the diagnostic energy we consider the evolution of the total, \Etot, in a volume $V$.
We consider a simple Newtonian description of the system starting from the mass conservation equation, momentum equation, and the first law of thermodynamics
\begin{eqnarray}
  \frac{d \rho}{d t} & = & -\rho \nabla \cdot \uvec \label{eq:masscon}\\
  \rho \frac{d \uvec}{d t} & = & -\nabla p - \rho\nabla \Phi \label{eq:newton}\\
  \frac{d}{d t} \left( \frac{\edint}{\rho} \right)& = & -p \frac{d}{d t} \left( \frac{1}{\rho} \right) + \qdot \label{eq:firstlaw}
\end{eqnarray}
where the symbols have the usual meaning and $\edint = \rho \, \eint$ is the volumetric internal energy density.
(For any energy density, \eden{x}, it can be written in terms of the density and specific energy density $\eden{x} = \rho \, \specen{x}$.)
In Equation~(\ref{eq:firstlaw}) we have included the total internal energy as evolved inside the simulations including the nuclear binding energy, as such, the source term $\qdot = \qdotnu$ contains only the neutrino heating component as the nuclear transformation term \qdotn\ that converts between the components of \edint, \edth\ and \edbind.
To compute the evolution of the thermal energy we decompose \edint\ into thermal and binding energy components, $\edint = \edth + \edbind$.
We can write a first law for the thermal energy
\begin{equation}
  \frac{d}{d t} \left( \frac{\edth}{\rho} \right) =  -p \frac{d}{d t} \left( \frac{1}{\rho} \right) + \qdotn + \qdotnu \label{eq:firstlawth}
\end{equation}
in which $\qdotn = d \edbind/dt$ is a now source term.
Our equations are written as Lagragian derivatives, $d/dt \equiv \partial / \partial t + \uvec \cdot \nabla$, and we prefer an Eulerian description for conservation laws in specified volumes.
The kinetic energy equation is obtained by taking the dot-product of \uvec\ with Equation~(\ref{eq:newton})
\begin{equation}
  \pardirt{\edkin} + \divv{\fluxkin} = -\uvec \cdot \nabla p - \rho \uvec\cdot\nabla\Phi, \label{eq:ekinevol}
\end{equation}
where $\edkin = \frac{1}{2}\rho\uvec\cdot\uvec$ is the kinetic energy density and $\fluxkin = \edkin\uvec$ is the kinetic energy flux density.
(We will construct the flux of any energy density, \eden{x}, as $\fluxX{x}=\eden{x}\uvec$.)
Using the mass conservation equation (\ref{eq:masscon}) we can rewrite the first law (Equation~\ref{eq:firstlaw}) as the evolution of the internal energy
\begin{equation}
  \pardirt{\left(\edth + \edbind\right)} + \divv{\fluxenth} + \divv{\fluxbind} = \uvec \cdot \nabla p + \rho \qdotnu, \label{eq:eintevol}
\end{equation}
where $h = \edth + p$ is the enthalpy density and $\fluxenth = h \uvec$ is the enthalpy flux density.
The binding energy contribution to \edint\ is again kept separate from the enthalpy so that we may track the binding energy flux density separately.
(Equation~\ref{eq:eintevol} can be derived from Equation~\ref{eq:firstlaw} without the \edbind\ and \fluxbind\ terms and the \qdotn\ term include, but we do not have \qdotn\ directly available in the NSE regions of our simulations so we must simultaneously construct an extraction of integrated nuclear heating.)
We obtain the gravitational energy equation by multiplying the mass conservation equation (\ref{eq:masscon}) by the gravitational potential $\Phi$ and assuming $\partial \Phi /\partial t = 0$ for simplicity, but without loss of generality, as
\begin{equation}
  \pardirt{\edgrav} + \divv{\fluxgrav} = \rho \uvec \cdot \nabla\Phi, \label{eq:egravevol}
\end{equation}
where we have defined the gravitational potential energy density as $\edgrav = \rho \Phi$.

From Equations~(\ref{eq:ekinevol}--\ref{eq:egravevol}) we find terms for the exchange of energy between components.
The exchange of energy between kinetic and gravitational potential energy is given by the $\rho \uvec\cdot\nabla\Phi$ term in Equations~(\ref{eq:ekinevol}) and~(\ref{eq:egravevol}) with opposite signs.
Likewise, the $\uvec \cdot \nabla p$ term appearing with opposite signs in Equations~(\ref{eq:ekinevol}) and~(\ref{eq:eintevol}) is an exchange of energy between kinetic and internal energy.
In Equation~(\ref{eq:eintevol}) this term emanates from the \pdV\ term in Equation~(\ref{eq:firstlaw}), however, parts of the enthalpy flux, \fluxenth, also emanate from the \pdV\ term rendering the concept of \pdV\ work ambiguous in the context of conservation laws.

To compute the evolution of the total fluid energy, $\edtot = \edint + \edkin + \edgrav$ we sum Equations~(\ref{eq:ekinevol}--\ref{eq:egravevol})
\begin{equation}
  \pardirt{} \left( \edth + \edkin + \edgrav \right) + \pardirt{\edbind} + \divv{\left(\fluxenth + \fluxkin + \fluxgrav \right)} + \divv{\fluxbind} = \rho \qdotnu,
\end{equation}
which reduces to an exact conservation law in the absence of external sources, $\qdot = 0$, and the first term is the derivative of the total energy, \edtot, used in the diagnostic energy computation.

Integrating over a fixed volume $V$ bound by the surface \dSurf, we obtain
\begin{equation}
  \dotEvol{V} \equiv \intvol{\pardirt{\edtot}} = \intvol{\rho\qdotnu} - \intsurf{\fluxenth} - \intsurf{\fluxkin} - \intsurf{\fluxgrav} -\intvol{\pardirt{\edbind}} - \intsurf{\fluxbind} \label{eq:postprocess}
\end{equation}
for the evolution of the total fluid energy in a fixed volume. The last two terms are volume integral of \qdotn\ which we would obtain if we had used Equation~(\ref{eq:firstlawth}) in the derivation with the volume integral of \qdotn\ equal to
\begin{equation}
  \intvol{\rho\qdotn} = -\intvol{\pardirt{\edbind}} - \intsurf{\fluxbind}.
\end{equation}
In our post-processing analysis we use Equation~(\ref{eq:postprocess}) directly as we do not have \qdotn\ for regions in NSE.
It is more illuminating to use the volume integral of the nuclear transformation energy so we rewrite Equation~(\ref{eq:postprocess}) as
\begin{equation}
  \dotEvol{V} \equiv \intvol{\pardirt{\edtot}} = \intvol{\rho\qdotnu} + \intvol{\rho\qdotn} - \intsurf{\fluxenth} - \intsurf{\fluxkin} - \intsurf{\fluxgrav}., \label{eq:dvolenergydt}
\end{equation}

Integrating from time $t_0$ to time $t$ for the analysis in Section~\ref{sec:volumeanalysis}, we obtain
\begin{equation}
\dEvol{V}(t) = \Qnu(t) + \Qnuc(t) + \FluxCum{enth}{V}(t) + \FluxCum{kin}{V}(t) + \FluxCum{grav}{V}(t), \label{eq:volenergy} 
\end{equation}
where we define the terms on the right hand side as
\begin{eqnarray}
 \dEvol{V}(t) & \equiv & \int_{t_0}^t \dotEvol{V} = \int_{t_0}^t \intvol{\pardirt{\edtot}}, \\
 Q_x(t) & \equiv & \int_{t_0}^t \intvol{\rho \, \dot{q}_x}, \, \mbox{and} \\
 \FluxCum{x}{V}(t)  & \equiv & - \int_{t_0}^t \intsurf{\fluxX{x}} \label{eq:fluxcum}.
\end{eqnarray}
We have included the sign in \FluxCum{x}{V} to match the sign of each component to the sign of its contribution.

\end{appendix}

\bibliographystyle{apj}

\end{document}